\documentclass[useAMS,usenatbib,times,letter,amssymb]{mnras}
\usepackage{epsfig,times,amssymb,amsmath,verbatim,xspace}
\usepackage[usenames,dvipsnames,svgnames,table]{xcolor}
\usepackage{tablefootnote}
\usepackage{todonotes}
\usepackage{tabularx}
\usepackage{hyperref}
\usepackage{float}
\usepackage{hhline}
\usepackage{lineno}
\usepackage{natbib,hyperref,ifthen,soul} 

\usepackage{eso-pic}

\AddToShipoutPictureBG*{
  \AtPageUpperLeft{
    \hspace{0.75\paperwidth}
    \raisebox{-9.\baselineskip}{
      \makebox[0pt][l]{\textnormal{DES-2018-0346}}
}}}

\AddToShipoutPictureBG*{
  \AtPageUpperLeft{
    \hspace{0.75\paperwidth}
    \raisebox{-10.\baselineskip}{
      \makebox[0pt][l]{\textnormal{FERMILAB-PUB-18-622-AE}}
}}}

\newcommand{\metacal}{\textsc{metacalibration}}

\newcommand{\est}{e}

\newcommand{\mcalRmean}{\mbox{\boldmath $\langle R \rangle$}}

\newcommand{\mcalRSmean}{\mbox{\boldmath $\langle R_S \rangle$}}
\newcommand{\mcalRg}{\mbox{\boldmath $R_\gamma$}}
\newcommand{\mcalRS}{\mbox{\boldmath $R_S$}}

\newcommand{\omegam}{\ensuremath{\Omega_\mathrm{m}}}
\newcommand{\omegab}{\ensuremath{\Omega_\mathrm{b}}}

\newcommand{\omeganu}{\ensuremath{\Omega_{\nu}h^2}}

\newcommand{\as}{\ensuremath{A_\mathrm{s}}}
\newcommand{\ns}{\ensuremath{n_\mathrm{s}}}
\newcommand{\lcdm}{$\Lambda$CDM}
\newcommand{\wcdm}{$w$CDM}

\newcommand{\Tb}{\ensuremath{T_\mathrm{BPZ}}}
\newcommand{\aia}{\ensuremath{A_\mathrm{IA}}}
\newcommand{\bpz}{\blockfont{bpz}}
\newcommand{\cosmolike}{\blockfont{CosmoLike}}
\newcommand{\rz}{\ensuremath{r - z}}

\newcommand{\shearshear}{\ensuremath{\gamma \gamma}}
\newcommand{\galaxyshear}{\ensuremath{\delta_g \gamma}}
\newcommand{\galaxygalaxy}{\ensuremath{\delta_g \delta_g}}

\title[Modelling Intrinsic Alignments in DES Y1]{Dark Energy Survey Year 1 Results: Constraints on Intrinsic Alignments and their Colour Dependence from Galaxy Clustering and Weak Lensing}

\author[S.~Samuroff et al]{
\parbox{\textwidth}{
\Large
S.~Samuroff,$^{1}$\thanks{ssamurof@andrew.cmu.edu}
J.~Blazek,$^{2,3}$\thanks{blazek@berkeley.edu}
M.~A.~Troxel,$^{2,4}$
N.~MacCrann,$^{2,4}$
E.~Krause,$^{5}$
C.~D.~Leonard,$^{1}$
J.~Prat,$^{6}$
D.~Gruen,$^{7,8\dagger}$
S.~Dodelson,$^{1}$
T.~F.~Eifler,$^{5,9}$
M.~Gatti,$^{6}$
W.~G.~Hartley,$^{10,11}$
B.~Hoyle,$^{12,13}$
P.~Larsen,$^{14}$
J.~Zuntz,$^{15}$
T.~M.~C.~Abbott,$^{16}$
S.~Allam,$^{17}$
J.~Annis,$^{17}$
G.~M.~Bernstein,$^{18}$
E.~Bertin,$^{19,20}$
S.~L.~Bridle,$^{21}$
D.~Brooks,$^{10}$
A.~Carnero~Rosell,$^{22,23}$
M.~Carrasco~Kind,$^{24,25}$
J.~Carretero,$^{6}$
F.~J.~Castander,$^{26,27}$
C.~E.~Cunha,$^{7}$
L.~N.~da Costa,$^{23,28}$
C.~Davis,$^{7}$
J.~De~Vicente,$^{22}$
D.~L.~DePoy,$^{29}$
S.~Desai,$^{30}$
H.~T.~Diehl,$^{18}$
J.~P.~Dietrich,$^{31,32}$
P.~Doel,$^{10}$
B.~Flaugher,$^{17}$
P.~Fosalba,$^{26,27}$
J.~Frieman,$^{17,33}$
J.~Garc\'ia-Bellido,$^{34}$
E.~Gaztanaga,$^{26,27}$
D.~W.~Gerdes,$^{35,36}$
R.~A.~Gruendl,$^{24,25}$
J.~Gschwend,$^{23,28}$
G.~Gutierrez,$^{17}$
D.~L.~Hollowood,$^{37}$
K.~Honscheid,$^{2,4}$
D.~J.~James,$^{38}$
K.~Kuehn,$^{39}$
N.~Kuropatkin,$^{17}$
M.~Lima,$^{40,23}$
M.~A.~G.~Maia,$^{23,28}$
M.~March,$^{19}$
J.~L.~Marshall,$^{29}$
P.~Martini,$^{2,41}$
P.~Melchior,$^{42}$
F.~Menanteau,$^{24,25}$
C.~J.~Miller,$^{35,36}$
R.~Miquel,$^{43,6}$
R.~L.~C.~Ogando,$^{23,28}$
A.~A.~Plazas,$^{42}$
E.~Sanchez,$^{22}$
V.~Scarpine,$^{17}$
R.~Schindler,$^{8}$
M.~Schubnell,$^{36}$
S.~Serrano,$^{26,27}$
I.~Sevilla-Noarbe,$^{22}$
E.~Sheldon,$^{44}$
M.~Smith,$^{45}$
F.~Sobreira,$^{46,23}$
E.~Suchyta,$^{47}$
G.~Tarle,$^{36}$
D.~Thomas,$^{48}$
and V.~Vikram$^{49}$
\begin{center} (DES Collaboration) \end{center}
}
\vspace{0.4cm}
\\
Author affiliations are listed at the end of the paper.
}

\def\gs{\mathrel{\raise1.16pt\hbox{$>$}\kern-7.0pt %
\lower3.06pt\hbox{{$\scriptstyle \sim$}}}}         %
\def\ls{\mathrel{\raise1.16pt\hbox{$<$}\kern-7.0pt %
\lower3.06pt\hbox{{$\scriptstyle \sim$}}}}         %

\newcommand{\blockfont}[1]{{\textsc{#1}}\xspace}

\begin{document}

\maketitle

\begin{abstract}
We perform a joint analysis of intrinsic alignments and cosmology using tomographic weak lensing, galaxy clustering and galaxy-galaxy lensing measurements from Year 1 (Y1) of the Dark Energy Survey. 
We define early- and late-type subsamples, 
which are found to pass a series of systematics tests, including for spurious photometric redshift error and 
point spread function correlations.
We analyse these split data alongside the fiducial mixed Y1 sample using a range of intrinsic alignment models.
In a fiducial Nonlinear Alignment Model (NLA) analysis, 
assuming a flat \lcdm~cosmology, we find a 
significant 
difference in intrinsic alignment amplitude, with early-type galaxies favouring
$A_\mathrm{IA} = 2.38^{+0.32}_{-0.31}$
and late-type galaxies consistent with no intrinsic alignments
at 
$0.05^{+0.10}_{-0.09}$. 
The analysis is repeated using a number of extended model spaces, including a physically motivated model that includes both tidal torquing
and tidal alignment mechanisms.
In multiprobe likelihood chains in which cosmology, intrinsic alignments in both galaxy samples
and all other relevant systematics are varied simultaneously,
we find the tidal alignment and tidal torquing parts of the intrinsic alignment
signal have amplitudes 
$A_1 = 2.66 ^{+0.67}_{-0.66}$, $A_2=-2.94^{+1.94}_{-1.83}$, respectively, for early-type galaxies 
and
$A_1 = 0.62 ^{+0.41}_{-0.41}$, $A_2 = -2.26^{+1.30}_{-1.16}$ 
for late-type galaxies.
In the full (mixed) Y1 sample the best constraints are $A_1 = 0.70 ^{+0.41}_{-0.38}$, $A_2 = -1.36 ^{+1.08}_{-1.41}$.
For all galaxy splits and IA models considered,
we report cosmological parameter constraints consistent with 
the results of the main DES Y1 cosmic shear and multiprobe cosmology papers.
\end{abstract}

\begin{keywords}
cosmological parameters - cosmology: observations - gravitational lensing: weak - galaxies: statistics
\end{keywords}


\section{Introduction}\label{sec:introduction}

Within a little over a decade the study of late-time cosmology has 
grown from a set of theoretically justified but empirically untested ideas,
to a rigorous experimental field.
With the current generation of surveys now in the process of cataloguing millions of galaxies
and new experiments planned to reach even larger cosmological volumes,
the ideas of the past half century are now finally being implemented. 
In many ways low redshift measurements are complementary to
other cosmological probes such as the 
Cosmic Microwave Background (CMB), the masses and abundances of galaxy clusters 
and cosmographic observables such as supernovae and strong lensing.
Cosmological lensing probes the large scale distribution of mass directly
and is also sensitive to geometric distance ratios, 
which define a window of sensitivity on the line of sight (see e.g. \citealt{weinberg13}).

Advances have come in part due to the sheer number
of galaxies imaged by modern surveys.
Since shape noise scales as the inverse root of the number
of galaxies, expanding datasets have afforded gradually better 
signal-to-noise on cosmic shear statistics. Though statistical power can be continuously improved, 
an additional floor to the precision of the resulting cosmological inferences
is imposed by systematic errors.
In order to codify this, it is typically necessary to introduce ``nuisance parameters" 
in any cosmological analysis, which are marginalised out.
In the systematics-limited regime 
the only way to achieve tighter cosmological constraints
is to improve one's understanding of the systematics in question.
One is left with a choice of acquiring 
information from external data or theory,
and incorporating it into the analysis via a prior,
or self-calibrating the systematics by including new measurements
in the likelihood calculation. 

Indeed, there has long been recognition that combining 
different measurements can improve the quality of cosmological constraints.
Even very similar measurements extracted from the same galaxy survey can
be complementary if their parameter degeneracies and their systematic errors
differ.
Combining lensing auto-correlations with 
galaxy-galaxy lensing and two-point galaxy clustering,
for example,
is powerful as a means to ``self-calibrate" redshift 
error and other systematic uncertainties (see e.g. \citealt{joachimi10a}).
Another idea is
to use cross correlations between lensing and
CMB maps as a way to check for residual errors
in the shape measurement process \citep{schaan17, alonso18, harnois_deraps17, abbott_cmb18}.

There are many possible sources of systematic uncertainty
in late-time datasets
(see \citealt{mandelbaum17, mandelbaum15} 
for cosmic shear-specific reviews and  
\citealt{ross11, mandelbaum13, leistedt16, kwan17}; \citealt{y1ggl} [their Section V]; \citealt{y1clustering}
for more detailed discussions of systematics that can occur in 
galaxy-galaxy lensing and galaxy clustering measurements).
One major class of systematics arises from local astrophysical
effects, which can mimic a cosmological shear signal.
Spurious (non-cosmological) correlations between galaxies, 
known as intrinsic alignments (IAs) have long been known to affect
cosmic shear estimates.
Such effects arise because galaxies are not independent point measurements 
of the large scale cosmic shear field,
but
rather extended astrophysical objects that interact with each other
and with their environment.
It was realised over a decade ago that galaxies hosted by a
common dark matter halo tend to align through shared tidal 
interactions \citep{catelan01} and rotational torquing \citep{mackey02}. 
This results in alignment in the intrinsic shapes of physically close pairs of
galaxies, known as II correlations.
An often more pervasive effect comes from the fact that the same foreground matter
experiences local gravitational interactions over short spatial scales,
and also induces lensing of background galaxies. This generates
correlations in shape between foreground galaxies and 
background sources \citep{hirata04}, which are
known as the GI contribution; this is often the dominant form of intrinsic 
alignments in lensing surveys.
It has been shown by \citet{croft00} and others 
that the total IA contamination to cosmological shear can be as high as $10\%$ in modern surveys,
and neglecting these effects can result in significant 
cosmological biases
\citep{kirk12, krause16}.

The particular challenge posed by IA modelling
is in large part down to the nature of the contamination;
biases in shear measurement, photo-$z$ estimation,
point spread function (PSF) modelling errors and instrumental systematics are all fundamentally
methodological problems. One can understand them using image simulations 
and mitigate them by devising new methods.
In contrast, IA correlations are a real astrophysical signal, which enters much 
the same angular scales as cosmic shear itself.
Indeed, it has been suggested that if correctly modelled they can in principle be used as a 
probe of cosmology \citep{chisari13, troxel15}, primordial non-Gaussianity \citep{chisari16}, or galaxy formation \citep{schmitz18}.
Given this context, if we are to avoid becoming limited by intrinsic alignments 
it is important that the lensing community develops a robust understanding of the nature
of this signal and techniques for dealing with it.
A number of mitigation techniques have been proposed, 
involving discarding physically close pairs of galaxies \citep{catelan01, kirk15},
downweighting \citep{king03, heymans03, heavens03, heymans05},
or nulling \citep{joachimi10b}.
All of these methods depend on the existence of accurate redshift information to
allow galaxies to be located relative to each other along the line of sight.
Significantly, they are also ineffective in mitigating GI correlations,
which are often dominant in galaxy samples typical of cosmic shear measurements.
Alternatively one could impose colour or morphology cuts designed to isolate
a subsample free of IA contamination \citep{krause16}. 
This approach, however, has a number of obvious drawbacks, not least that one has no
theoretical grounds for believing any given population of galaxies to be perfectly 
without intrinsic alignments.

The issues with modelling IAs can broadly be separated into two problems.  
First, the models are known to perform poorly on small physical scales, 
where intra-halo interactions dominate the galaxy two-point correlations.
Progress on these scales requires an understanding of how galaxies populate
and interact within their host halos
(see, for example, \citealt{schneider10} for a halo model-based treatment
of the small scale IA power spectra).
Halo models have the advantage of mathematical elegance, and can be (validly) 
extended down to nonlinear scales.
They do, however, require calibration using numerical simulations, and 
are thus only as reliable as the simulations in question.
A similar idea is to use ``semi-analytic" modelling, based on cosmological simulations,
as discussed in \citet{joachimi13}.
Model testing on these scales is further complicated by the influence of other
poorly understood effects such as baryonic feedback.  
The second problem is the existence of known deficiencies in IA modelling on two-halo scales.
These occur primarily because the most common large scale alignment models are
based on a population of galaxies that
is highly unrepresentative of the typical samples used for lensing studies. 
Recent years have seen the emergence of a small handful 
of more complete physically motivated models, which seek to build a unified
IA prescription in a mixed galaxy population (\citealt{blazek15, tugendhat17, blazek17};
see also \citealt{svcosmology, y1cosmicshear} for practical implementations).
Similarly, \citet{larsen16} use perturbation theory to model 
scale dependence of CMB - intrinsic shape cross correlations, 
which they argue should match the GI term in cosmic shear on large scales.   
They predict that IAs due to tidal torquing should exhibit a very similar 
scale dependence to the commonly used linear alignment model.

It has been noted in both simulations and data that
the choice of galaxy shape estimation method can alter the magnitude of the IA signal
by an overall scale-independent factor \citep{singh16, hilbert17}.
One interesting idea devised by \citet{leonard18} 
takes advantage of
this concept, using multiple shape measurement techniques to measure the scale dependence of the IA signal in
the nonlinear regime, 
a subject that is poorly understood at a theoretical level at the present time.
This method
carries the advantage of being relatively robust to photometric redshift error compared with conventional measurements. 

Notably several authors have found the intrinsic alignment correlations
measured in hydrodynamic simulations to be dependent on
galaxy type, mass and magnitude; these dependencies are also poorly understood 
at the theoretical level \citep{joachimi13,chisari15, hilbert17}.
In recent years there have been attempts to place observational constraints on
the alignment properties of galaxy
samples more representative of the sort used for cosmological lensing measurements 
\citep{mandelbaum11, blazek12, tonegawa17}. 
Despite these efforts, given limitations in the sample selection and size, 
we still have little clear information about the expected values of the 
free parameters in our IA models.
It is thus common to choose what is known to be an incomplete model and 
to marginalise over it using uninformative priors.

This work sits alongside a series of other DES studies based on the same data.
\citet{y1shearcat} describe the construction of the Y1 shape catalogues
and provide a basic usage guide.
In \citet{y1ggl} and \citet{y1clustering} the
galaxy-galaxy lensing and galaxy clustering measurements and their
potential systematics are examined in detail.
The cosmological analysis choices and the robustness of the Y1 pipeline to various
forms of systematic error are tested using noiseless synthetic data in \citet{y1methodology}
and N-body simulations in \citet{y1cosmosims}.
Cosmology constraints from cosmic shear alone and 
shear, galaxy-galaxy lensing and clustering
are set out in \citet{y1cosmicshear} and \citet{y1keypaper} respectively.
More recent follow-on work has included a methodology paper for a future 
analysis combining $3\times2$pt measurements with CMB cross-correlations \citep{y15x2ptmethodology},
a joint constraint on the local Hubble parameter using DES alongside external Baryon Acoustic Oscillation and
Big Bang Nucleosynthesis data \citep{y1h0}
and, most recently, a study setting out
a series of cosmological
modelling extensions
\citep{y1extensions}. 
This paper seeks to explore a significant cosmological systematic using the same Y1 lensing dataset:
intrinsic alignments and their colour dependence.

In Section \ref{sec:theory} we outline the theory of modelling intrinsic
alignments and introduce the formalism adopted in this study.
We describe the DES Y1 data in Section \ref{sec:data}
and define a number of galaxy samples, which are selected to
separate differences in the underlying IA signal.
Section \ref{sec:measurements} sets out the measurements used in this
work, which include real space two-point correlations of 
cosmic shear, galaxy-galaxy lensing and galaxy clustering.
In Section \ref{sec:results} we present the main results of this analysis,
using a range of intrinsic alignment models and
three different galaxy samples.
We conclude and provide a brief summary in Section \ref{sec:conclusion}.


\section{Theory \& Background}\label{sec:theory}
 
\subsection{Observational Constraints on Intrinsic Alignments}

Attempts to constrain intrinsic shape correlations between galaxies fall broadly into
two categories.
The first are \emph{direct} constraints, which typically use
galaxies at low to intermediate redshift 
and often impose colour cuts to isolate well-measured red galaxies,
and assume some fixed known cosmology.
Correlation statistics used in these measurements are explicitly designed to 
maximise the IA signal 
(e.g. \citealt{hirata07}, \citealt{faltenbacher09}, \citealt{okumura09}, 
\citealt{mandelbaum11}, \citealt{blazek11}, \citealt{blazek12}).
Since IA correlations are a fundamentally local phenomenon it is common to focus on
samples for which high quality spectroscopic data is available,
allowing three-dimensional reconstruction of the physical field.
In such studies it is also common to restrict measurements to the
low redshift regime, where the amplitude of cosmological lensing is low.

The second class of measurements are indirect,
or \emph{simultaneous} constraints.
Generally they measure statistics designed to be sensitive to cosmic shear
such as $\xi_\pm$ and use faint high-redshift galaxies in 
which the cosmological signal is strongest.
While some studies attempt to remove the lensing signal to 
obtain a clearer picture of IAs (e.g.\ \citealt{blazek12,chisari14}),
cosmic shear and galaxy-galaxy lensing analyses must necessarily address the questions
of intrinsic alignments and lensing together. Any investigation that involves
marginalising over IAs rather than suppressing them directly falls into this category
\citep{heymans13,svcosmology,jee16,hildebrandt17,koehlinger17,y1cosmicshear,chang18,hikage18}. 
The assumptions about IAs differ slightly between studies, but they
all assume the same basic model (the nonlinear alignment model),
sometimes with a multiplicative scaling in redshift or luminosity. 

There is some direct evidence for differences in the IA contamination,
depending on the nature of the galaxy sample \cite{heymans13,y1cosmicshear}.
Broadly there are two paradigms: 
early-type ellipticals, which tend to be redder and structurally pressure dominated;
and late-type spirals, which tend to be bluer and rotation dominated.
The former are thought to align through tidal interactions
with the background large scale structure of the Universe.
If a dark matter halo sits in a local gradient in the gravitational field,
it will be sheared along that gradient and
nearby galaxies will become aligned with their common background tidal field.
If the distortion is small, the induced ellipticity can be assumed to be linear
in the gravitational potential.
A handful of direct studies over the past decade have sought to place constraints 
on IAs in red galaxies
(see e.g. \citealt{mandelbaum06,hirata07,okumura09,joachimi11,li13,singh15}).
In each case, a strong IA signal is reported, with no statistically significant 
detection of redshift dependence.

The picture for late-type galaxies is rather different.
These objects form galactic discs, which, depending on the orientation,
will have an apparent ellipticity. 
One common picture is that galaxy spin
(which ultimately decides the disc orientation)
is generated by tidal torquing,
exerted on a halo in its early stages of development.
Direct constraints on blue galaxy IAs 
are generally relatively weak.
Measurements have been made on blue samples from 
SDSS \citep{york00} and WiggleZ \citep{parkinson12}
at low to mid redshifts,
but impose only upper limits on the intrinsic alignment
amplitude
\citep{hirata07,mandelbaum11}.
\citet{blazek12} use a blue sample from SDSS to make such a measurement,
but place an upper limit 
only on the IA signal at $z \sim 0.1$.
A similar analysis by \citet{tonegawa17},
using Emission Line Galaxies from FastSound and 
the Canada France Hawaii Lensing Survey (CFHTLenS),
also reports a null detection, showing no evidence of either
non-zero amplitude or redshift dependence.  

For a more detailed overview of the theory and observational history of intrinsic alignments
we direct the reader to a number of extensive reviews on the subject 
\citep{troxel15, joachimi15, kirk15, kiessling15}

\subsection{Theory}

Theory modelling and parameter estimation for this study are performed within the 
\blockfont{CosmoSIS}
framework \citep{zuntz15}. We use 
the \blockfont{Multinest} nested sampling package \citep{feroz13}
to sample the joint model space of cosmology, intrinsic alignment
and systematics parameters.
For consistency with previous publications, 
our choices regarding sampler settings follow those used by
\citet{y1methodology}.
The dark matter power spectrum is estimated at each cosmology using 
\blockfont{camb}\footnote{http://camb.info/}, 
with nonlinear corrections generated by \blockfont{Halofit} \citep{takahashi12}.
We do not explicitly model baryonic effects
and the intrinsic alignment prescriptions considered do not attempt to model the one-halo regime,
but as noted in the next section our choice of scale cuts is relatively conservative.
Except in Section \ref{sec:beyond_lcdm}, where we explicitly set out to extend the
cosmological model space, we assume a flat \lcdm~cosmology
with six free parameters 
$\mathbf{p}_\mathrm{cosmology} = (h, \omegam, \omegab, \as, \ns, \Omega_\nu h^2)$.

The following paragraphs describe how each of the three types of observable correlation,
and their intrinsic alignment contribution, is modelled for the purposes of parameter inference.

\subsubsection{Cosmic Shear}

For cosmic shear we use real-space angular correlation
functions in four tomographic bins.
The measurements map onto the angular shear power spectrum via Hankel 
transforms:

\begin{equation}\label{eq:hankel_shear}
\xi^{ij}_\pm(\theta) = \frac{1}{2\pi} \int \ell J_{0/4}(\ell \theta) C_{\gamma\gamma}^{ij}(\ell) d\ell
\end{equation}

\noindent
where the indices $ij$ indicate a pair of tomographic bins, and $J_0$ and $J_4$ are Bessel functions of the first kind.
For the moment we will assume no intrinsic alignments, 
and so the shear-shear angular power spectrum $C_{\gamma\gamma}$
is interchangeable with the signal predicted from cosmological
lensing only $C_\mathrm{GG}$.
$C_\mathrm{GG}$ is related to the dark matter power spectrum 
under the Limber approximation as,
\begin{equation}\label{eq:c_GG}
C^{ij}_\mathrm{GG} = \int^{\chi_\mathrm{hor}}_0 \frac{g^i(\chi)g^j(\chi)}{\chi^2}  P_\delta \left ( k=\frac{\ell}{\chi}, z \right ) d\chi.
\end{equation}

\noindent
We assume a flat universe,
such that the transverse angular diameter distance $S_K(\chi)=\chi$.
The term $\chi_\mathrm{hor}$ is the comoving horizon
distance
and the lensing kernel in each bin is given by

\begin{equation}\label{eq:lensing_kernel}
g^i(\chi) = \frac{3}{2} \frac{H^2_0 \omegam}{c^2} \frac{\chi}{a(\chi)} \int ^{\chi_\mathrm{hor}}_{\chi} n^i(\chi')  \frac{\chi' - \chi}{\chi'} d\chi'
\end{equation}

\noindent 
The redshift distributions $n(z)$ are assumed to be normalised over the depth of the survey,
and defined such that
$n(z)dz = n(\chi)d\chi$.
Likelihoods for trial cosmologies are calculated by generating theory angular spectra,
which are integrated over with the Bessel kernels, resampled at the appropriate angular scales,
and then compared with the measurements of $\xi_\pm^{ij}$.

\subsubsection{Galaxy Clustering}\label{sec:theory:gg}

The formalism for predicting galaxy clustering observables
follows by close analogy to the previous section.
The spatial distribution of lens galaxies traces out the
underlying dark matter, albeit via some unknown galaxy bias.
In this work we adopt a simple scale-independent linear bias model,
with the overdensity of
galaxies at a particular scale related to the dark matter density
as $\delta_g(k) = b_g(z) \delta(k)$.
We adopt the same scale cuts used in the DES Y1 key paper
\citep{y1keypaper},
under which it has been demonstrated that higher-order bias terms
have negligible impact on cosmology \citep{y1methodology}.
The correlation function of galaxy density has the form

\begin{equation}
w^{ij}(\theta) = \frac{1}{2\pi} \int \ell J_{0}(\ell \theta) C_{\delta_g\delta_g}^{ij}(\ell) d\ell,
\end{equation}

\noindent
where the galaxy-galaxy angular power spectrum between tomographic bins $i$ and $j$ is given by

\begin{equation}
C^{ij}_{\delta_g\delta_g}(\ell) = \int^{\chi_\mathrm{hor}}_0 \frac{n_l^i(\chi)n_l^j(\chi)}{\chi^2} b_g^i b_g^j P_\delta \left ( k=\frac{\ell}{\chi}, z \right ) d\chi.
\end{equation}

\noindent
Since we have no good first-principles model for the galaxy bias and its redshift evolution
we allow $b_g^i$ to vary independently in each redshift bin.
Within each bin $b_g^i$ is scale and redshift independent
and can thus be taken outside of the integral.
The subscript $l$ in the $n_l^i(\chi)$ terms denotes lens galaxies,
for which we use the DES Y1 \blockfont{redMaGiC} sample as presented by 
\citet{y1clustering}.

\subsubsection{Galaxy-Galaxy Lensing}

The final part of the $3\times2$pt combination of late-time probes
is galaxy-galaxy lensing.
As the cross correlation between galaxy shapes and number density, the
galaxy-galaxy lensing formalism follows similar lines to the two 
auto-correlations described above.
A commonly used observable, $\gamma_t(\theta)$, is given by the Hankel transform

\noindent
\begin{equation}\label{eq:hankel_ggl}
\gamma^{ij}_t(\theta) = \frac{1}{2\pi} \int \ell J_{2}(\ell \theta) C_{\delta_g\gamma}^{ij}(\ell) d\ell,
\end{equation}

\noindent
where the angular spectrum (again assuming zero IAs for the moment) is

\begin{equation}\label{eq:c_gG}
C^{ij}_{\delta_g\mathrm{G}}(\ell) = \int^{\chi_\mathrm{hor}}_0 \frac{n_l^i(\chi)g^j(\chi)}{\chi^2} b_g^i P_\delta \left ( k=\frac{\ell}{\chi}, z \right ) d\chi.
\end{equation}

\noindent
Again, we assume linear galaxy bias, allowing the $\delta_g \mathrm{G}$ power spectrum to be 
expressed as the matter power spectrum modulated by a scale-independent bias coefficient 
$b_g^i$.
The lensing kernel $g(\chi)$ is defined by equation \ref{eq:lensing_kernel}.
It is worth bearing in mind that a small handful of different galaxy-galaxy lensing 
estimators exist in the literature,
most notably $\Delta \Sigma$ (related to $\gamma_t$ via a factor of the critical density $\Sigma_\mathrm{c}$; see \citealt{mandelbaum13})
and $\Upsilon$ (devised to remove contributions from small scales; see \citealt{baldauf10}).

\subsubsection{Modelling Intrinsic Alignments}\label{sec:theory:ia_models}

Even a perfectly unbiased measurement of the ellipticity-ellipticity 
two-point function in a set of galaxies is not a pure estimate
of the cosmic shear spectrum.
Correlations between the intrinsic (pre-shear) shapes 
contribute unknown additive terms of the form

\begin{equation}
C^{ij}_{\gamma\gamma}(\ell) = C^{ij}_\mathrm{GG}(\ell) + C^{ij}_\mathrm{II}(\ell) + C^{ij}_\mathrm{GI}(\ell) + C^{ji}_\mathrm{GI}(\ell),
\end{equation}

\noindent
where we make the distinction between the observable estimate for the shear 
correlation $C_{\gamma\gamma}$ and the cosmological GG component.
Note that it is $\gamma\gamma$, not GG that appears in equation \ref{eq:hankel_shear}.
The spectra with subscripts GI and II are intrinsic alignment correlations,
and arise via the mechanisms described in Section \ref{sec:introduction}.
The IA contribution to galaxy-galaxy lensing follows a similar form,
but is insensitive to II correlations:

\begin{equation}
C^{ij}_{\delta_g\gamma}(\ell) = C^{ij}_{\delta_g \mathrm{G}}(\ell) + C^{ij}_{\delta_g \mathrm{I}}(\ell),
\end{equation}

\noindent
A number of different prescriptions for calculating the GI and II terms
exist in the literature.

These Limber projections in bins $ij$ are simply expressed 
in terms of the IA power spectra in the form

\begin{equation}
C^{ij}_\mathrm{II}(\ell) = \int \frac{n^i(\chi)n^j(\chi)}{\chi^2} P_\mathrm{II}\left ( k=\frac{\ell}{\chi}, \chi \right ) d\chi
\end{equation}

\noindent
and 

\begin{equation}\label{eq:limber_GI}
C^{ij}_\mathrm{GI}(\ell) = \int \frac{g^i(\chi)n^j(\chi)}{\chi^2}  P_\mathrm{GI}\left ( k=\frac{\ell}{\chi}, \chi \right ) d\chi,
\end{equation}

\noindent
where the GI and II power spectra $P_\mathrm{GI}$ and $P_\mathrm{II}$
are generic, and can be generated by any of the IA models discussed below.
Similarly, the galaxy-intrinsic term, which appears in
galaxy-galaxy lensing correlations is given by:

\begin{equation}\label{eq:limber_gI}
C^{ij}_{\delta_g \mathrm{I}}(\ell) = \int \frac{n_l^i(\chi)n^j(\chi)}{\chi^2}  b^i_g P_\mathrm{GI}\left ( k=\frac{\ell}{\chi}, \chi \right ) d\chi,
\end{equation}

\noindent
under the assumption of linear galaxy bias. Note that though they are both
sensitive to the GI power spectrum $P_\mathrm{GI}$,
the relation between 
$C_\mathrm{GI}$ and $C_{\delta_g \mathrm{I}}$
is non-trivial because the projection kernels in equations 
\ref{eq:limber_GI} and \ref{eq:limber_gI}
differ.

Under the common family of ``tidal alignment'' models, in which the intrinsic galaxy shapes are assumed to
be linearly related to the local tidal field, the IA power spectra
are assumed to be of the same shape as the matter power spectrum,
but subject to a redshift-dependent rescaling:

\begin{equation}\label{eq:p_gi}
P_\mathrm{GI}(k,z) = A(z) P_\delta(k,z),
\end{equation}

\noindent
and

\begin{equation}\label{eq:p_ii}
P_\mathrm{II}(k,z) = A^2(z) P_\delta(k,z).
\end{equation}

\noindent
Owing to its good performance in matching data and simulations,
one prescription, known as the nonlinear alignment (NLA) model
\citep{bk07} has become particularly popular.
This is an empirical modification to the linear alignment model of 
\citet{catelan01} and \citet{hirata04}, whereby the linear matter power spectrum is
replaced by the nonlinear spectrum.
The normalisation in the NLA model is typically expressed as

\begin{equation}\label{eq:nla}
A(z) = -A_\mathrm{IA} \bar{C_1} \frac{3H^2_0 \omegam}{8\pi G} D^{-1}(z) \left( \frac{1+z}{1+z_0} \right )^{\eta_\mathrm{IA}}.
\end{equation}

\noindent
The dimensionless amplitude $A_\mathrm{IA}$ is an unknown scaling parameter governing the strength of the
IA contamination for a particular sample of galaxies,
and is generally left as a free parameter to be constrained. 
Here $G$ is the gravitational constant and $D(z)$ is the linear growth factor.
The normalisation constant $\bar{C_1}$ is typically fixed at a value obtained from the SuperCOSMOS Sky
Survey by \citet{brown02} of $\bar{C_1}= 5 \times 10^{-14} M_{\odot}^{-1} h^{-2}$ Mpc$^3$.
The redshift evolution is expressed by a power law index $\eta_\mathrm{IA}$,
which has been measured in low redshift samples of luminous red galaxies
\citep{joachimi11}. The value of $\eta_\mathrm{IA}$ can capture underlying evolution of 
the alignment or evolution within a given sample
of other galaxy properties that impact alignment, such as luminosity and 
morphology.\footnote{Luminosity dependence could also be explicitly included in the normalisation.}
The denominator $1+z_0$ sets a pivot redshift, for which we assume $z_0=0.62$
whenever equation \ref{eq:nla} is used in this paper.
Note that the same value was used in the previous Y1 analyses
of \citet{y1cosmicshear} and \citet{y1keypaper}.

In addition to the baseline NLA model, one could conceivably add flexibility 
to the IA model by allowing the amplitudes entering the GI and II power spectra
(equations \ref{eq:p_gi} and \ref{eq:p_ii})
to behave as independent free parameters.
For the purpose of this study, we will treat this as a separate IA model with four free parameters
(the fourth row of Table \ref{tab:models}).
Alternatively, one could maintain the link between the II and GI spectra,
and instead allow $A$ to vary independently in each redshift bin.
This approach, analogous to the treatment of galaxy bias in this paper,
has four free parameters and
is referred to as the `Flexible NLA' model (row 3 of Table \ref{tab:models}).

The NLA model, defined by the equations above, is physically
motivated and found to match observational data well in specific circumstances.
That is, on linear scales, in bright red low-redshift populations where 
intrinsic alignments have been measured with high signal-to-noise
\citep{hirata07, blazek11}. 
Unfortunately, there is neither \emph{prima facie} theoretical motivation nor
strong observational evidence to suggest this model applies equally well to the
type of galaxies sampled by modern lensing surveys.
Moreover, the picture is further complicated by the fact that
galaxies used for lensing cosmology are typically mixed 
(i.e. with no explicit colour or morphology based cuts),
going from a predominantly elliptical population at low redshifts
to one dominated by rotation-dominated spirals at high $z$.
There is evidence from both theoretical studies \citep{catelan01,mackey02}
and from hydrodynamic simulations \citep{chisari15,hilbert17}
that the alignment mechanisms at play in these different galaxy types are very different. 

The standard approach to this question is to assume that red galaxies can be
modelled using the NLA model and blue galaxies have no
intrinsic shape correlations.
In this picture the observed IA contribution in cosmic shear data
is a pure NLA signal, but scaled by an effective IA amplitude,
which absorbs the dilution due to randomly
oriented blue galaxies.
This strategy will, however, be effective only in the limit of zero alignments 
in blue galaxies.

In addition to the NLA model, 
we will also employ a model intended to address this concern. 
Based on perturbation theory, the model of \citet{blazek17} combines
alignment contributions from tidal torquing 
(quadratic in the tidal field; thought to dominate in blue galaxies)
and from tidal alignments 
(linear in the tidal field; dominant in red galaxies).
In this model, the intrinsic galaxy shape $\gamma_{ij}^I$ can be expressed as an expansion in the tidal field $s_{ij}$ and the density field $\delta$, with the subscripts denoting components of spin-2 tensor quantities.

\begin{equation}\label{eq:tatt_exp}
\gamma_{ij}^I  = C_1 s_{ij} + C_2 \left( s_{ik} s_{kj} - \frac{1}{3}s^2 \right) + C_{1\delta} \left(\delta s_{ij} \right) + \cdots
\end{equation}

\noindent
In this expansion, $C_1$ captures the tidal alignment contribution. Using the full nonlinear density field to calculate $s_{ij}$ yields the NLA model. $C_2$ captures the quadratic contribution from tidal torquing.
Finally, $C_{1\delta}$ can be seen as a contribution from ``density weighting'' the tidal alignment contribution: we only observe IAs where there are galaxies, which contributes this additional term at next-to-leading order.
While these coefficients can be associated with tidal alignment and tidal torquing mechanisms, as done here,
these can also be considered ``effective'' parameters capturing any relevant astrophysical processes that produce IA with the given dependence on cosmological 
fields.\footnote{This approach is general up to a given order in perturbation theory, although one must in principle include additional contributions from higher derivative terms, which become relevant at roughly the halo scale (e.g.\ \citealt{desjacques17}). As discussed in \citet{blazek17,schmitz18}, the TATT model used here is not fully general at next-to-leading order, since it neglects two potential nonlinear contributions.}
Furthermore, we note that $A_1 \neq 0$ can potentially arise from tidal torquing combined with nonlinear structure growth \citep{larsen16,blazek17}. Despite this potential complication, in the following discussion we assume the standard mapping between these parameters and the underlying IA formation mechanisms.

As implemented in this work, this formalism has four adjustable parameters:
an amplitude and a redshift power law governing each of the tidal alignment ($C_1$) and 
tidal torque ($C_2$) power spectra. Following \citet{blazek17}, 
we assume $C_{1\delta} = b^\mathrm{src}_g C_1$, i.e.\ the density weighting is given by the bias of the source sample.
The source bias can be then either be fixed
(as in \citealt{y1cosmicshear}, which assumed $b^\mathrm{src}_g = 1$),
or marginalised over a plausible range of values.
For the main section of this paper we fix source bias.
Note that the model requires no explicit assumptions about the fraction of red
galaxies or its evolution with redshift.
We have the following parameterization:
\begin{equation}\label{eq:tatt_c1}
C_1(z) = -A_1 \bar{C_1}\rho_\mathrm{crit} \frac{\omegam}{D(z)} \left ( \frac{1+z}{1+z_0}\right )^{\eta_1} 
\end{equation}
\noindent
for the tidal alignment part. For the tidal torque contribution,

\begin{equation}\label{eq:tatt_c2}
C_2(z) = 5 A_2 \bar{C_1}\rho_\mathrm{crit} \frac{\omegam}{D^2(z)} \left ( \frac{1+z}{1+z_0}\right )^{\eta_2},
\end{equation}

\noindent
with the four IA parameters $\mathbf{p_\mathrm{IA}} = (A_1, \eta_1, A_2, \eta_2)$.

The corresponding IA power spectra (GI and II) are $k$-dependent functions derived from perturbation theory
and are given by integrals over the matter power spectrum;
for the full expressions and visual comparison see \citet{blazek17} Sections A-C.
These alignment power spectra define what we will refer to as the `Complete TATT' model.
We will also treat the pure tidal alignment and tidal torque scenarios
as models in their own right (Table \ref{tab:models}, third and fourth from last rows).

In the most na{\"i}ve theoretical picture of intrinsic alignments,
galaxies are either pressure-supported ellipticals, whose shapes respond linearly to the
background tidal field,
or rotation-dominated spirals, whose alignment is quadratic in the tidal field.  
For comparison with previous theoretical studies
we will, then, consider TA and TT cases, 
with power spectra
obtained from the equations above,  
but with fixed amplitudes $A_2=0$ and $A_1=0$ respectively.

For computational reasons we assume negligible B-mode IA contribution.
These analysis choices have been tested and shown to
have no significant effect on our conclusions 
in Appendix \ref{app:source_bias} and Appendix \ref{app:bmodes}.

The $k$ dependent terms in these equations are computed using the \blockfont{FAST-PT} code
\citep{mcewan16, fang17}.
For both the mode-coupling integrals and the TATT model predictions,
we use code implementations within \blockfont{CosmoSIS},
which are common to \citealt{y1cosmicshear} and the forecasts in \citealt{blazek17}.

The intrinsic alignment models discussed in the above paragraphs 
and their free parameters are summarised in Table \ref{tab:models}.
For reference we also include the ranges over which 
the various parameters are allowed to vary.
The prescription referred to as the `Complete TATT Model' in this work, 
which includes $C_1$ and $C_2$ contributions and has fixed $b^\mathrm{src}_g = 1$
is identical to the `Mixed Model' of \citealt{y1cosmicshear},
the `Complete Model' (Section D) of \citealt{blazek17}
and the `TATT Model' of \citealt{y1extensions}.
It is worth noting that \citealt{y1cosmicshear} also present
constraints with the baseline and flexible NLA models,
but with cosmic shear alone. 
Both \citet{y1cosmicshear} and \citet{y1keypaper}
opt to marginalise over the two-parameter NLA model as their fiducial
IA treatment;
their headline cosmology constraints come from such treatment.

\begin{table}
\begin{center}
\begin{tabular}{cccc}
\hline
IA Model                     &   Free Parameters                  & Priors                \\
\hhline{===} 
No Alignments                & None                                 & None      \\
NLA (fiducial)               & $A_\mathrm{IA}$                      & $U[-6,6]$ \\
NLA (fiducial)               & $\eta_\mathrm{IA}$                   & $U[-5,5]$ \\
Flexible NLA                 & $A^{(i)}, i\in(1,2,3,4)$             & $U[-6,6]$ \\
NLA (separate GI $+$ II)     & $A_\mathrm{GI}, A_\mathrm{II}$       & $U[-6,6]$ \\
                             & $\eta_\mathrm{GI}, \eta_\mathrm{II}$ & $U[-5,5]$ \\
Tidal Alignment              & $A_1$                                & $U[-6,6]$ \\
Tidal Torque                 & $A_2$                                & $U[-6,6]$ \\
TATT                         & $A_1$                                & $U[-32,32]$ \\
                             & $A_2$                                & $U[-6,6]$ \\
TATT ($z$ power law)         & $A_1$                                & $U[-32,32]$ \\
                             & $A_2$                                & $U[-6,6]$ \\
                             & $\eta_1, \eta_2$                     & $U[-32,32]$ \\
\hline
\end{tabular}
\end{center}
\caption{Summary of the intrinsic alignment models used in this paper.
The right-hand column shows the parameters varied under each model.
In principle the galaxy bias in the source population $b_g^\mathrm{src}$ also
enters the TATT model (both variants) and the TA models \citep{blazek17} .
Other than in Appendix \ref{app:source_bias}, 
where we explicitly test its impact, however, we fix $b_g^\mathrm{src}=1$.
}\label{tab:models}
\end{table}

\subsubsection{Other Systematics}

In addition to five cosmological parameters and the IA model parameters
we marginalise over thirteen nuisance parameters. The point
here is to encapsulate residual systematic errors entering the measurement
due to a number of effects.
Following \citet{y1keypaper}, we marginalise over an offset in the mean
of the photometric redshift distributions in each of the four lensing bins.
At least in the context of $3\times2$pt cosmology at current precision
there is evidence in the literature that a shift in the ensemble mean
of the redshift distribution is the most 
salient form of redshift error
(see e.g. Figure 20 of \citealt{y1keypaper}).
This transforms the $n(z)$ entering into equation \ref{eq:lensing_kernel} as
$n^i(z)\rightarrow n^i(z-\Delta z^i)$, where $\Delta z^i$ is the redshift error for bin $i$.
There is reason for caution here, however, particularly if one wishes to draw conclusions about 
less well-understood effects such as intrinsic alignments: 
photo-$z$ modelling errors can easily be absorbed into an apparent IA signal (see, for example, 
Section 6.6 of \citealt{hildebrandt17}).
We seek to test the impact of photo-$z$ modelling insufficiency 
in Section \ref{sec:results:systematics} and find our results are robust to 
reasonable changes in the shape of the  $n(z)$s.  
In addition, there is some level of uncertainty in the treatment of shear estimation
bias, for which it is necessary to include an additional nuisance parameter $m^i$ per source bin.
This modulates the angular spectra in equations \ref{eq:hankel_shear} and \ref{eq:hankel_ggl} by factors of 
$(1+m^i)(1+m^j)$ and $(1+m^i)$ respectively. Finally, there are five nuisance parameters to 
account for lens redshift errors and five for lens galaxy bias. The redshift parameters act in the 
same way as the source errors, but on the clustering sample $n_l(z)$. Our treatment of lens bias is 
discussed in Section \ref{sec:theory:gg}.

Since the clustering sample is unchanged relative to that set out in \citet{y1keypaper} we adopt the
priors on lens redshift error and galaxy bias used in that paper.
Similarly, the uncertainty in $m^i$ is dominated by limitations in
how the shear measurement handles blending. This is not expected to differ significantly with
galaxy type, and so for all of the samples described in the next section we adopt the fiducial
Gaussian prior on $m^i$ recommended by \citet{y1shearcat}.
The source redshift error, however, could very easily differ between galaxy samples of different
colour. We recompute priors on $\Delta z^i$ for the different
samples using galaxies from the COSMOS field, a calculation discussed further in Section \ref{sec:data:pz}.


\section{Data \& Sample Selection}\label{sec:data}

In this section we define the galaxy samples used in this paper.
The subsamples are disjoint populations from the DES Y1 weak lensing
catalogue\footnote{For the public release of the data see 
\url{https://des.ncsa.illinois.edu/releases/y1a1}},
intended to isolate morphological differences relevant to IA. 
The following paragraphs discuss the practical details of the split,
including how we manage selection effects.

\subsection{The Dark Energy Survey Y1 Data}

The Dark Energy Survey (DES) has now completed its five-year observing
campaign, covering a footprint of around 
5000 square degrees to a depth of $r\sim24.1$ magnitudes. 
The observing program made use of the 570 megapixel DECam
\citep{flaugher15},
which is mounted on the Victor Blanco telescope 
at the Cerro Tololo Inter-American Observatory
(CTIO) in northern Chile. 
Its five-band grizY photometry spans a broad region of the 
optical and near infrared spectrum between 0.40 and 1.06 microns.
Each $griz$ exposure is 90 seconds in duration and the final mean tiling depth 
will be ten exposures over the full footprint.

The wide-field observations for Y1 encompass a large region
completely overlapping the footprint of the South Pole Telescope 
(SPT; \citealt{carlstrom11}) CMB experiment and extends roughly 
over the range $\delta=[-60, -40]$ degrees. A significantly smaller 
region in the north of the Y1 footprint also overlaps with the 
Stripe 82 field of the Sloan Digital Sky Survey (SDSS); data from 
this region are excluded from this analysis, as they were from the 
main Y1 cosmology papers. In total the Y1 cosmology dataset encompasses an 
area of 1321 square degrees of the southern sky with a mean depth of three
exposures. This includes masking for potentially bad regions deemed to be of 
unsuitable quality for cosmological inference.
A more detailed description of the final Gold sample can be found in \citet{y1gold}.
These data were collected between 31st August 2013 and
9th February 2014 during the first full season of DES operations.

For lensing measurements we make use of the larger of the two DES Y1 shape catalogues
(see \citealt{y1shearcat}), which contains $\sim 26$ million galaxies in the final 
cosmology selection. This dataset, known as the \metacal~catalogue, relies on the 
eponymous technique for correcting shear measurement bias. We discuss how these 
corrections, which include sample selection effects, are computed in 
Section \ref{sec:measurements}.

The catalogue used for two-point clustering measurements
comprises a set of luminous red galaxies selected by the \blockfont{redMaGiC} 
algorithm \citep{rozo16} using a method designed to minimise photometric 
redshift error. The sample contains roughly 0.66 M galaxies at constant 
comoving density over the range $z=0.15-0.9$ \citep{y1clustering}.

\subsection{Blinding}

This analysis was doubly blinded, following the same protocol outlined in
\citet{y1shearcat} and implemented in \citet{y1cosmicshear}. 
First, the early stages of this analysis were performed using modified shear
catalogues, wherein each measured ellipticity was multiplied by a blinding
factor. The factor was constructed such that the mathematical bounds of the ellipticity
were unchanged by the transformation.
This catalogue-level blinding was maintained until shortly after 
the point at which the fiducial Y1 results \citep{y1keypaper} were unblinded. 
By this time the basic methodology of
the analysis had been decided and the selection criteria for the galaxy samples were fixed.

Second, higher level blinding was imposed by the authors throughout the course of
this analysis. The axis labels and range of any figures showing cosmological 
parameter constraints were omitted during the blinded period. This was intended 
to prevent unconcious bias from entering the analysis, for example, if the split 
samples were seen to be exhibit significant tensions. The bulk of the analysis, 
including running chains, comparing constraints from colour samples and creating 
figures, and all basic methodological decisions was carried out prior to lifting 
either form of blinding. A small number of notable changes were made after unblinding, 
namely:
(a) generating and validating the multicolour covariance matrix,
(b) running and analysing the chains shown in Figure \ref{fig:results:multicolour}.
Though this could conceivably lead to expectation bias.
We do, however, carry out a series of validation tests, which involve comparing
subsections of the new covariance matrix (and the derived constraints) with the
single colour matrices used in the earlier sections of this paper. The cosmology
contours in Figure \ref{fig:cosmology:multicolour_cosmology} were also generated 
only after the multicolour covariance matrix had been finalised.
These steps, while not comprehensive, guard to some extent against such bias.

\subsection{Splitting the Y1 Shape Catalogue}\label{sec:sample_definition}

There are a number of terms used in the literature to classify galaxies,
which are broadly analogous but non-identical.
This paper primarily focuses on two, both of which are ultimately derived
from differences in the flux of a galaxy in different optical bands.
Though these names are often used somewhat interchangeably in the literature,
in the following analysis the terms 
`early-type', `red', `late-type' and `blue' have distinct meanings, as set out below.
The characteristics of these samples are summarised in Table \ref{tab:data:statistics}.
In both cases, we use these flux-based categories as a proxy for galaxy morphology and kinematics, which affect which alignment mechanism(s) are most relevant.

\subsubsection{Spectral Class}

\begin{figure}
\centering
   \includegraphics[width=\columnwidth, angle=0]{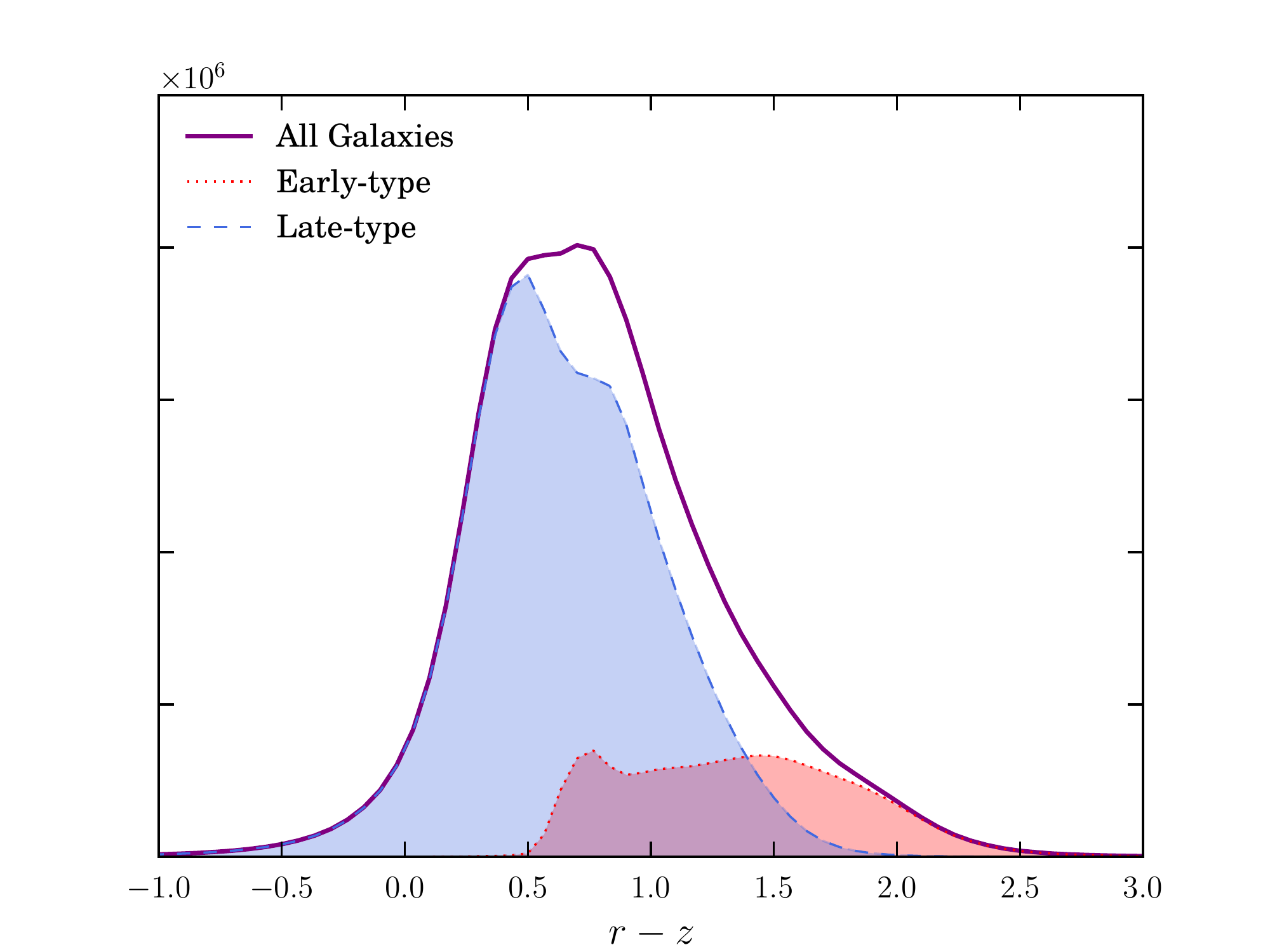} 
   \includegraphics[width=\columnwidth, angle=0]{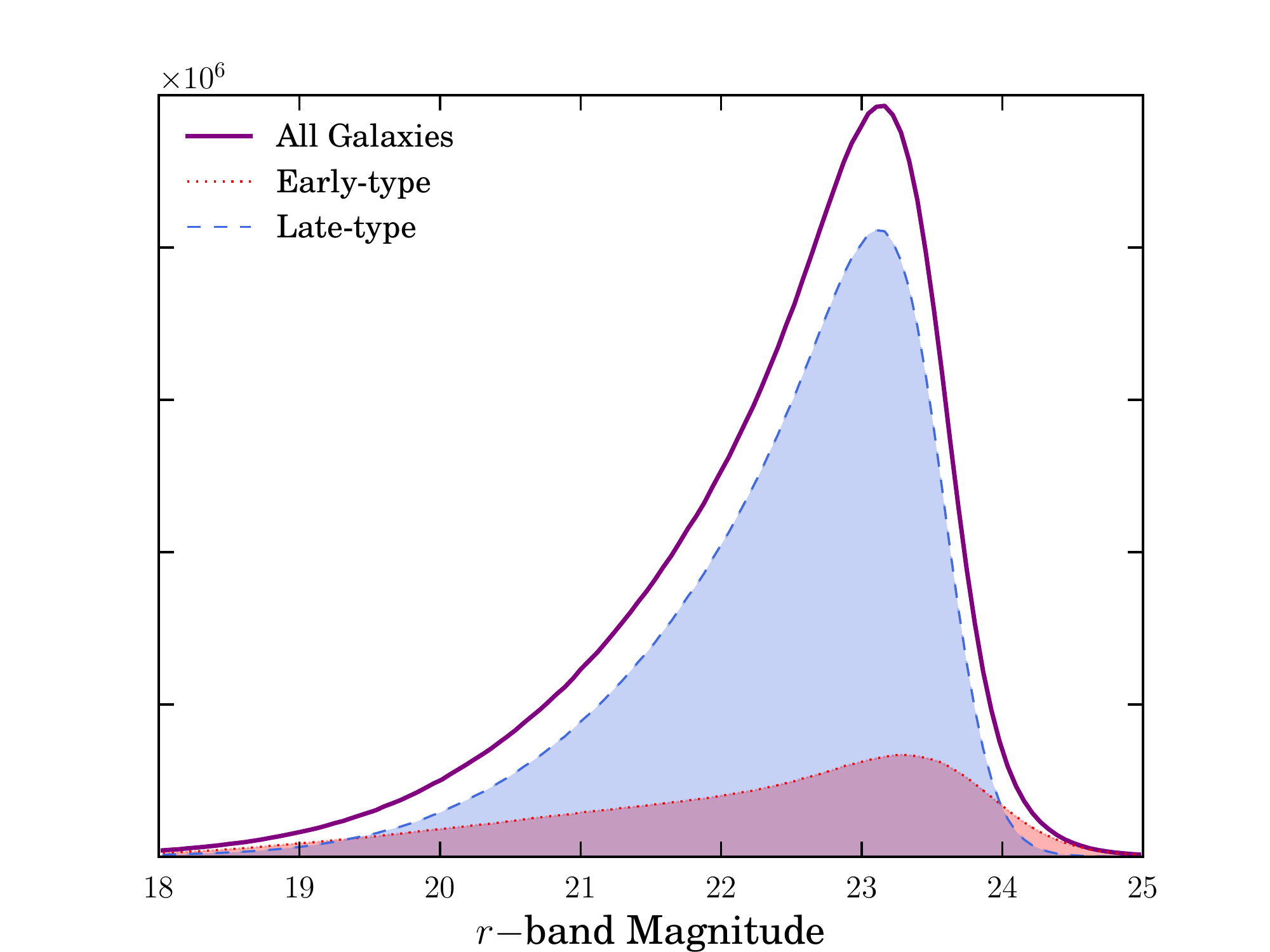} 

   \caption{\emph{Top}: The one dimensional distribution of photometric $r-z$ colour in the
   samples described in this study. The solid purple line shows the histogram of the
   full sample, and the dashed blue and dotted red lines show late-type and early-type
   galaxies only. 
   \emph{Bottom}: $r$-band magnitude distributions of the galaxy samples defined in this paper.
   Note that the solid purple line here is defined by the \metacal~selection flag,
   and corresponds to the dark red histogram in Figure 3 of \citet{y1shearcat}.}
   \label{fig:colour_definitions:mag_histograms}
\end{figure}

A quantity commonly used to split galaxy populations is spectral class.
Template-based photo-$z$ codes such as \bpz\
work by redshifting a library of spectral templates repeatedly.
Fits are performed to produce a likelihood as a function of redshift
for each galaxy, assuming each of the discrete library templates.
The conditional likelihoods are interpolated to produce a single $p(z)$ and a non-integer
best-fitting spectral class \Tb,
which represents an interpolated blend of templates
and acts as a morphological class for each galaxy. 
This quantity has been used in previous studies to divide galaxies expected to have
different systematics \citep{simon13,heymans13}.
We follow those papers and define a boundary at $\Tb=1$
to separate ``early-type" and ``late-type" galaxies. 
Imposing this split on the DES Y1 cosmology sample of \citet{y1cosmicshear},
we obtain early- and late-type samples containing 4.8 M and 28.8 M galaxies respectively.
In Figure \ref{fig:colour_definitions:mag_histograms} we show the distributions of
photometric colour, defined by the difference in magnitudes
between the $r$ and $z$ bands, and $r$-band magnitude in these two populations.

\subsubsection{Photometric Colour}

Another quantity frequently used as a proxy for morphological type
is photometric colour, defined by differences between the measured
brightness of a galaxy in different bands.
The 2D histogram of galaxies in colour magnitude space is expected to be 
bimodal \citep{wyder07}.
In the following we use a boundary in the \rz~plane to define red and blue galaxies,
defined by the equation

\begin{equation}\label{eq:colour_boundary}
r-z = a^{i}_\mathrm{rz} \times r + b^{i}_\mathrm{rz}.
\end{equation} 

\begin{figure}
\centering
\includegraphics[width=0.78\columnwidth, angle=0]{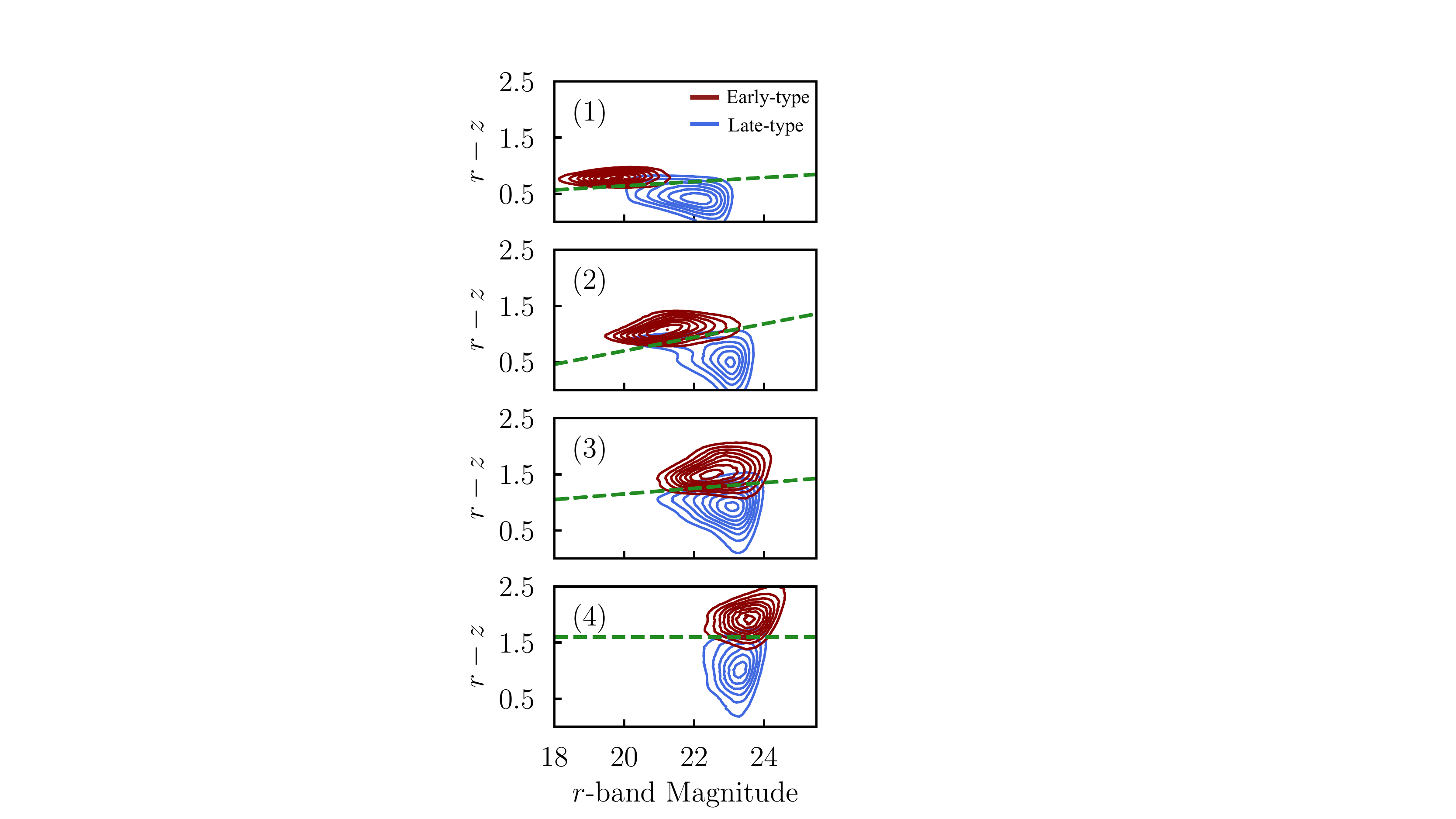}
\caption{The distribution of galaxies in $r-z$ colour-magnitude space in the Y1 \metacal~catalogue.
The panels show galaxies in four tomographic bins, which are labelled in parentheses.
The red contours indicate early-type $(\Tb<1)$ objects only, and the blue contours are the equivalent for 
late-type galaxies. Each is independently normalised to unity.
The green dashed line shows the divisions used to define the red and blue samples described in 
Section \ref{sec:sample_definition}, and are placed such that they roughly mimic the 
split between the red and blue contours. 
}   
\label{fig:colour_definitions:2d_histograms_tomographic}
\end{figure}

\noindent 
Unlike previous studies, we do not have reliable $k$-corrected magnitudes,
nor do we impose selection criteria designed to produce a homogenous low-redshift sample.
To account for the fact that the observed colour-magnitude diagram
is redshift dependent we adjust the values of
the parameters $a_\mathrm{rz}$ and $b_\mathrm{rz}$ 
in each tomographic bin (denoted by the index $i$). The boundary is shifted manually in each bin
to roughly follow the green valley division between peaks, 
and is shown in Figure \ref{fig:colour_definitions:2d_histograms_tomographic}.
In the four DES Y1 source redshift bins we obtain
$\mathbf{a}_\mathrm{rz} = (0.04,0.12,0.05,0.00)$
and
$\mathbf{b}_\mathrm{rz} = (-0.1,-1.7,0.15,1.6)$.

It is worth finally bearing in mind
that there are several similar sets of photometric measurements derived from DES Y1, which are used 
by different authors in slightly different contexts. 
In summary, three useful sets of galaxy fluxes are available to us:
(a) those obtained from the source detection algorithm \blockfont{SExtractor},
(b) the best-fitting fluxes from running our shape measurement code (known as \metacal; see Section \ref{sec:data:shapes}) on the raw galaxy images,
(c) those obtained using \metacal~from reprocessed images with neighbour light subtracted away,
using a technique called Multi-Object Fitting (MOF).
Though (a) are included in the \blockfont{Gold} catalogue,
they are not used in this work.
We use type (b) photometry, and products derived thereof, for the catalogue splits described in this section
as well as for dividing galaxies into redshift bins.
Though MOF partially mitigates the effects of blending
and so is thought to produce more accurate fluxes, type
(c) fluxes are used only for estimating
the galaxy redshift PDFs (see Section \ref{sec:data:pz} below).
This detail arises from an oddity of DES Y1: for computational 
reasons, at the time of writing only one MOF shape run was carried out. 
To allow us to split on (c) type photometry and correctly treat the selection effects induced,
we would require additional MOF runs on several sets of artificially sheared images (see Section \ref{sec:data:shapes}).

\begin{table}
\begin{center}
\begin{tabular}{c|cccc}
\hline
Sample & $N_\mathrm{gal}$  & $z^\mathrm{med}$ & $\bar{r}$ & $r-z$\\   
\hhline{-|----}
All galaxies            & 25.7 M     & 0.57 & 22.2 & 0.79 \\
Early                   & 4.8 M      & 0.65 & 21.9 & 1.31  \\
Late                    & 20.8 M     & 0.55 & 22.3 & 0.64 \\
Red                     & 6.5 M      & 0.61 & 21.8 & 1.25 \\
Blue                    & 19.2 M     & 0.55 & 22.4 & 0.64 \\
Early $\cap$ Red        & 2.3 M      & 0.66 & 21.9 & 1.37  \\
Late $\cap$ Blue        & 18.5 M     & 0.55 & 22.4 & 0.62 \\
\hline
\end{tabular}
\caption{Observational characteristics of the sub-populations defined in
this paper. Note that the mean and median values shown are
weighted by the mean galaxy response
$R=(R_{11}+R_{22})/2$.
\label{tab:data:statistics}}
\end{center}
\end{table}

\begin{figure}
\includegraphics[width=\columnwidth]{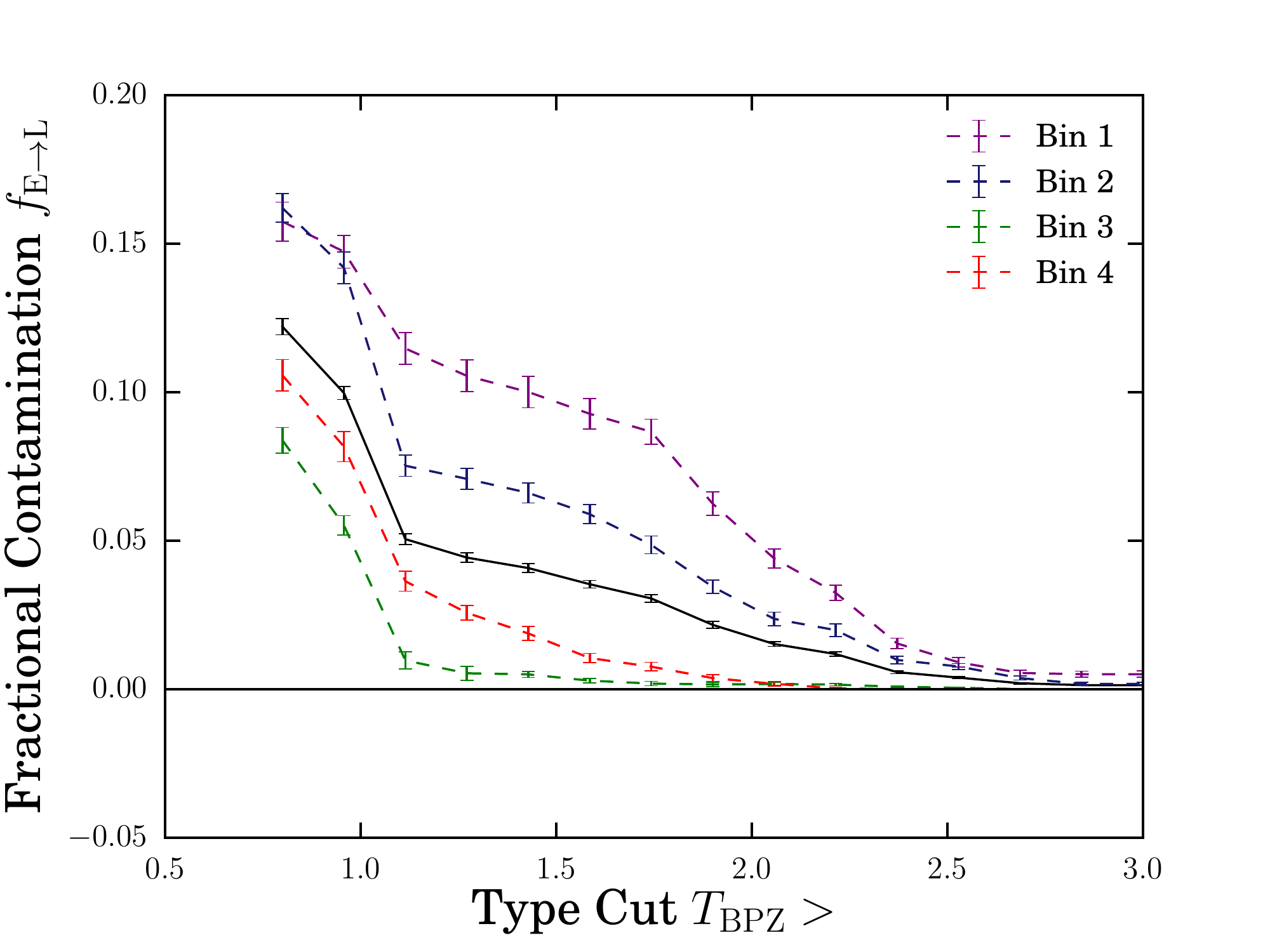}
\includegraphics[width=\columnwidth]{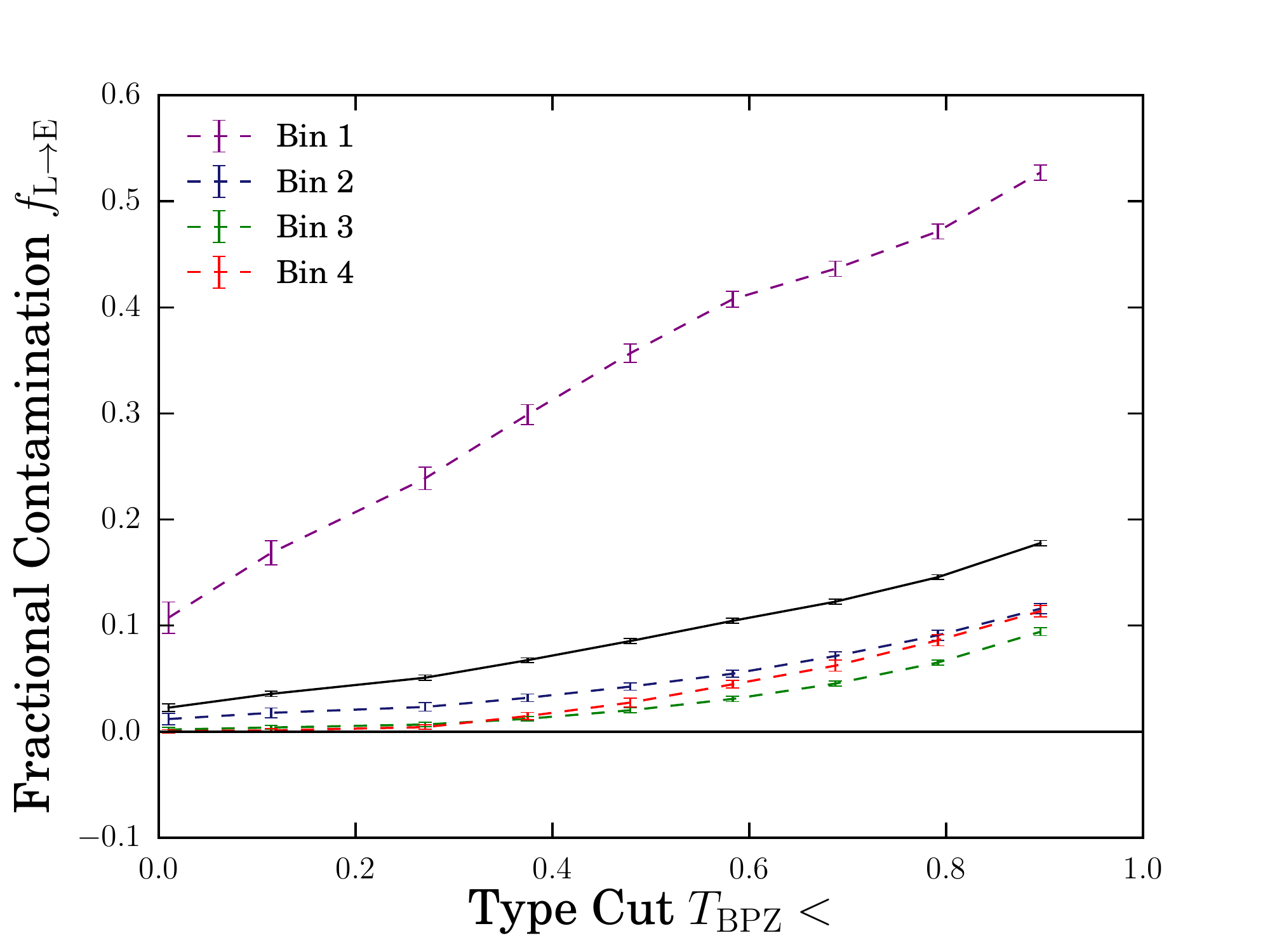}
\caption{Galaxy colour leakage as a function of the upper/lower cut imposed on \Tb. 
The upper panel shows the magnitude of misclassified early-type galaxies that appear in our
late-type sample. The lower panel shows the equivalent late-type to early-type contamination.
The leakage estimator $f_\mathrm{Type 1 \rightarrow Type 2}$ here is defined as the fraction of the lensing weight in each bin 
coming from such misclassified galaxies.
In each case we show the four tomographic bins, as well as the whole unbinned sample in black.}\label{fig:data:type_leakage}
\end{figure}

Finally we attempt to gauge the level of leakage between our galaxy samples.
Since we define our samples about fixed boundary in noisy measured quantities
it is inevitable that there will be some cross-contamination. 
That is, a population of galaxies that, if measured under ideal noiseless conditions
would be classified as one type, but which in reality end up being classified as the other.
We test this as follows.
We re-run the \bpz~algorithm twice on a matched COSMOS sample (described in Section \ref{sec:data:pz}),
(a) using a set of degraded galaxy fluxes designed to mimic DES-like noise levels
and 
(b) using the original fluxes measured with DECam from deeper observations than in the DES wide-field.
This exercise provides a redshift PDF and a best-estimate \Tb~value per COSMOS galaxy.
We then define an early-type sample based on the noisy \Tb~from run (a)
and compute the fraction
of the lensing weight in that sample that is contributed by galaxies where the value from 
run (b) is $\Tb>1$.
The results are shown in Figure \ref{fig:data:type_leakage}.

The leakage is relatively small in most tomographic bins, with the mis-allocated lensing
weight at or below $\sim15\%$. 
The notable exception is the lowest tomographic bin in the early-type sample,
which exhibits a strong fractional contamination.
This can be rationalised in simple terms, as follows;
there is some degeneracy between colour and redshift.
That is, galaxies assigned to the red sample and the lowest redshift bin
can be (a) inherently red, low redshift galaxies
or (b) bluer objects, which have been redshifted and thus appear red.
A similar logic applies, such that a fraction of the 
blue sample galaxies in the upper tomographic bin will actually be inherently
red low redshift objects mistakenly identified. 
The key difference is that the quality of the photo-$z$ for the
red low $z$ objects tends to be superior than for more distant galaxies.
The leakage of blue galaxies into the lowest bin is thus stronger than
the converse.
The significance of this feature for our results is tested 
by rerunning a subset of the
chains in Section \ref{sec:results:main} with the lowest redshift bin removed.
As discussed in that section, the omission of the high-leakage bin
does not produce a significant shift in either the favoured cosmology,
nor the best-fitting IA parameters.

\subsection{Photometric Redshifts}\label{sec:data:pz}

\begin{figure}
\centering
   \includegraphics[width=\columnwidth, angle=0]{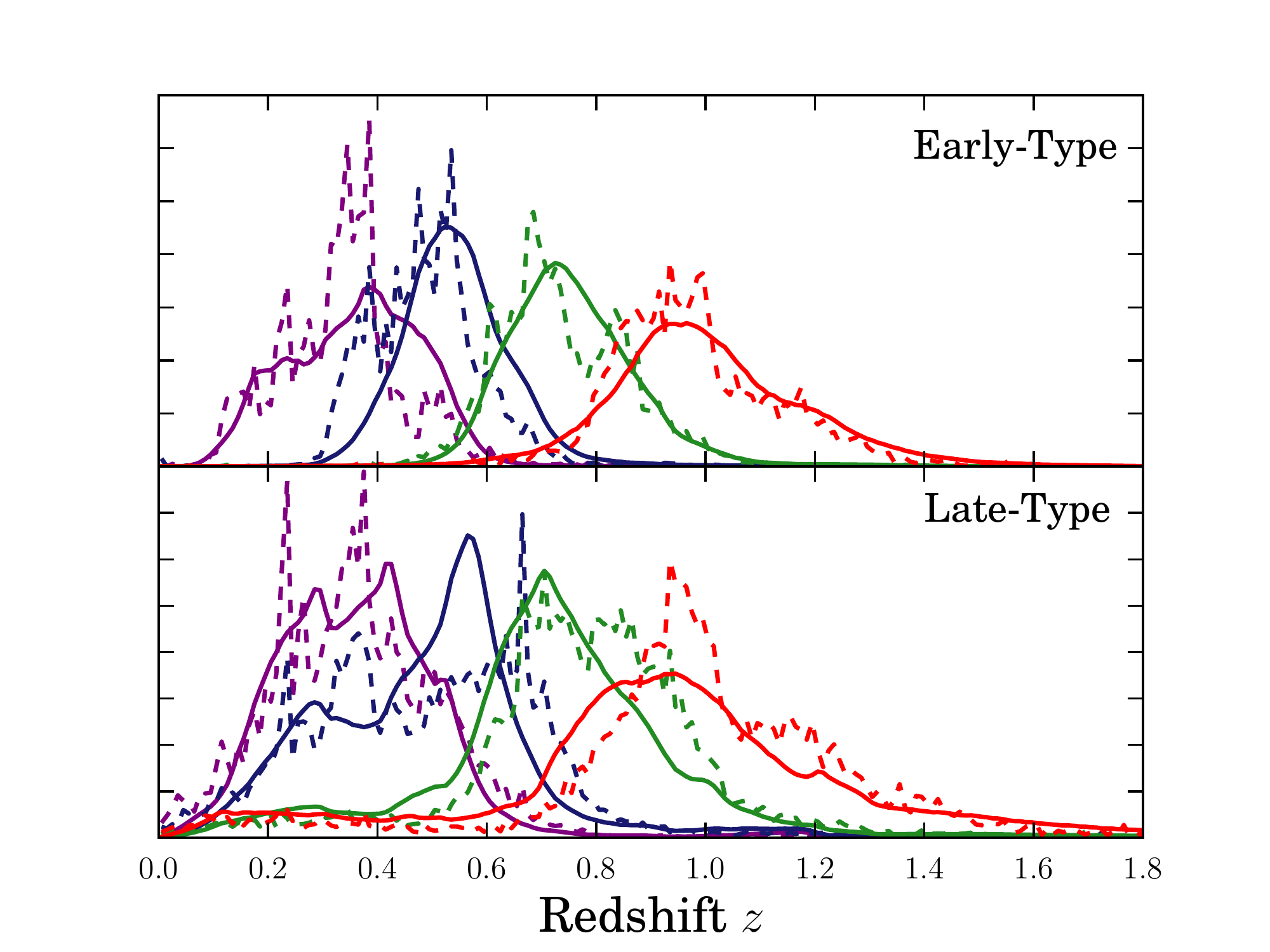}
   \caption{Estimated redshift distributions of the galaxy samples described in this paper.
   The upper and lower panels show our two Y1 type samples, 
   split according to best-fitting template type \Tb.
   The solid curves show estimates derived from the photometric redshift code \bpz.
   The dashed lines are histograms of photo-$z$s for a sample of galaxies from the
   32-band COSMOS data, which has been reweighted to match the DES flux and size distribution. 
   Each distribution is independently normalised to unity over the redshift range shown.
    }
   \label{fig:colour_definitions:f_R}
\end{figure}

We derive estimates for the redshift distribution of our samples
using the \bpz~code \citep{benitez00}. 
The results have been tested using simulations, against a limited
spectroscopic sample and against an alternative redshift algorithm \citep{y1photoz}.
For each sample used in this study, the per-galaxy PDFs are stacked in four tomographic bins
with bounds $z=[0.2,0.43,0.63,0.9,1.3]$.
Galaxies are assigned to bins using the expectation value of 
the $p(z)$ estimated with \metacal~photometry.
The run of \bpz~on the more optimal MOF photometry then provides the $p(z)$
stacked to generate the ensemble $n(z)$ estimates.
We show the measured $n(z)$ 
obtained using \bpz~for $\Tb<1$ and $\Tb>1$ galaxies in 
Figure \ref{fig:colour_definitions:f_R}. 

The main shear selection defined by \citet{y1shearcat} has been subjected 
to a rigorous set of tests designed to constrain this redshift bias
\citep{y1photoz, y1crosscorr, y1crosscorrmethod}.
This information is incorporated into cosmic shear analyses via (non-zero centred)
priors on redshift nuisance parameters.
Unfortunately, one cannot guarantee that these priors will be robust to
arbitrary division of the data.
If we propose to use any subset of the catalogue for tomographic shear measurements,
it is necessary to re-derive appropriate photo-$z$ priors.
To do this we use galaxies from the partially overlapping COSMOS field.
The low-noise 32-band photometry provides high-quality 
point redshift estimates for these galaxies.
In the following we will take these as ``true" redshifts.
In principle we can test for bias in a particular sample 
by comparing the distribution of the COSMOS redshifts
to the ensemble redshift distribution estimates for the
same set of galaxies in the DES images.
Selecting the galaxies in the COSMOS overlap,
however, can itself induce selection effects, since
the COSMOS galaxies are somewhat unrepresentative of DES in magnitude,
colour and size. The COSMOS catalogue is thus resampled such that 
the resulting sample matches the DES Y1 data.
The process results in a set of 200,000 DES galaxies matched to 
COSMOS counterparts with similar flux in four bands $griz$ and size 
(see \citealt{y1photoz} for a full description of the algorithm).

We divide these galaxies into four tomographic bins according to mean redshift,
as estimated from a re-run of \bpz~on the artificially noisy COSMOS $griz$ \metacal\ fluxes.
In each bin we compute a weighted mean

\begin{equation}
\left < z \right >^{(i)} = 
\frac{\sum^{N^{(i)}_\mathrm{gal}}_{j=1} w_j z^\mathrm{C}_j }{ \sum^{N^{(i)}_\mathrm{gal}}_{j=1} w_j},
\end{equation}

\noindent
where $z^\mathrm{C}_j$ is the COSMOS redshift estimate for galaxy $j$ and 
the sum runs over all galaxies placed in redshift bin $i$.
The weight is $w_j$ is given by the mean response (averaged over the two 
ellipticity components; see Section \ref{sec:measurements}).

The offset between the mean COSMOS redshift
and the equivalent weighted mean using the \bpz~Monte Carlo samples from artificially noisy MOF photometry
provides a constraint on the level of systematic bias in the latter.
We derive $\delta z$ in this way for our early, late and full samples, as defined by \Tb.
The result is shown in Table \ref{table:nofz:prior_centres}.

\begin{table*}
\begin{center}
\begin{tabular}{c|cccc}
\hline
Selection & $ \delta z ^{(1)}$ & $ \delta z ^{(2)}$ & $ \delta z ^{(3)}$ & $ \delta z ^{(4)}$ \\
\hhline{-|----}
All Galaxies & $-0.006 \pm 0.018$  & $-0.014 \pm 0.018$  & $0.018 \pm 0.017$   & $-0.018 \pm 0.018$     \\
Early-Type   & $-0.022 \pm 0.020$  & $-0.040 \pm 0.012$  & $-0.008 \pm 0.012$  & $-0.044 \pm 0.014 $  \\
Late-Type    & $-0.003 \pm 0.020$  & $-0.007 \pm 0.023$  & $0.030 \pm 0.020$  & $-0.010 \pm 0.023$    \\
Red          & $-0.034 \pm 0.012$  & $-0.075 \pm 0.011$  & $-0.015 \pm 0.011$  & $-0.060 \pm 0.013$ \\
Blue         & $0.000  \pm 0.030$  & $0.013 \pm 0.025$  & $0.032 \pm 0.024$  & $0.005 \pm 0.027$ \\
\hline
\end{tabular}
\end{center}
\caption{Priors on the redshift error derived from a matched sample of galaxies from the COSMOS field. Photometric redshift estimates for this matched sample are derived from 32-band photometry, as described by \citet{laigle16}.
Note that since we do not attempt a clustering-based estimate of the photo-$z$ error on our colour samples,
the numbers for the full sample are similar but non-identical to the priors on redshift error used in \citet{y1keypaper}.
}
\label{table:nofz:prior_centres}
\end{table*}

These values set the central values of the redshift priors.
In order to decide on an appropriate prior width we must consider a number of sources of uncertainty in this measurements.
We subject the reweighted COSMOS dataset to a series of tests,
outlined in Section 4 of \citet{y1photoz}, which are
designed to constrain systematic uncertainties.
This includes redshift error contributions for statistical uncertainty,
cosmic variance,
and the limited matching process using flux and size only.
The resulting prior widths in each sample are also shown in Table \ref{table:nofz:prior_centres}.

\noindent
In the following we adopt fiducial Gaussian priors for each sample centred according to Table \ref{table:nofz:prior_centres} and with widths
given by the above calculation.

\section{Measurements}\label{sec:measurements}

In this section we outline the measurements needed to set up the parameter inference
detailed in the following section. 
This section seeks to highlight the new measurements and changes in the
Y1 measurement pipeline implemented for this work.
Given that the Y1 lens catalogue used here is identical to that in previous
work, we simply refer the reader to \citet{y1clustering} and \citet{y1keypaper}
for details of the sample selection, binning and two-point measurement. 

\subsection{Galaxy Shapes}
\subsubsection{Measurement and Selection Bias}\label{sec:data:shapes}

To date, two validated science-ready shear catalogues have been built using the DES Y1 data.
The smaller of the catalogues, \blockfont{im3shape}, 
takes a conventional approach to calibrating shear
biases,
relying on a suite of complex image simulations.
A detailed discussion of the processes involved in constructing and
testing such a calibration is presented in \citet{y1shearcat}.
As we point out in that paper, additional selection can very easily induce multiplicative
shear bias.

For this analysis, however, we use the larger of the two shape catalogues.
The measurements are made using a technique called  \metacal, the basis of which is to derive
the calibration from the data itself using counterfactual copies of each galaxy
with additional shear applied.
The algorithm remeasures the shear and computes a quantity known as the \emph{response}:

\begin{equation} \label{eq:Rnum}
\mcalRg_{i,j} = \frac{\est_i^+ - \est_i^-}{\Delta \gamma_j},
\end{equation} 

\noindent
where $\est^+$ and $\est^-$ are the measured values of the ellipticity obtained from images
of the same object sheared by $+\gamma$ and $-\gamma$,
and $\Delta \gamma = 2\gamma$.
The galaxy response must be included whenever a shape-derived statistic is calculated.  
We refer the reader to \citet{sheldon17} and \citet{huff17} for a full explanation of the
algorithm and to \citet{y1shearcat} for details of the implementation used in
DES Y1 and a recipe for applying response corrections.

It is also possible to correct for selection bias using a similar
calculation. 
To do this we must measure the response of the mean ellipticity to the selection function.
Imagine for example, we wish to make a cut on galaxy type \Tb.
Since the photometry, and thus \Tb, 
are not independent of ellipticity the raw cut may induce shear
selection bias.
The photometry must be estimated five times per
galaxy: once in the original images,
and in four counterfactual sheared images.
From each set of photometry we re-evaluate \Tb~and thus derive a slightly different selection mask. 
A mean response \mcalRSmean\ contributed by a selection alone is then defined as the change in 
ellipticity

\begin{equation} \label{eq:RSmean}
\mcalRSmean_{i,j} \approx
\frac{\langle \est_i \rangle^{S+} - \langle \est_i \rangle^{S-}}{\Delta \gamma_j},
\end{equation}

\noindent
where $\left < e \right > ^{S\pm}$ denotes the mean ellipticity measured from
the unsheared images, 
after selection based on quantities measured from the sheared images.
The full response for the mean shear is then given by the sum
of the shear and selection parts,

\begin{align} \label{eq:Rmean}
\mbox{\mcalRmean} = \langle \mbox{\mcalRg} \rangle + \langle \mbox{\mcalRS} \rangle.
\end{align}

\noindent
This must be recalculated each time galaxies are split in any way, including for tomographic binning.
For the fiducial early- and late-type samples (divided about $\Tb=1$) we obtain a mean selection
response of $\left \langle R_S \right \rangle_\mathrm{early} = 0.0018$ and $\left \langle R_S\right \rangle_\mathrm{late} = -0.0006$ respectively.
We obtain a mean response in each sample of 
$\left \langle R_\gamma \right \rangle_\mathrm{early} = 0.6282$
and
$\left \langle R_\gamma \right \rangle_\mathrm{late} = 0.6458$
(compared with $\left \langle R_\gamma \right \rangle = 0.6416$ 
for the unsplit Y1 catalogue).

\subsubsection{Shear Systematics}\label{sec:measurements:shear_systematics}

In this section we repeat a raft of systematic tests designed to ensure the
(sub-)samples used in the following sections are of sufficient
quality for cosmology at the precision of DES Y1.
Although the full catalogue has been subjected to a rigorous set of tests in
\citet{y1shearcat},
it is conceivable that cuts ultimately derived from the observed fluxes
could introduce spurious correlations between ellipticity and galaxy properties.
The most straightforward diagnostic would simply be to measure the mean shear in bins of
observable properties and fit for correlations.

\begin{table}
\begin{center}
\begin{tabular}{c|cc}
\hline
Correlation & Early-Type & Late-Type \\
\hhline{-|--}
PSF $e1$                           & (-0.0340, 0.0031) & (-0.0270, 0.0017) \\
PSF $e2$                           & (0.0014,-0.0338) & (-0.0004,-0.0223)\\
PSF Size $T_\mathrm{PSF}^{1/2}$    & (0.0012, 0.0006) & (0.0001, -0.0004) \\
$S/N$                              & (0.0001, 0.0008) & (-0.0000(5), 0.0004) \\
Galaxy Size $T^{1/2}$              & (0.0006, 0.0009) & (0.0004, 0.0000(3)) \\
\hline
\end{tabular}
\caption{Residual correlations between the galaxy ellipticities and observable properties in our fiducial galaxy samples, after weighting by the mean \metacal~derived response. The numbers in each set of parentheses are the correlations between each quantity and the two ellipticity components $(e_1,e_2)$. The PSF/galaxy size and signal-to-noise ratio follow the definitions in \citet{y1shearcat} and \citet{svshearcat}.
}\label{tab:data:correlations}
\end{center}
\end{table}

The results of this exercise are shown in Table \ref{tab:data:correlations}.
We test for correlations with a number of observable properties, including
seeing (PSF size) and the signal-to-noise of the measurement.
As in the unsplit catalogue, the measured correlations are comfortably at the 
sub-percentage level. We do not consider these to be of concern for cosmological 
analyses at the precision afforded by our data.

Although we do see a significant non-zero correlation between PSF ellipticity
and galaxy shape, the magnitude does not appear to vary significantly as a 
function of galaxy type. This offers some reassurance that there are not
significant selection-based systematics introduced by our cuts.
As in \citet{y1cosmicshear}, we measure the mean shear directly in each tomographic
bin. In both early- and late-type split samples we report 
$|e_{1,2}|\lesssim 10^{-4}$
in all redshift bins.

\subsection{Two-Point Correlations}

\begin{figure*}
\centering
 \includegraphics[width=1.8\columnwidth, angle=0]{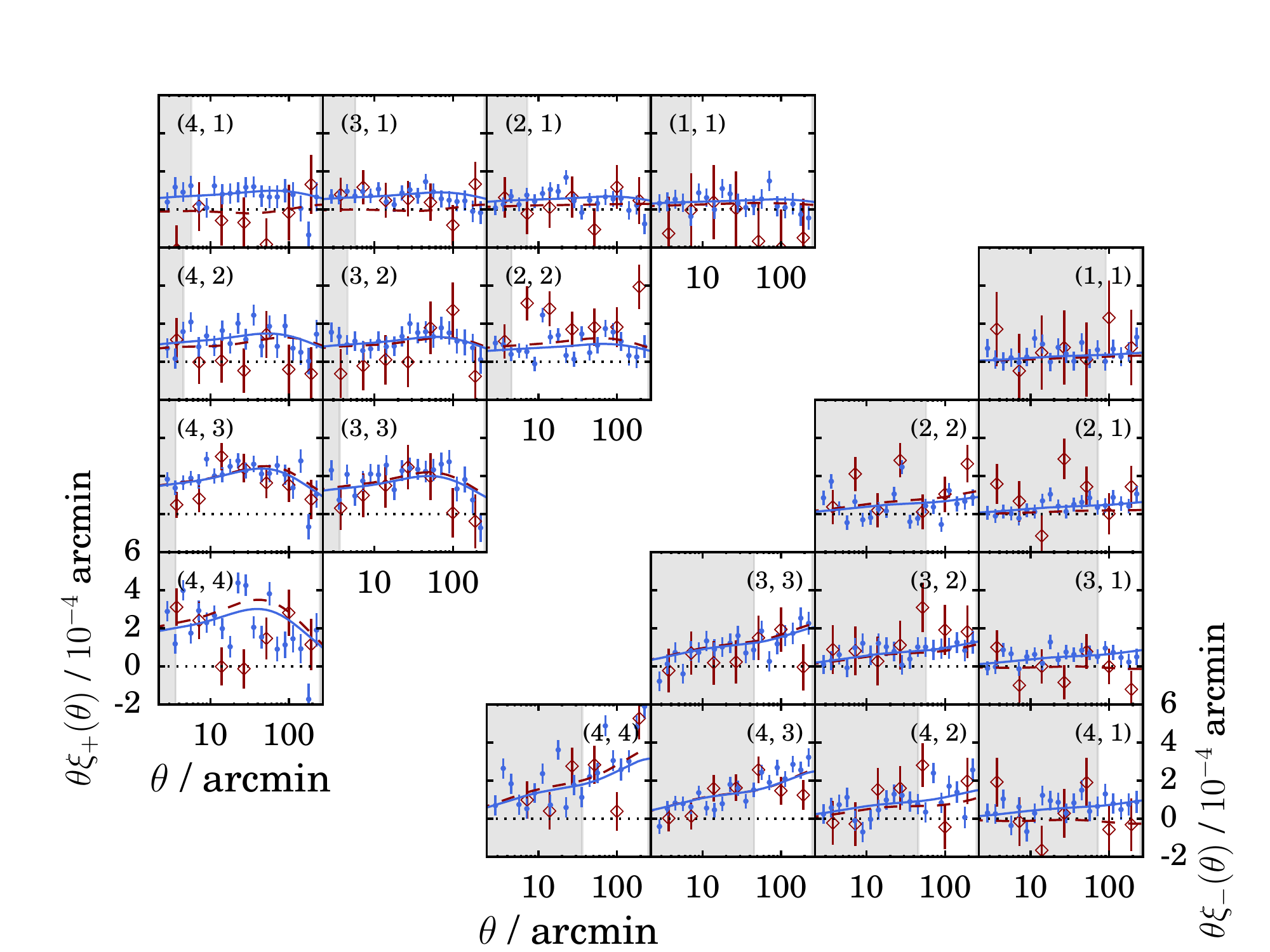}
   \caption{The fiducial cosmic shear datavectors.
   The filled blue points show the correlation functions measured from the late-type sample described in Section \ref{sec:sample_definition}. Open red diamonds show the same, but for our early-type sample. 
   The solid and dashed lines show the best fitting theoretical prediction for each galaxy sample,
   as obtained from a $3\times2$pt analysis.
   The shaded grey regions indicate the scale cuts for each bin pair, with all
   points within these areas discarded prior to parameter inference. 
   The reduced $\chi^2$ obtained from all points outside the shaded bounds is 1.74 with 59 degrees of freedom for the early-type sample,
   and 1.28 with 201 degrees of freedom for the late-type sample.}
   \label{fig:data:xipm}
\end{figure*}

\begin{figure*}
\centering
 \includegraphics[width=1.8\columnwidth, angle=0]{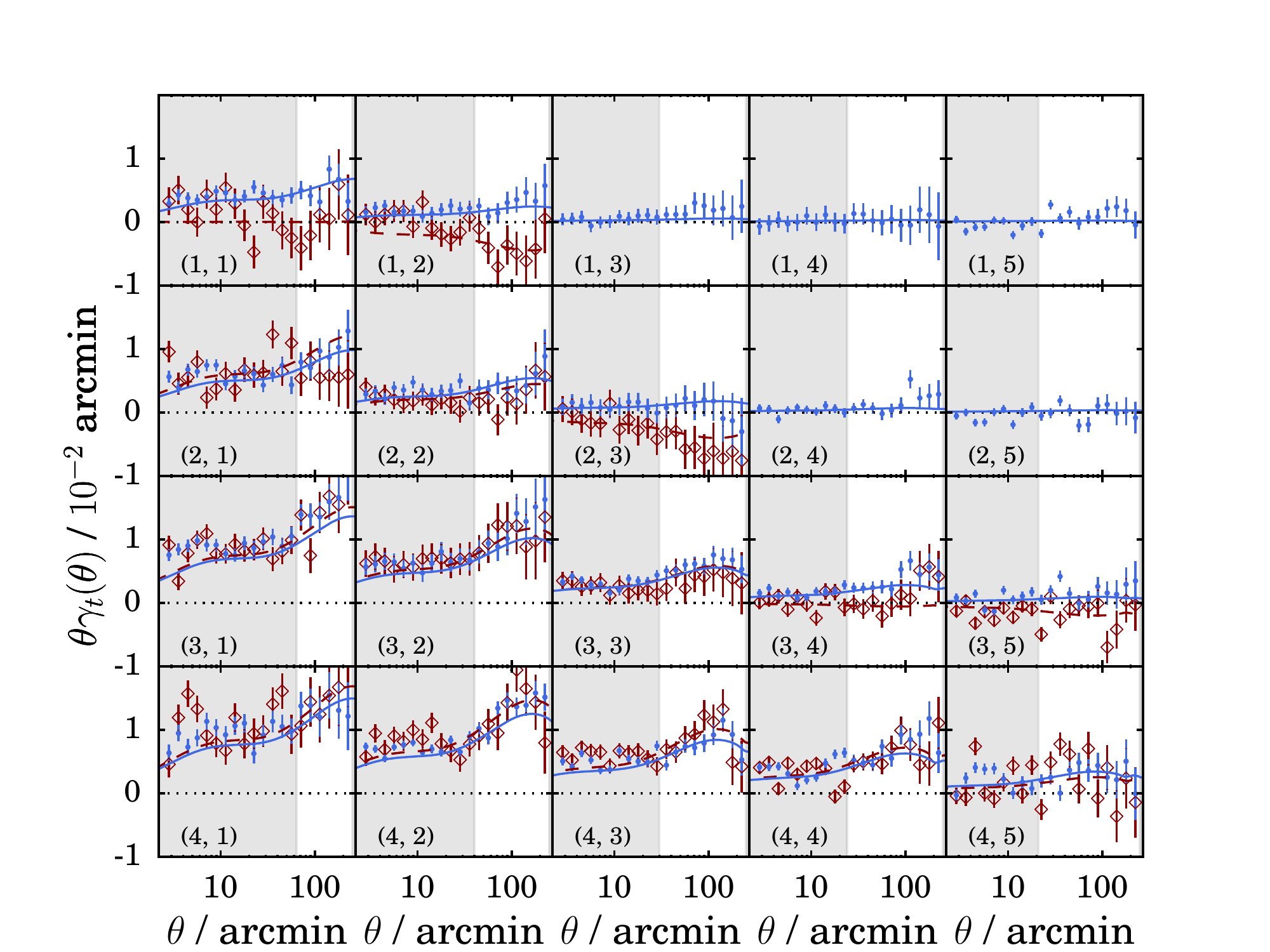}
   \caption{The fiducial galaxy-galaxy lensing datavector. 
   As in Figure \ref{fig:data:xipm}, filled blue points show measurements
   on our late-type split sample and open red points show those on early-type galaxies.
   The numbers in parentheses indicate a pair of redshift bins (source bin, followed by lens).
   The solid and dashed lines show the best fitting theoretical prediction for late-type and early-type galaxies respectively.
   The $\chi^2$ per degree of freedom obtained from all points outside the shaded bounds 
   (plus the galaxy clustering data $w(\theta)$, which are not shown here)
   is 1.25 (153 degrees of freedom) for the early-type sample,
   and 1.22 (204 degrees of freedom) for the late-type sample. 
   A small number of correlations were discarded (and so are missing from this figure)
   because the estimated redshift distribution of lenses had significant weight above the
   equivalent distribution of source galaxies. }
   \label{fig:data:gammat}
\end{figure*}

This work makes use of three sets of correlation function measurements:
between galaxy ellipticities, between galaxy positions, and the cross correlation of the two.
All two-point measurements presented in this paper make use of 
\blockfont{TreeCorr}\footnote{https://github.com/rmjarvis/TreeCorr}.
To manage calls to \blockfont{TreeCorr} and handle sample selection
and binning we make use of a DES-specific python wrapper,
which is also publicly 
available\footnote{https://github.com/des-science/2pt\_pipeline}.

The Y1 shear catalogues are used to construct two-point correlation functions
of cosmic shear.
Our method and choice of statistics and redshift binning follows \citet{y1cosmicshear}.
The shear-shear correlations $\xi_+$ and $\xi_-$ are measured
in log-spaced bins in angular scale.
To achieve roughly comparable signal-to-noise, 
measurements on the late-type and blue samples use 20 separation bins,
but those on the early-type and red samples use only seven.   
Galaxy ellipticities are rotated, weighted and averaged in each bin
as

\begin{equation}\label{eq:data:xipm}
\xi^{ij}_{\pm}(\theta) = \frac{\sum_{\alpha,\beta} w^\alpha w^\beta (e_+^\alpha e_+^\beta \pm e_\times^\alpha e_\times^\beta)}{\sum_{\alpha,\beta} w^\alpha w^\beta (1+m^\alpha)(1+m^\beta)}
\end{equation}
where the sums run over pairs of galaxies $(\alpha,\beta)$,
which are drawn from redshift bins $(i,j)$
and 
whose angular separation falls within a bin of some finite width $\theta \pm \Delta \theta$.
The correlation functions for the fiducial early- and late-type samples used in this paper
are shown in Figure \ref{fig:data:xipm}. Shaded regions corresponding to angular scales
discarded in subsequent likelihood calculations.

To avoid the effects of theoretical uncertainties on small scales 
we impose a lower angular scale cut in each bin.
These bounds are relatively stringent compared with contemporary shear
analyses and are set out in more detail in \citet{y1cosmicshear}.
No angular scales smaller than $\theta_+ = 3.61$ arcmin and
$\theta_- = 36.06$ arcmin are used respectively 
for $\xi_+$ and $\xi_-$ correlations.
Although designed to remove the potential contamination of baryonic effects,
this minimum scale cut also reduces the impact of IA on small scales not captures in the NLA or TATT models.
An upper cut of $\theta < 250$ arcmin is also imposed
to remove scales on which additive shear biases become dominant.
The correlation is corrected with an average scale-independent
selection response, as outlined by \citet{sheldon17} and \citet{y1cosmicshear}.

Very similar expressions can be constructed for the other 
two-point  correlations used in this work.
Following \citet{y1ggl}, we use tangential shear about galaxy positions as 
an estimator for the galaxy-galaxy lensing signal:

\begin{equation}\label{eq:data:gammat}
\gamma_t^{ij} (\theta) = \frac{\sum_{\alpha,\beta} w^\alpha e_t(\alpha | \beta)}{\sum_{\alpha,\beta}  (1+m^{\alpha}) w^\alpha}.
\end{equation}

\noindent
The ellipticity notation $e_t(\alpha|\beta)$ represents the $+$ component
of source galaxy $\alpha$ relative to the position of lens galaxy $\beta$.
Due to the stronger signal-to-noise of the galaxy-galaxy lensing signal,
we use 20 bins for both the early- and late-type samples.
We make an empirical correction for additive systematics, which commonly affect large scale
galaxy-galaxy lensing correlations, 
by evaluating $\gamma_t$ around random points drawn from the Y1 footprint and subtracting
the result from the estimated signal around galaxies.
The random points are drawn from the DES Y1 footprint,
excluding masked regions. For a longer discussion of the random 
subtraction and the impact it has on the galaxy-galaxy lensing measurement
see \citet{y1ggl} (their Sec IV A and Appendix B).
We do not incorporate boost factors into this analysis, but rather
follow \citet{y1ggl} and apply a scale cut at 
$12 h^{-1}$ Mpc comoving separation 
(corresponding to the grey shaded portions of Figure \ref{fig:data:gammat} ).
This is designed to remove scales thought to be significantly
impacted by nonlinear bias,
and comfortably removes the sections of the data where
source-lens contamination is non-negligible.
Similarly to with cosmic shear, these minimum scale cuts also reduce potential contamination from
IA on fully nonlinear scales.

This analysis explicitly excludes galaxy-galaxy lensing measurements where
there is a significant probability that the source galaxy is in front of the
lens. That is, we reject correlations where the estimated lens redshift distribution
is peaked significantly higher than the source redshift distributions.
Due to slight differences in the early- and late-type $n(z)$, 
this cut removes $\gamma_t$
correlations between the lowest early-type redshift bin and the upper three
lens bins, but leaves the late-type datavector unchanged.

Finally, the angular clustering auto correlation is constructed, 
mirroring the choices of \citet{y1clustering}, 
from a mixture of galaxy positions $D$ and random points $R$ using the Landy Szalay estimator 
\citep{landy93},

\begin{equation}\label{eq:data:gammat}
w^{ij} (\theta) = \frac{D^i D^j - D^iR^j - R^iD^j + R^iR^j}{R^iR^j}.
\end{equation}

\noindent
The positions in the galaxy catalogue $D$ are sorted into tomographic bins,
denoted by the Roman index $i,j$.
The random points $R$ are also assigned randomly to tomographic bins,
such that the number of randoms per bin matches the number of galaxies.
As the sample used for galaxy clustering measurements is the same as that 
described in \citet{y1clustering}, we do not show the resulting correlation functions,
but refer the reader to Figure 3 of that paper.

The three measurements on the unsplit sample have passed a raft of null tests 
(\citealt{y1shearcat}, \citealt{y1cosmicshear}, \citealt{y1ggl}, \citealt{y1clustering}),
and show no indication of significant B-modes.
We measure the two-point correlations separately in the full catalogue, and 
also in our fiducial early-type and late-type samples.

\subsection{Covariance Matrix}\label{sec:data:covariance}

The covariance matrix of the two-point data is estimated using the
\cosmolike~software package \citep{krause17}.
The calculation employs a halo model to generate four-point correlations,
which are then used to calculate an analytic non-Gaussian approximation of the multiprobe covariance.
For this calculation we assume a flat $\Lambda$CDM
universe with cosmological parameters
$(\omegam, \omegab, \sigma_8, \ns, h) = (0.286, 0.05, 0.82, 0.96, 0.7)$.
Though the covariance matrix is cosmology dependent, 
\citet{y1keypaper} have shown that
rerunning the likelihood 
chains with covariance matrices recomputed at the best fitting cosmology
does not induce any significant change in the best fitting parameters obtained from the Y1 data.
The \cosmolike~covariance code has been tested against log-normal
simulations which include the DES survey mask \citep{y1methodology}.
Like almost all previous studies of cosmic shear, our covariance matrix does not include
the impact of intrinsic alignments.
In a similar analysis based on CFHTLenS, \citet{heymans13} justify this in two ways.
First, the galaxy catalogues used in cosmic shear measurements are typically not dominated
by low redshift red population objects, in which IAs are known to be strong
(in absolute terms, and relative to the lensing signal).
Constraints using mixed samples from contemporary shear surveys
have found alignment amplitudes in the range $\aia \sim 0 - 1$.
The impact on the true covariance of the data  
due to the presence of IAs is thus expected to be small.
Second, the red fraction is typically $\sim20\%$ or less.
Imposing a colour split will leave one with a relatively small red sample,
and it is likely its covariance matrix will be dominated by shot noise.

Since the survey properties of DES Y1 are significantly different to those of CFHTLenS,
we seek to verify these assumptions.
To test this we use a fast analytic 
code\footnote{https://ssamuroff@bitbucket.org/ssamuroff/combined\_probes\_cosmosis-standard-library}
to generate Gaussian covariances for the shear-shear
angular power spectrum $C_{\gamma\gamma}$ in DES Y1-like tomographic bins.
The IA power spectra are modelled using the NLA model with a range of amplitudes.

We proceed by inspecting the shift in diagonal elements of the covariance matrix.
Unsurprisingly (since the dominant GI term will tend to surpress power in the cosmic shear signal)
on most scales ignoring IA in the covariance matrix leads one to overestimate the uncertainties.
This is particularly true in the autocorrelation of the lower redshift bins. 
On the largest scales (small $\ell$) this exercise suggests a potential slight underestimation of 
our errorbars.
Mapping this onto a change in parameter space constraints is, however, a non-trivial exercise.
We test this explicitly by running a series of MC forecasts on noise-free simulated $C(\ell)$
data using Gaussian covariance matrices with $A_\mathrm{IA}=[0,1,3,8]$.
The parameter space is identical to that described in Section \ref{sec:theory}
(all cosmological and nuisance parameters).
Using 20 multipole bins in the range $\ell=[5,2000]$ we find no significant change in the
marginalised parameter contours between these four cases.


\section{Results}\label{sec:results}

This section describes the main results of this paper. 
We outline the baseline constraints obtained from the colour-split samples
described in the earlier sections. The robustness of our results to redshift error and
galaxy colour leakage is tested using a series of high-level validation exercises.
For comparing IA models run on the same data, we make use of two single-number metrics:
the difference in the reduced $\chi^2$ at the respective means of the parameter 
posteriors\footnote{Due to a subsequent correction to the cosmic shear part of the \cosmolike~covariance calculation, 
our $\chi^2$ results differ slightly from those presented in later versions of \citet{y1keypaper} and \citet{y1cosmicshear}
(see \citealt{troxel18} for details). This accounts for the apparently
poor stand-alone $\chi^2$ values shown in Table \ref{tab:results_summary}. 
This is not thought to affect the comparison 
between galaxy samples, or between different models on the same data.}
$\Delta \chi ^2$
\citep{krause16},
and the Bayes Factor $B$ (the ratio of evidence values; 
see \citealt{marshall06} for a functional definition and discussion of its usage for cosmological model comparison).
The evidence ratios quoted are evaluated using \blockfont{Multinest},
but are also tested using the 
Savage-Dickey approximation, outlined by \citet{trotta07}. 
In all cases the two values are seen to agree to $\sim 50 \%$
of the \blockfont{Multinest} estimate.

\subsection{Simultaneous Constaints on Cosmology and Intrinsic Alignments}\label{sec:results:main}

\begin{figure}
\centering
\includegraphics[width=\columnwidth, angle=0]{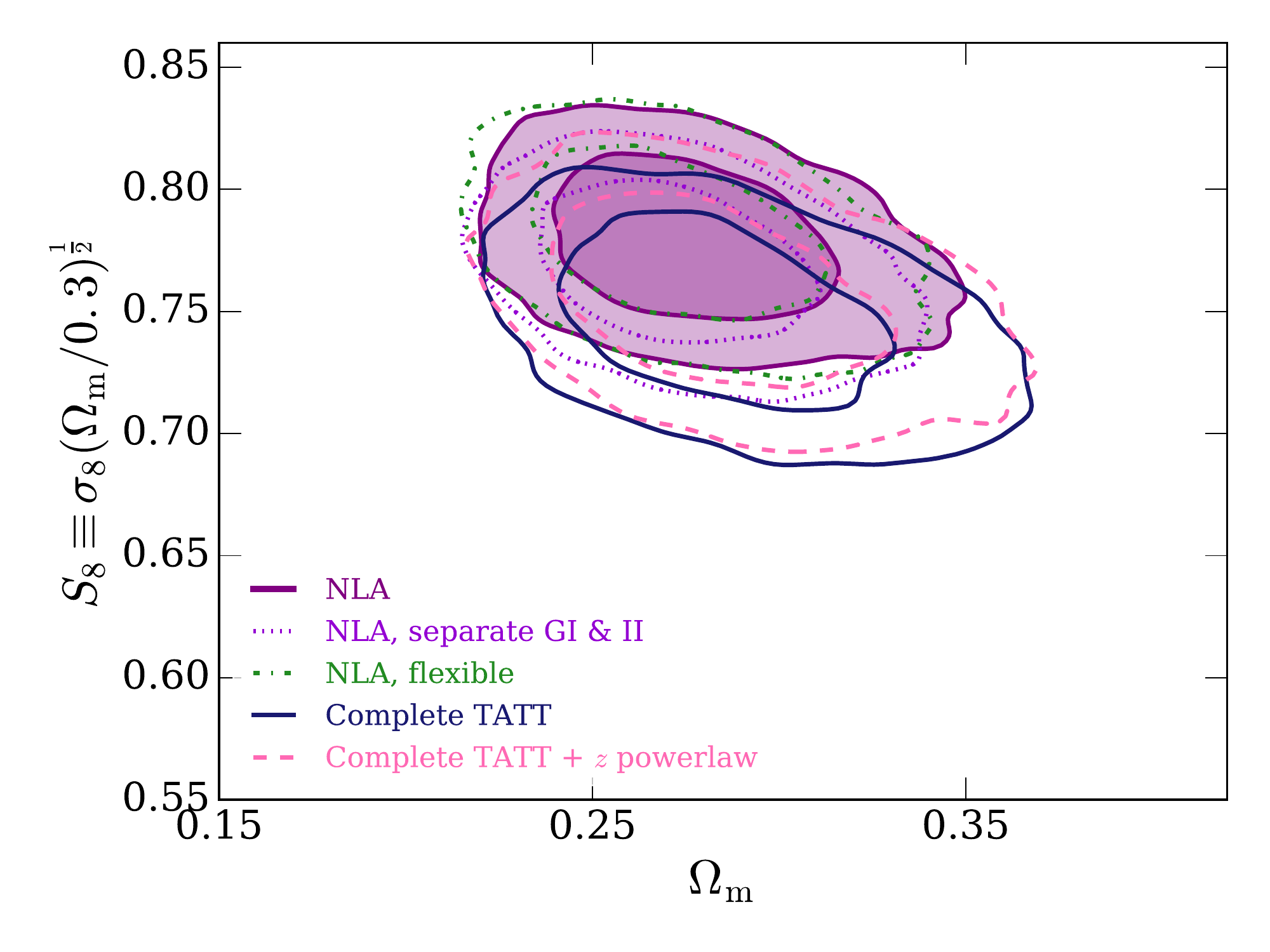}
\caption{\lcdm~constraints on $S_8$ and \omegam from 
cosmic shear, galaxy clustering and galaxy-galaxy lensing,
using the unsplit Y1 cosmology sample. The solid filled (purple)
contours show the baseline analysis, which assumes the nonlinear
alignment model for intrinsic alignments, and is equivalent to the 
blue contours in Figure 11 of \citet{y1keypaper}.
The other lines show extended IA models, the parameters of
which are listed in Table \ref{tab:models}.
}\label{fig:simultaneous_constraints:models}
\end{figure}

Our baseline analysis fits three samples independently
(early-type, late-type and mixed) using the NLA model
for intrinsic alignments in each,
and assuming a \lcdm~cosmology. 
We will, however, consider a number of more complex IA treatments
in the following sections. For reference, the mixed sample
$3\times2$pt cosmology constraints under each of these models 
are shown in Figure \ref{fig:simultaneous_constraints:models} 
(see also Table \ref{tab:models}).
In all cases the posterior constraints on $S_8$
are statistically consistent, though
there are small downwards shifts in some of the models. 
These individual cases are discussed in more detail below.

\begin{figure*}
\centering
\includegraphics[width=\columnwidth, angle=0]{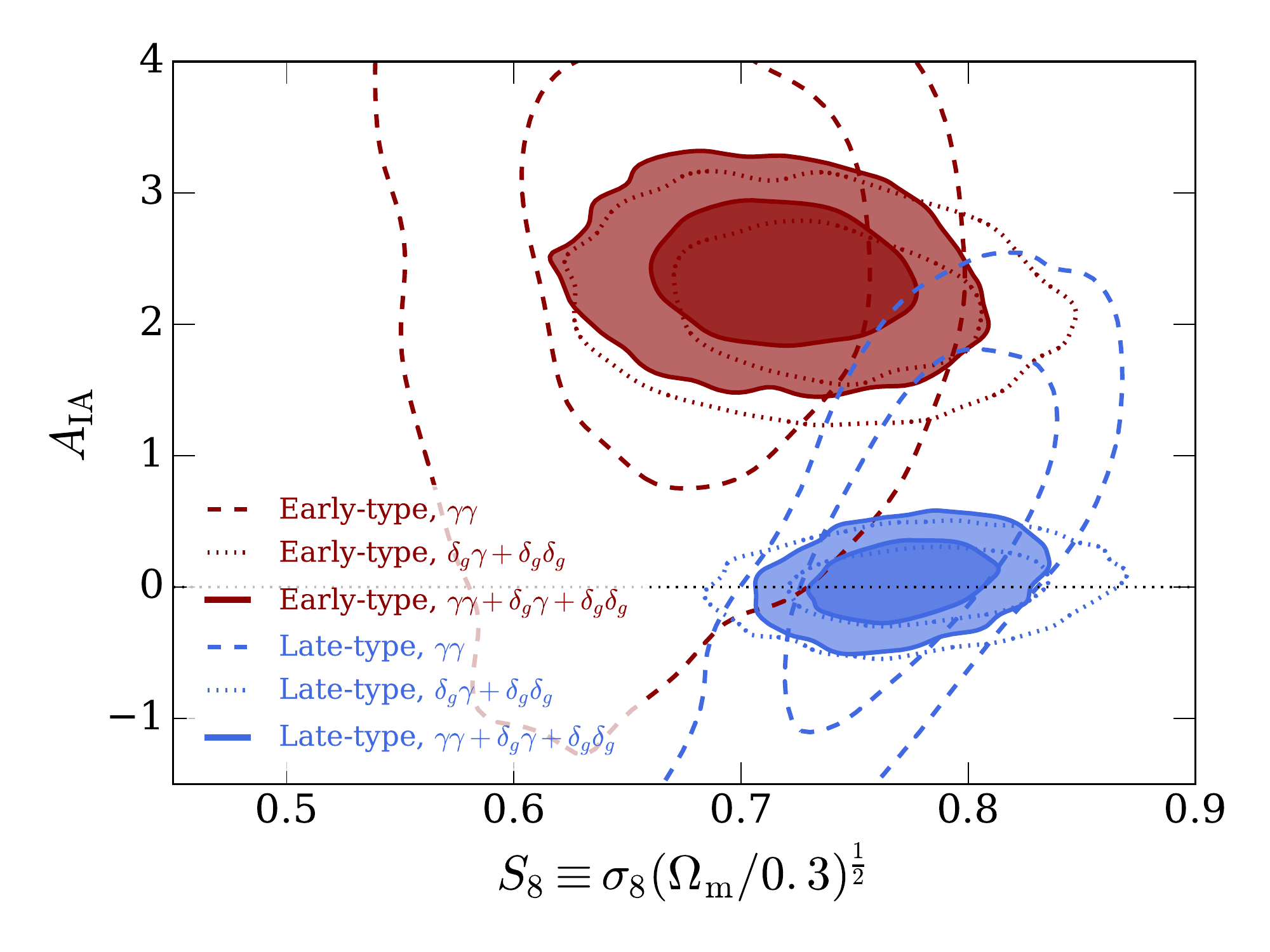}
\includegraphics[width=\columnwidth, angle=0]{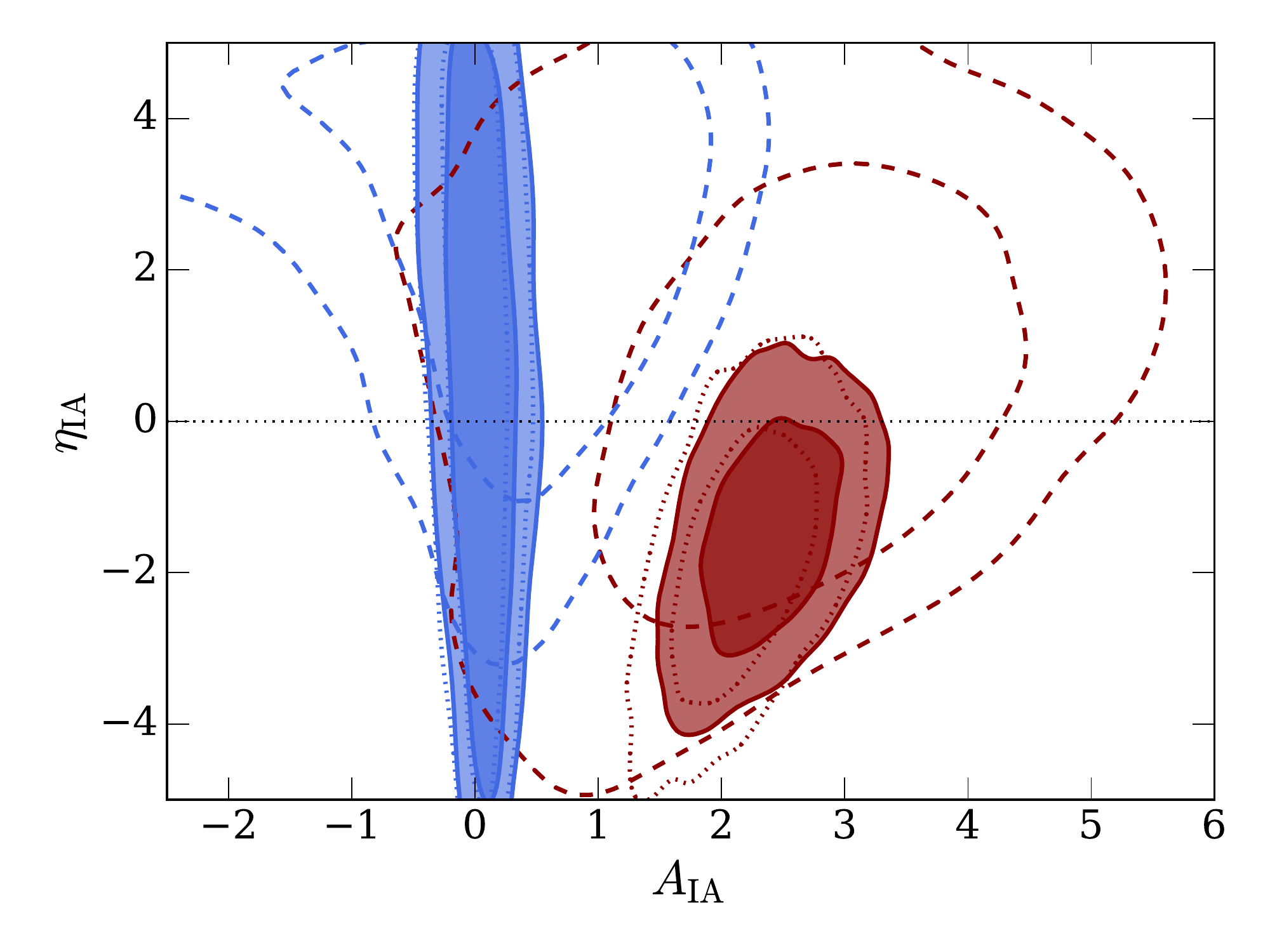} 
\caption{Joint constraints on cosmology 
and a single NLA model intrinsic alignment amplitude from subpopulations of the DES Y1
fiducial shear catalogue.
The two sets of confidence contours are defined by a split according to best-fitting
SED, roughly corresponding to early (red) and late (blue) type galaxies. }\label{fig:simultaneous_constraints:multiprobe}
\end{figure*}

The parameter constraints resulting from the basic analysis are
shown in Figure \ref{fig:simultaneous_constraints:multiprobe}.
The dashed contours show shear alone, 
the dotted show the combination of galaxy-galaxy lensing and two-point clustering
and the solid (filled) contours show the joint constraints from all three probes.
Strikingly, much of the constraining power on the IA model parameters comes from
galaxy-galaxy lensing.
This can be understood as follows:
the II contribution, to which $\gamma_t$ is insensitive,
is generally subdominant in the NLA model.
Combined with the fact that the signal-to-noise on $\gamma_t$
is high (compared with the equivalent shear-shear correlations),
this allows a relatively strong IA constraint from galaxy-galaxy lensing data.
The choice of lens sample is relevant here;
the redshift distributions of the 
\blockfont{redMaGiC} lenses overlap strongly with the
lower source bins, which boosts
the $C_{\delta_g \mathrm{I}}$ alignment term.
The level of sensitivity of a galaxy-galaxy lensing measurement
to IAs will clearly depend on the details of the lens and source
redshift distributions. 
It is, finally, also true that the $\galaxyshear+\galaxygalaxy$ data allows some
level of self calibration,
effectively breaking the degeneracy between intrinsic alignments
and, for example, photometric redshift error.

One notable feature of Figure \ref{fig:simultaneous_constraints:multiprobe}
is the apparent lack of a constraint on the redshift evolution
in late-type galaxies.
Though it is counterintuitive that the $3\times2$pt
analysis should result in a weaker constraint 
on $\eta_\mathrm{IA}$ than cosmic shear alone,
it is understandable in the context of an extended parameter space.
The $\delta_g \gamma + \delta_g \delta_g$ data greatly restricts the allowed range of \aia~about zero,
which reduces the signal-to-noise of the IA contribution 
(in the limit $\aia \rightarrow 0 $ one has no ability to constrain $\eta_\mathrm{IA}$),
resulting in an expansion of the uncertainty on $\eta_\mathrm{IA}$.

Under this model all our results are consistent with zero alignments in late-type
galaxies at any redshift. 
In contrast, 
the IA constraints from the early-type sample are non-zero at the level of
$\sim6.6 \sigma$ with the full $3\times2$pt data.
We also find hints of redshift evolution, with negative $\eta_\mathrm{IA}$
resulting in a signal that diminishes at high redshifts. 
It is worth being cautious here, however, given that 
(a) the deviation from zero is still only just over $1 \sigma$,
and (b) direct comparison with previous null measurements 
(e.g. \citealt{hirata07}, \citealt{joachimi11}) are complicated by a basic difference in analysis method.
Unlike those studies, we do not explicitly model luminosity dependence 
in equation \ref{eq:nla}.
The index $\eta_\mathrm{IA}$ should thus be interpreted as an effective
parameter, which absorbs both genuine evolution of the IA contamination in the same galaxies
and the changing composition of the sample along the line of sight.

Considering the final two columns in Table \ref{tab:results_summary},
we see a slight improvement in the $\chi^2$ of the 
NLA fit to the early-type sample relative to a case with $A_\mathrm{IA}=0$.
More noticeably, the Bayes factor appears to strongly disfavour the reduced model
in this sample.
Though the $\Delta \chi^2$ is close to zero,
perhaps unsurprisingly, the Bayes factors appear to favour the unmarginalised
zero alignment scenario in the late-type sample.  

\subsection{Robustness to Systematic Errors}\label{sec:results:systematics}

In this sub-section we seek to demonstrate that our results do,
in fact, provide meaningful information about IAs and are not
the result of residual systematic errors in our analysis pipeline.

\subsubsection{Shape of the Redshift Distributions}

Though it has been shown \citep{y1cosmicshear} 
that DES Y1 shear-only cosmology constraints
are insensitive to the precise shape of the redshift distributions,
this is not trivially true for IA constraints from sub-divisions of the data.
The kernels entering the IA spectra differ significantly from those in cosmic shear alone;
it is not inconceivable that the favoured IA parameters derived from these spectra
are more sensitive to the details of the $n(z)$ shape than the cosmological parameters.    
To test this we rerun our six fiducial analysis chains, 
replacing the smooth PDFs obtained from \bpz~with histograms of 
COSMOS redshifts (shown in Figure \ref{fig:colour_definitions:f_R}).
Since the means of the two sets of distributions 
per redshift bin are the same
by construction, the comparison gives us an estimate for
how far reasonable changes to the shape of the $n(z)$ might impact upon
our results.
The constraints from this test are not shown, but 
we find only minor changes in the contour size, position and shape for each sample.

\subsubsection{Colour Leakage}

\begin{figure}
\includegraphics[width=\columnwidth]{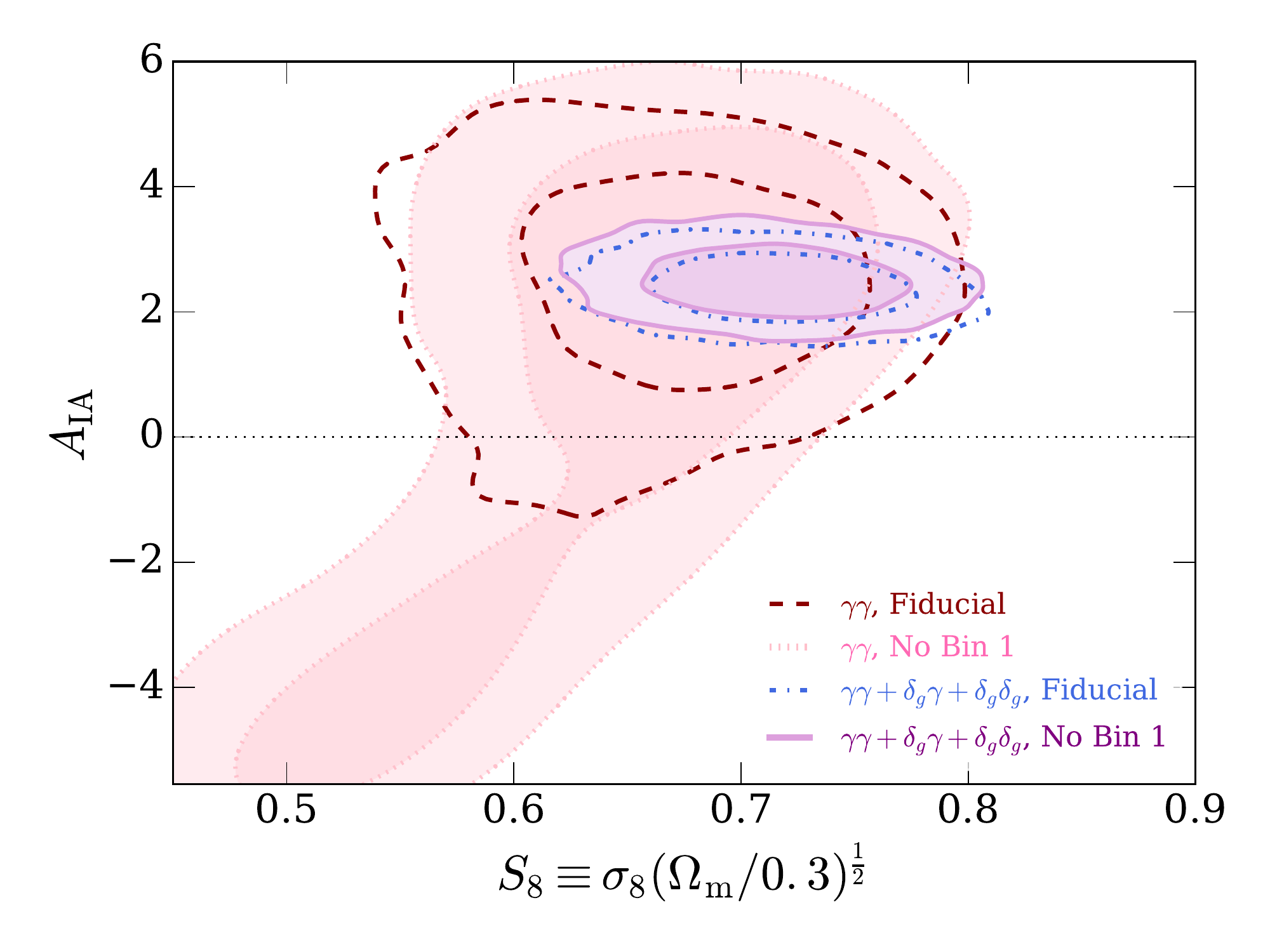}
\caption{The impact of colour leakage on our fiducial results. 
The dashed red and dot-dashed blue lines show the baseline 
$\shearshear$ and $\shearshear+\galaxyshear+\galaxygalaxy$ NLA results for the early-type sample.
These are identical to the red dashed and solid lines in Figure \ref{fig:simultaneous_constraints:multiprobe}.
The filled pink (dotted) and purple (solid) contours show the equivalent constraints in
this parameter space when all two-point correlations involving the lowest lensing redshift bin,
which was found to exhibit potentially strong galaxy type cross-contamination,
are excluded.}\label{fig:simultaneous_constraints:sample_contamination}
\end{figure}

The previous test offers some reassurance that our photo-$z$
error parameterisation is sufficient.
It does not, however, say anything about potential cross
contamination between galaxy samples.
We next seek to test the impact of potential colour leakage.
In Section \ref{sec:measurements:shear_systematics} we saw 
leakage affecting the lowest tomographic bin of the early-type sample more strongly
than any other selection of the data.
To gauge the importance of this we rerun the 
\shearshear~and $\shearshear+\galaxyshear+\galaxygalaxy$ early-type chains,
now explicitly excluding any parts of the data vector involving 
the lowest redshift bin.
The result is shown in Figure \ref{fig:simultaneous_constraints:sample_contamination}.
The best fit of the multivariate posterior is not significantly altered by
these cuts, though we see a degradation in statistical power
in the shear-only case.
In the case of shear alone we also see some level of bimodality
about $A_{\rm IA}=0$.
We note, however, that similar behaviour has been seen before
when adding flexibility to the IA model, particularly in redshift
(see, for example, Figure 8 in \citealt{svcosmology}
and to a lesser extent Figure 9 in \citealt{joudaki17}
).
We thus view the opening up of the parameter space
as an indication of insufficient information
to properly constrain the IA signal without the lowest redshift bin,
not as a cause for concern in itself.

A significant caveat here is that the removal of the 
lowest bin will naturally change the composition of the 
galaxy sample, which in turn could result in a shift in the
IA signal. Unfortunately, it is very difficult to devise a
test of leakage that does not.
Despite this, the fact that the $3\times2$pt constraints
are almost unchanged by this test is reassuring.
It implies that our IA model constraints are not dominated by 
galaxies in the lowest bin, which in turn implies the leakage
seen in that bin is unlikely to be systematically biasing 
our results from the early-type sample. 

Overall, this test does not give us reason to suspect our results are systematically
biased by type-leakage. 

\subsubsection{Splitting Method}

\begin{figure}
\centering
\includegraphics[width=\columnwidth]{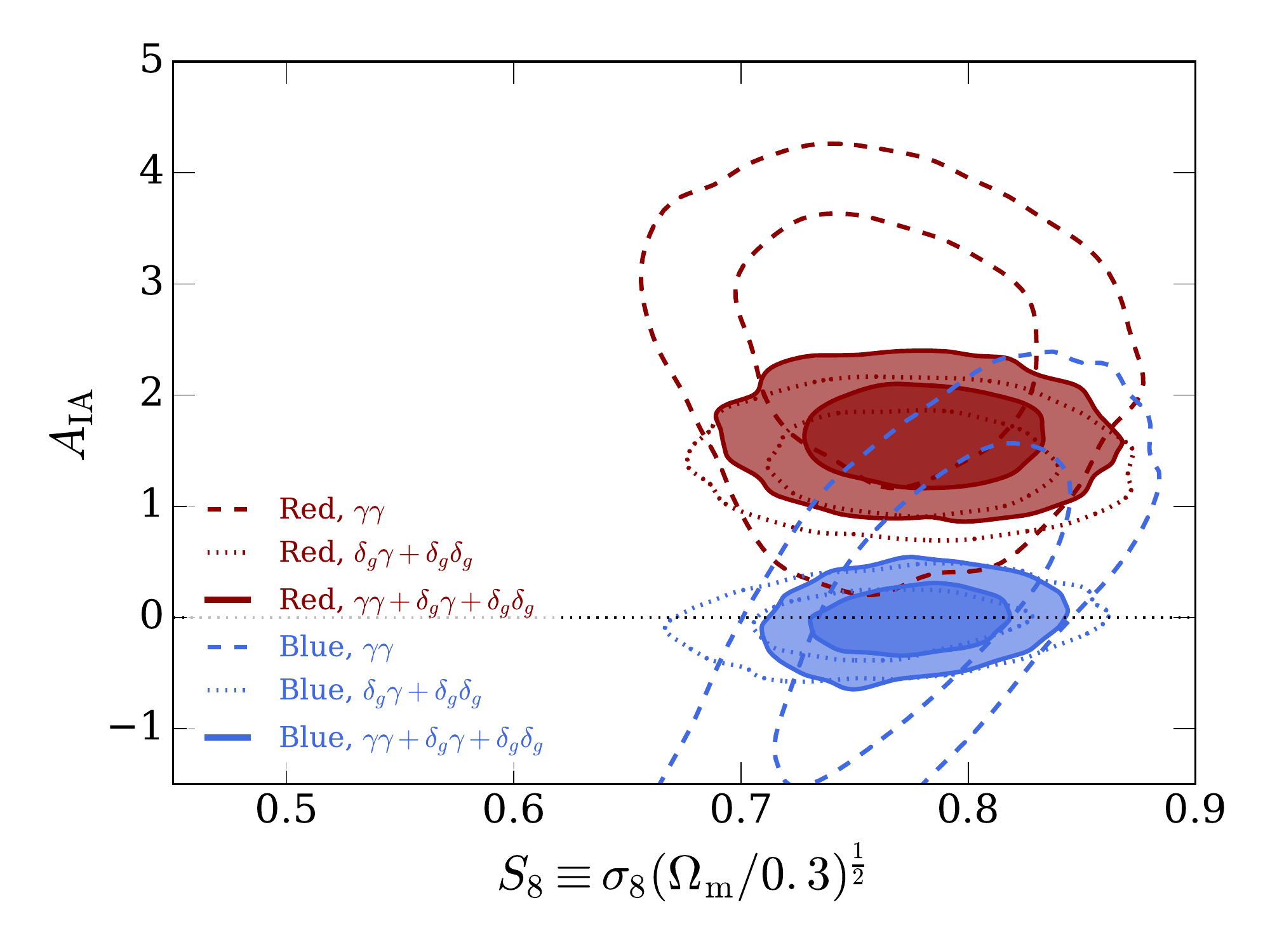}
\includegraphics[width=\columnwidth]{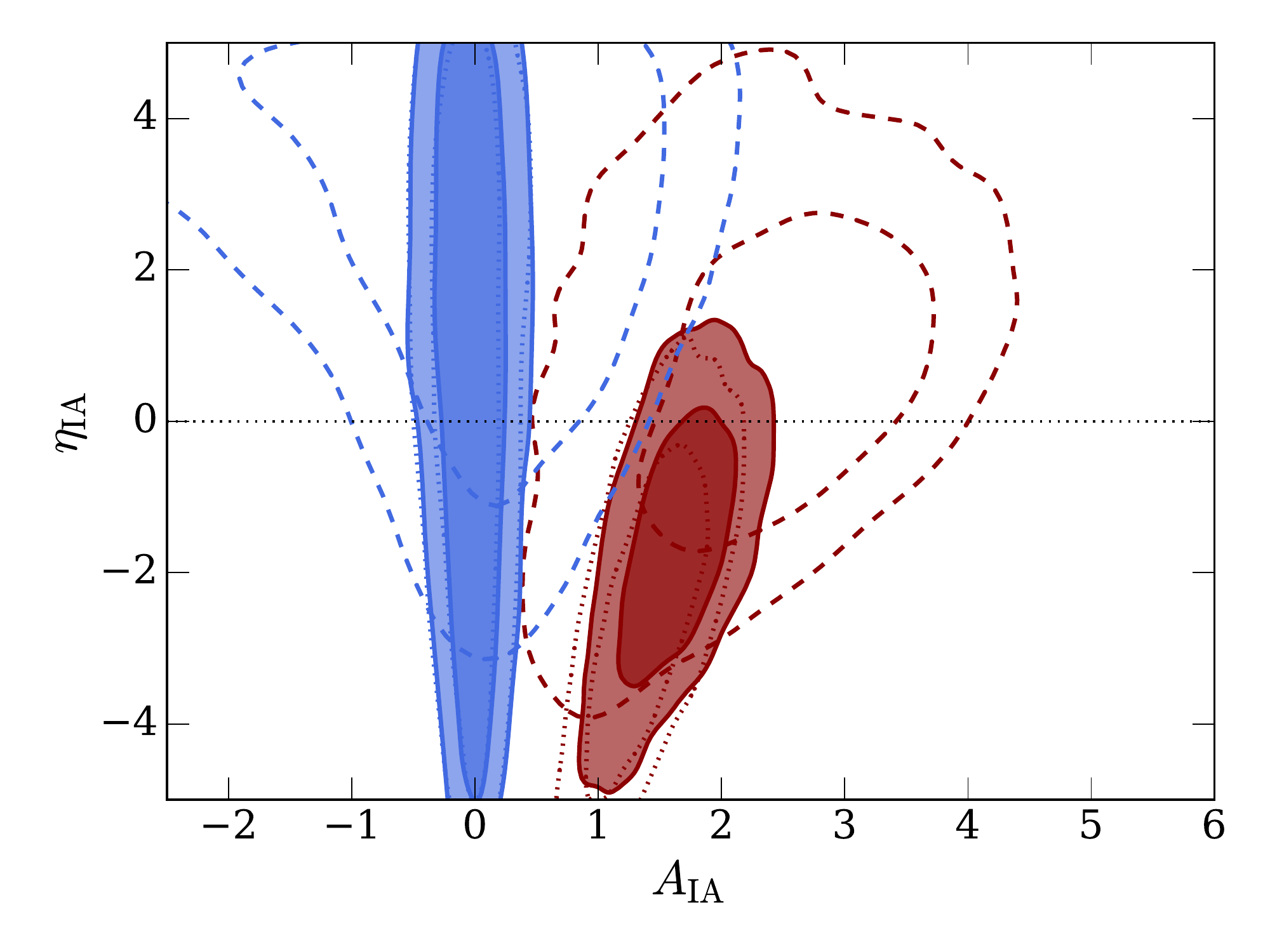}
\caption{Joint constraints on cosmology and intrinsic alignments using galaxy samples split by photometric colour.
The split is implemented independently in each redshift bin using equation \ref{eq:colour_boundary}
and is designed to approximate the evolution of the green valley bimodality in colour-magnitude space.
}\label{fig:colour_split}
\end{figure}

Since we are using a measured quantity (in our case SED type)
as a proxy for galaxy morphology,
one would ideally like the result to be independent (within reason)
of how that proxy is defined.
To test the level at which this is true, we rerun our baseline analysis using
the alternative catalogue split described earlier in this paper
(`red' and `blue' samples; see Section \ref{sec:sample_definition}). 
The constraints from this alternative split sample are shown in Figure \ref{fig:colour_split}.
Though the early-type and late-type samples do not map exactly onto the red and blue
populations, our results here are very similar to those in the fiducial analysis.
The most notable difference is a slight downwards shift in the favoured amplitude 
\aia~for the red sample compared with early-types.  
One interpretation for this might be that the early-type sample is a 
purer population of elliptical pressure-supported galaxies. 
That is, the red sample suffers from 
contamination by objects that appear red in colour
(e.g. due to dust reddening),
but which are morphologically closer to spiral galaxies and
more akin to them in their alignment properties.
The IA signal is thus diluted and the effective amplitude of the sample is shifted downwards.
The qualitative picture is, however, consistent between the two splitting methods. 

\subsection{Model Extensions}

\begin{table*}
\begin{center}
\begin{tabular}{c|cccccc}
\hline
Sample        &     IA Model                              &   Probe                                       &  $A_1$ & $A_2$  & Bayes Factor & $\chi^2 /$ dof  \\
\hhline{=|======} 
              &                                           &   \shearshear                                 &   0  & 0       &  2.00   &  $281.4/203=1.39$  \\
Full DES Y1   &     No IA                                 &   $\galaxyshear+\galaxygalaxy$                &   0  & 0       & $1.74$  &  $275.1 / 206 = 1.34$   \\
              &                                           &   $\shearshear+\galaxyshear+\galaxygalaxy$    &   0  & 0       & 0.83     &  $582.7 / 433 = 1.35$  \\
              &                                           &   \shearshear                                 &   0  & 0       &  1.37 & $  110.7 / 61 = 1.81$\\
Early         &     No IA                                 &   $\galaxyshear+\galaxygalaxy$                &   0  & 0       &  0.0 & $275.1 / 206 = 1.34$    \\
              &                                           &   $\shearshear+\galaxyshear+\galaxygalaxy$    &   0  & 0       &  0.0 & $570.7/382=1.49$   \\
              &                                           &   \shearshear                                 &   0  & 0       &  11.29 & $261.1 / 203=1.29$    \\
Late          &     No IA                                 &   $\galaxyshear+\galaxygalaxy$                &   0  & 0       &  28.76  & $249.7 / 206=1.21$  \\
              &                                           &   $\shearshear+\galaxyshear+\galaxygalaxy$    &   0  & 0       &  33.25  & $530.6 / 433=1.23$  \\
\hline
              &                                           &   \shearshear                                 & $1.03^{+0.45}_{-0.57}$& 0  &  1  & $276.0/201=1.37$ \\
Full DES Y1   &     NLA (fiducial)                                &   $\galaxyshear+\galaxygalaxy$                & $0.38^{+0.13}_{-0.14}$ & 0 & 1  & $270.9/204=1.33$ \\
              &                                           &   $\shearshear+\galaxyshear+\galaxygalaxy$    & $0.49^{+0.15}_{-0.15}$ & 0 &  1  & $575.3/431=1.33$ \\
              &                                           &   \shearshear                                 & $2.37^{+1.16}_{-0.95}$ & 0 &  1  & $102.6/59 = 1.74$ \\
Early         &     NLA (fiducial)                                 &   $\galaxyshear+\galaxygalaxy$                & $2.17^{+0.33}_{-0.32}$& 0  &  1  & $191.0/153=1.24$ \\
              &                                           &   $\shearshear+\galaxyshear+\galaxygalaxy$    & $2.38^{+0.32}_{-0.31}$ & 0 &  1  & $512.4/380=1.35$ \\
              &                                           &   \shearshear                                 & $0.07^{+0.78}_{-1.40}$ & 0&  1  & $256.3/201 = 1.28 $ \\
Late          &     NLA (fiducial)                                &   $\galaxyshear+\galaxygalaxy$                & $0.01^{+0.16}_{-0.17}$ & 0&  1  & $249.6/204 = 1.22$ \\
              &                                           &   $\shearshear+\galaxyshear+\galaxygalaxy$    & $0.05^{+0.17}_{-0.13}$ & 0 &  1  & $531.1/431 = 1.23$ \\
\hline
              &                                           &   \shearshear                               & $2.20 ^{+0.85}_{-1.01}$ & 0 & 1.40 & $103.5 / 60 = 1.72$ \\
Early         &     TA                       &   $\galaxyshear+\galaxygalaxy$              &  $2.05 ^{+0.27}_{-0.25}$ & 0 & $1.35$ &   $192.9 / 154 = 1.35$ \\
              &                                           &   $\shearshear+\galaxyshear+\galaxygalaxy$  & $2.17 ^{+0.27}_{-0.25}$ & 0  & $1.36$ &  $515.9 / 381 = 1.36 $    \\
              &                                           &   \shearshear                               & 0 & $0.08 ^{+0.47}_{-0.51}$ &  $9.44$ & $260.5 / 202 = 1.29$ \\
Late          &     TT                       &   $\galaxyshear+\galaxygalaxy$               & 0 & $0.04 ^{+0.41}_{-0.42}$ &    17.20  & $ 250.2 / 205 = 1.22$  \\
              &                                           &   $\shearshear+\galaxyshear+\galaxygalaxy$  & 0 & $-0.11 ^{+0.45}_{-0.44}$ & $15.58$ & $532.8 / 432 = 1.23$      \\

              &                                           &   \shearshear                               & $0.95 ^{+0.24}_{-0.29}$, & $-2.25 ^{+0.65}_{-0.57}$ & $0.52$ & $266.0 / 201 = 1.32$   \\
Full DES Y1   &     TATT                         &   $\galaxyshear+\galaxygalaxy$              & $0.45 ^{+0.29}_{-0.28}$, & $-0.42 ^{+1.03}_{-1.05}$  &   $0.10$  &   $271.3 / 204 = 1.33$  \\
              &                                           &   $\shearshear+\galaxyshear+\galaxygalaxy$  & $0.97 ^{+0.16}_{-0.16}$, & $-2.28 ^{+0.49}_{-0.47}$ & $2.58$     & $ 569.6 / 431 = 1.32 $   \\

             &                                           &   \shearshear                               & $2.46 ^{+0.87}_{-1.01}$, &  $-3.16 ^{+2.26}_{-1.44}$ & $0.76$ & $101.0 / 59 = 1.71 $   \\
Early   &     TATT                         &   $\galaxyshear+\galaxygalaxy$              & $2.07 ^{+0.30}_{-0.29}$, &  $-0.02 ^{+0.53}_{-0.51}$  & $0.91$   & $192.3 / 153 = 1.26 $    \\
              &                                           &   $\shearshear+\galaxyshear+\galaxygalaxy$  &  $2.21 ^{+0.29}_{-0.28}$, &  $-0.13 ^{+0.56}_{-0.48}$  & 1.20 & $515.6 / 380 = 1.36 $   \\

             &                                           &   \shearshear                               &  $0.53 ^{+0.37}_{-0.51}$, &  $-1.08 ^{+1.45}_{-0.92}$ & $0.23$ & $255.1 / 201 = 1.27 $      \\
Late   &     TATT                         &   $\galaxyshear+\galaxygalaxy$     & $-0.00 ^{+0.20}_{-0.20}$, &  $ 0.03 ^{+0.51}_{-0.53}$,   & 0.75 & $249.9 / 204 = 1.23$ \\         
              &                                           &   $\shearshear+\galaxyshear+\galaxygalaxy$  & $0.34 ^{+0.28}_{-0.39}$, &  $ -1.17 ^{+1.35}_{-0.90}$ &   $0.20$   & $530.6 / 431 = 1.23 $   \\
              &                                           &   \shearshear                               & $0.70 ^{+0.64}_{-1.51}$, & $-1.46 ^{+1.46}_{-1.25}$  &  $0.03$ & $272.1 / 199 = 1.37$   \\
Full DES Y1   &     TATT ($z$ power law)          &   $\galaxyshear+\galaxygalaxy$              & $0.27 ^{+0.23}_{-0.20}$, & $0.13 ^{+0.83}_{-0.81}$  &  $0.01$ &  $271.4 / 202 = 1.34$ \\
              &                                           &   $\shearshear+\galaxyshear+\galaxygalaxy$  & $0.70 ^{+0.21}_{-0.19}$, & $-1.36 ^{+0.54}_{-0.70}$ & 0.07 & $568.0 / 429 = 1.32$     \\
              &                                           &   \shearshear                               & $1.07 ^{+1.61}_{-3.77}$, &  $ -1.68 ^{+2.06}_{-1.97}$  &  $0.03$ & $ 104.6 / 57 = 1.83$    \\
Early   &     TATT ($z$ power law)         &   $\galaxyshear+\galaxygalaxy$                    & $1.89 ^{+0.30}_{-0.28}$ & $-0.04 ^{+0.52}_{-0.50}$ & $0.32$ & $200.3/151 = 1.27$  \\
              &                                           &   $\shearshear+\galaxyshear+\galaxygalaxy$  & $2.17 ^{+0.40}_{-0.38}$, &  $ -0.57 ^{+1.29}_{-1.30}$ & 0.01 &  $515.4 / 378 = 1.37$    \\
              &                                           &   \shearshear                        &  $-1.42 ^{+1.37}_{-1.51}$, &  $ -0.51 ^{+1.48}_{-1.73}$ & $35.68$  & $253.6 / 199 = 1.28$   \\        
Late   &     TATT ($z$ power law)          &   $\galaxyshear+\galaxygalaxy$            & $-0.00 ^{+0.23}_{-0.21}$, &  $ -0.09 ^{+0.87}_{-0.91}$ & $0.01$ & $250.6 / 202 = 1.25 $ \\  
              &                                           &   $\shearshear+\galaxyshear+\galaxygalaxy$  & $0.14 ^{+0.25}_{-0.27}$, &  $ -0.66 ^{+0.94}_{-0.93}$ & $0.02$ & $532.9 / 429 = 1.24$ \\
\hline
\end{tabular}
\end{center}
\caption{Best-fitting parameters and fit metrics for a selection of the analyses discussed in this paper.
The Bayes Factor is defined in this case as the ratio of the evidence obtained from the chain in question
relative to that from the NLA model analysis of the same sample.
}\label{tab:results_summary}
\end{table*}

\subsubsection{Separating GI \& II}

In order to better understand the nature of the IA signal
we next introduce a slight generalisation to our fiducial model.
Although the linear alignment paradigm has II and GI spectra modulated by the same
amplitude, it could argued that one should allow the data to speak for itself where
possible. 
In this spirit, we allow the amplitude and power law index applied to the two IA spectra
to vary independently.  
Our two free alignment parameters are then expanded to four:
$\mathbf{p_\mathrm{IA}} = (A_\mathrm{GI}, A_\mathrm{II}, \eta_\mathrm{GI}, \eta_\mathrm{II})$.
The increased flexibility degrades the $S_8$ constraint somewhat
(see the purple dotted contours in Figure \ref{fig:simultaneous_constraints:models}),
and is accompanied by a small downwards shift in $S_8$.
From the $3\times2$pt analyses we obtain marginalised amplitudes of
$A^\mathrm{early}_\mathrm{GI}=2.64^{+0.59}_{-1.20}$, 
$A^\mathrm{early}_\mathrm{II}=0.02^{+3.35}_{-3.34}$
and
$A^\mathrm{late}_\mathrm{GI}=0.06^{+0.33}_{-0.34}$, 
$A^\mathrm{late}_\mathrm{II}=0.00^{+2.20}_{-2.16}$
for early- and late-type samples respectively.
As expected, the GI term correlates with $S_8$
(as the GI contribution increases the shear signal becomes increasingly diluted,
and so $S_8$ must increase to compensate).
The II amplitude shows a weaker negative correlation. 
With no information about the II part coming from the galaxy-galaxy lensing data,
the constraint on $A_\mathrm{II}$ and $\eta_\mathrm{II}$ is relatively
weak.

\begin{figure}
\centering
\includegraphics[width=\columnwidth]{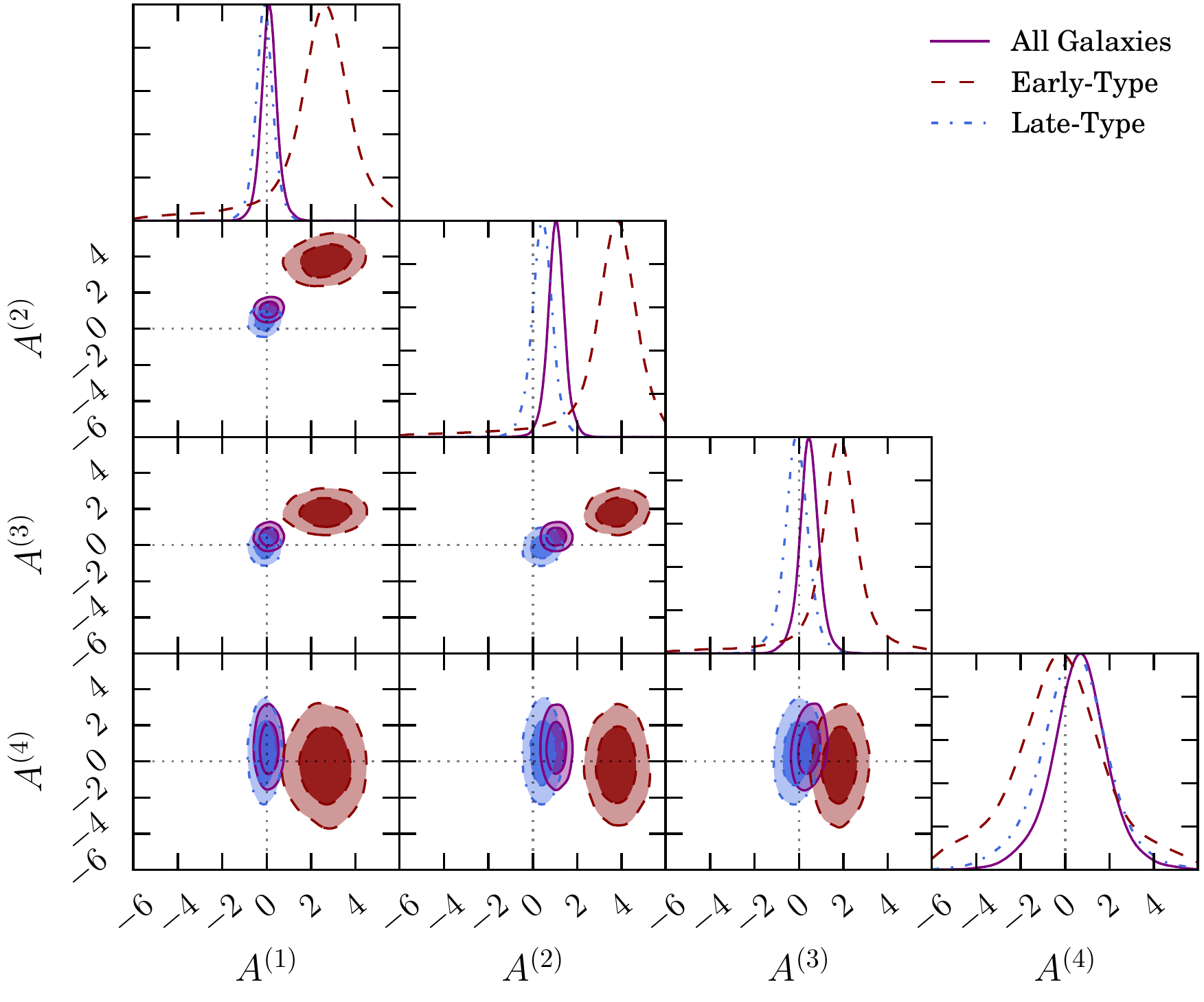}
\caption{Joint constraints on the amplitudes of the IA spectrum in four tomographic bins for the
$\shearshear+\galaxyshear+\galaxygalaxy$ combination.
In each case, the red dashed contours show early-type galaxies, 
the dot-dashed blue show late-types and the shaded contours
show the mixed Y1 cosmology sample.}
\label{fig:results:redshift}
\end{figure}

\subsubsection{Flexibility in Redshift}

We next rerun our fiducial analyses with a free IA amplitude
in each redshift bin. This is analogous to our treatment of galaxy bias,
and simply modulates the IA power spectra used in the projection integrals.
In reality the IA contamination in adjacent redshift slices will, of course, 
be correlated but we expect the impact to be small and 
do not attempt to model this here.
We show the result of this analysis in Figure \ref{fig:results:redshift}.
Although the late-type signal is consistent with zero at all redshifts,
the \aia~amplitude inferred from the early-type sample drops from $\sim3-4$
in the lower bins to consistent with zero in the upper-most bin.
This is consistent with the mildly negative value of $\eta_\mathrm{IA}$
seen in the fiducial analysis.
As before, it is not possible to separate the effects of the changing
composition of the sample from changes in the IA signal for a 
given set of galaxies. The unsplit sample is relatively stable, with 
$A^{(i)}\sim0.5$ in all four redshift bins. As in \citet{y1cosmicshear}, 
we see a mild degradation of constraints on $S_8$ compared with the 
fiducial model. Where in that paper shifting to the more flexible 
alignment model was seen to result in a downwards shift in $S_8$, 
however, in the $3\times2$pt case presented here we find no 
corresponding shift.

\subsubsection{TATT Model (Perturbation Theory)}\label{sec:results:tatt}

\begin{figure}
\centering
\includegraphics[width=\columnwidth]{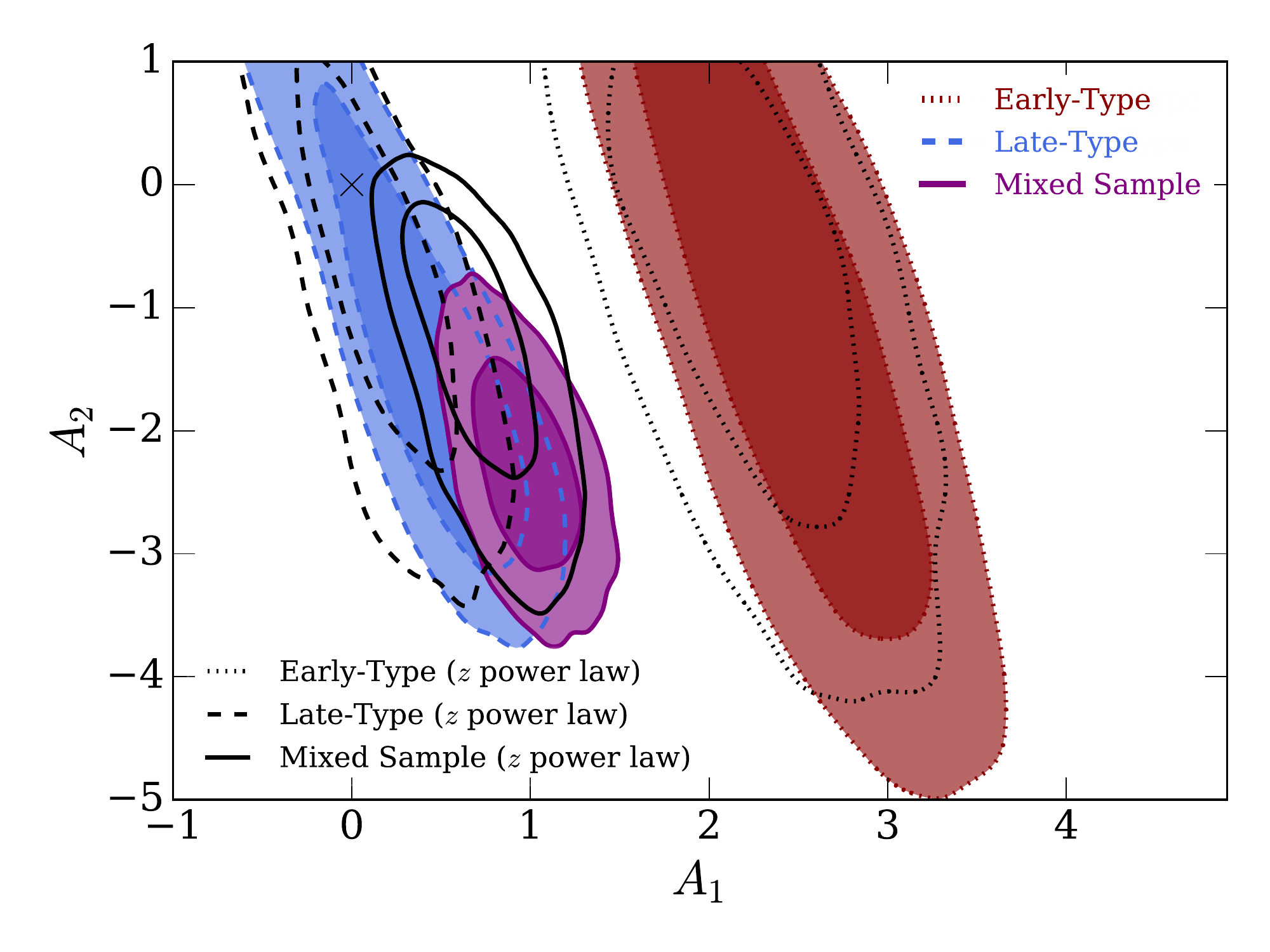}

\includegraphics[width=\columnwidth]{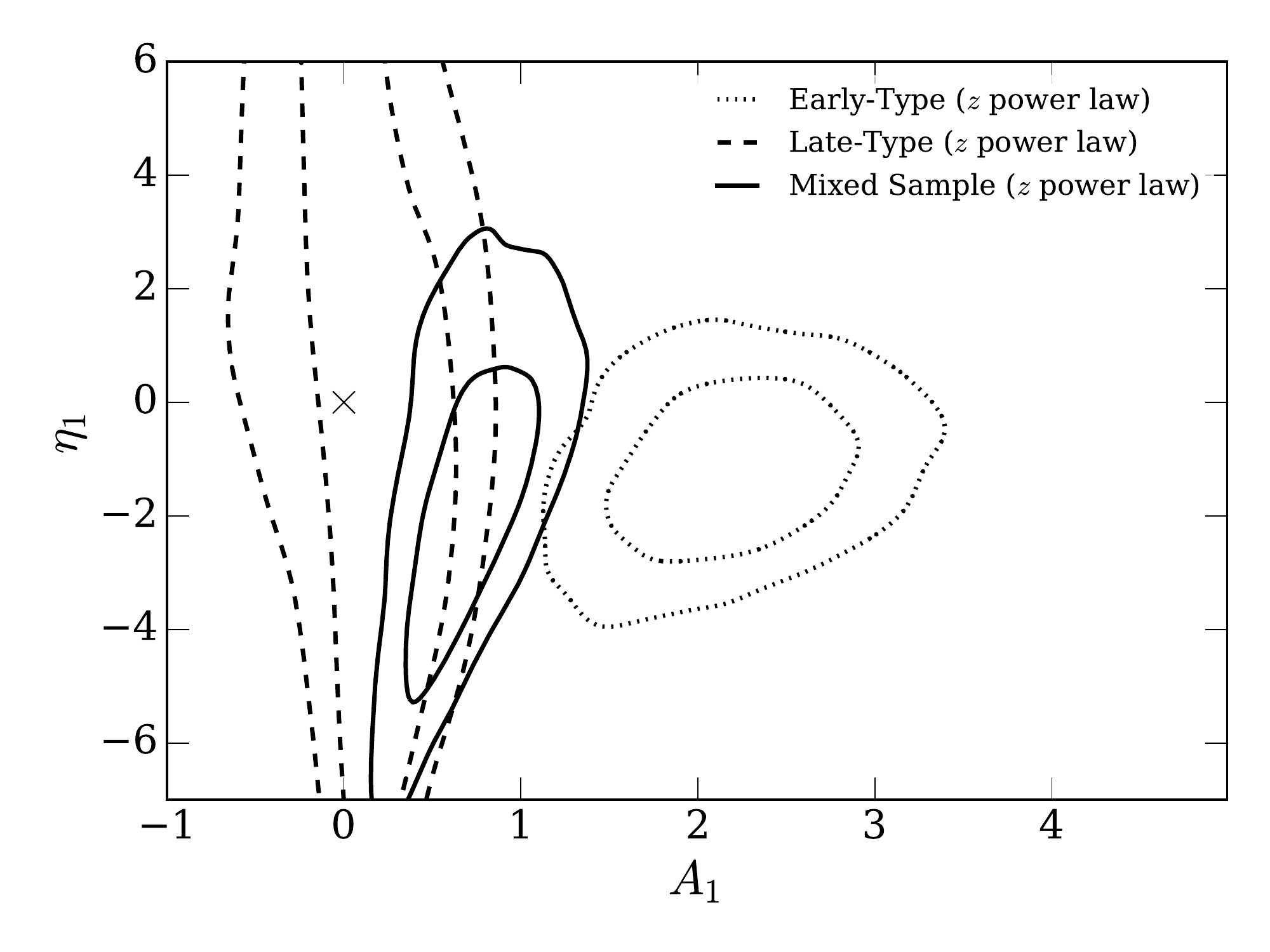}
\includegraphics[width=\columnwidth]{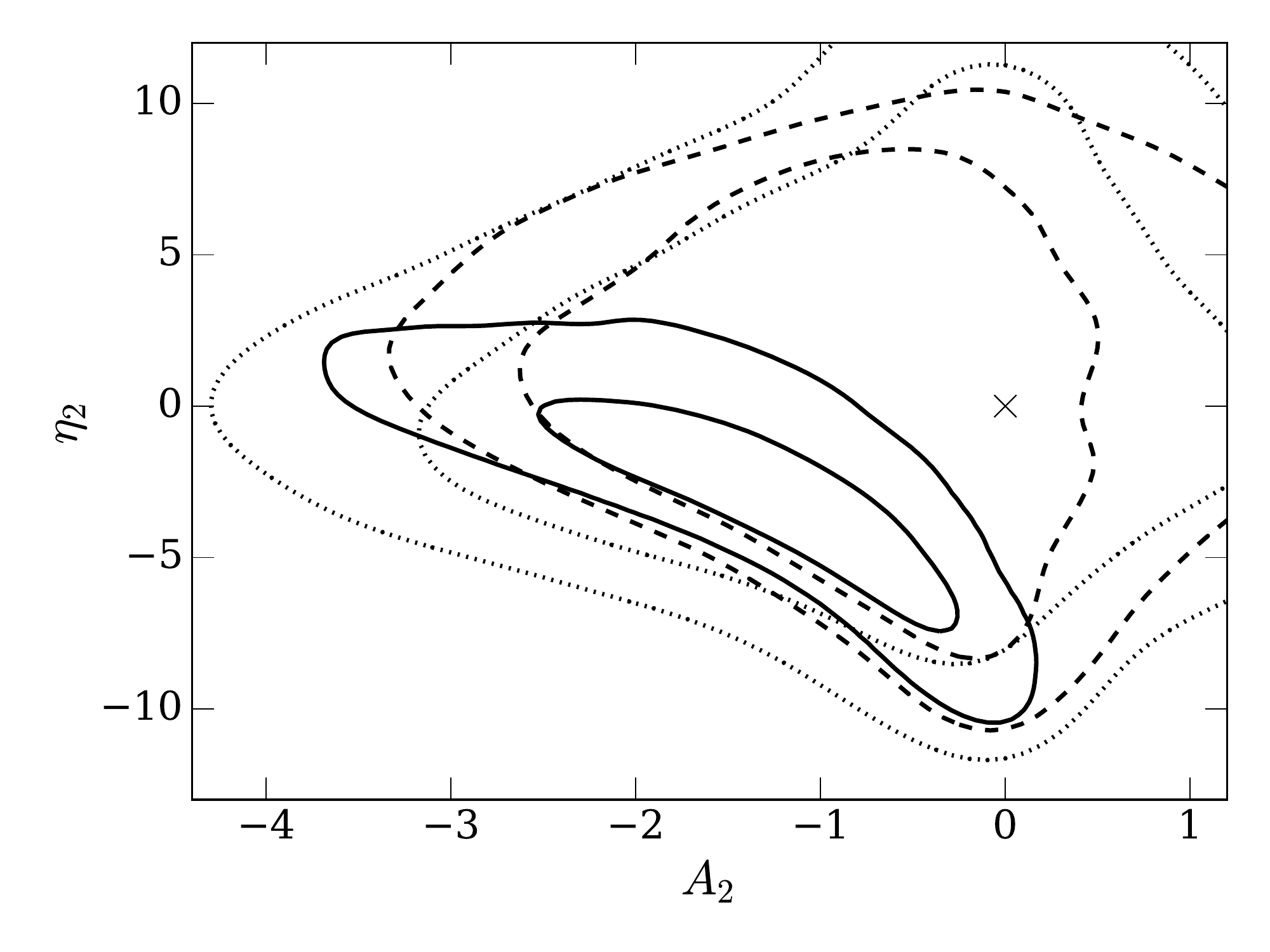}
\caption{Joint constraints on tidal alignment and tidal torque amplitudes in the TATT model.
The three sets of filled contours (dotted red, dashed blue and solid purple)
show the results of fitting the baseline TATT model to each of the 
fiducial early-type, late-type and mixed samples used in this analysis.
The unfilled black contours show the same, but with additional power laws in
redshift $\eta_1$ and $\eta_2$, which are also marginalised.}\label{fig:cosmology:tatt_sample_comparison}
\end{figure}

Prior to this point all of the IA models we have considered have been
permutations of the Nonlinear Alignment Model.
Using this approach for a mixed popluation of galaxies relies on the 
assumption that the IA contribution to the data is 
a pure NLA signal, scaled by an effective IA amplitude,
which absorbs the dilution due to randomly
oriented blue galaxies.
In this section we instead employ the
TATT model described in Section \ref{sec:theory:ia_models},
which includes linear and quadratic contributions.
There are various physically useful variants of this model, 
with different parameters fixed.
For clarity, in the following we will consider, in ascending
order of complexity: 
(a) the TA and TT models, fit to the early-type and late-type
samples respectively;
(b) the TATT model with no redshift scaling;
(c) the TATT model with a free redshift scaling of the
form $[(1+z)/(1+z_0)]^{\eta_i}$, $i\in(1,2)$.

In the simplest case (a), the IA model has only one free parameter
(either $A_1$ or $A_2$ for TA and TT respectively),
but results in 
a significantly different IA power spectra (see Figure 2 of \citealt{blazek17}).
In the TA fit on the early-type sample,
the results closely mirror those from the NLA analysis in 
Section \ref{sec:results:main}; 
this is unsurprising, given that these models are the same up to 
the galaxy density weighted term (in the TA model but not NLA),
and a redshift scaling (included in the NLA model but not TA).
Our results are consistent with the TT IA amplitude in blue galaxies 
being zero (and also with mildly positive or negative values).
The constraints under these models are not shown, 
but are summarised in Table \ref{tab:results_summary}.

\begin{figure*}
\centering
\includegraphics[width=0.99\columnwidth]{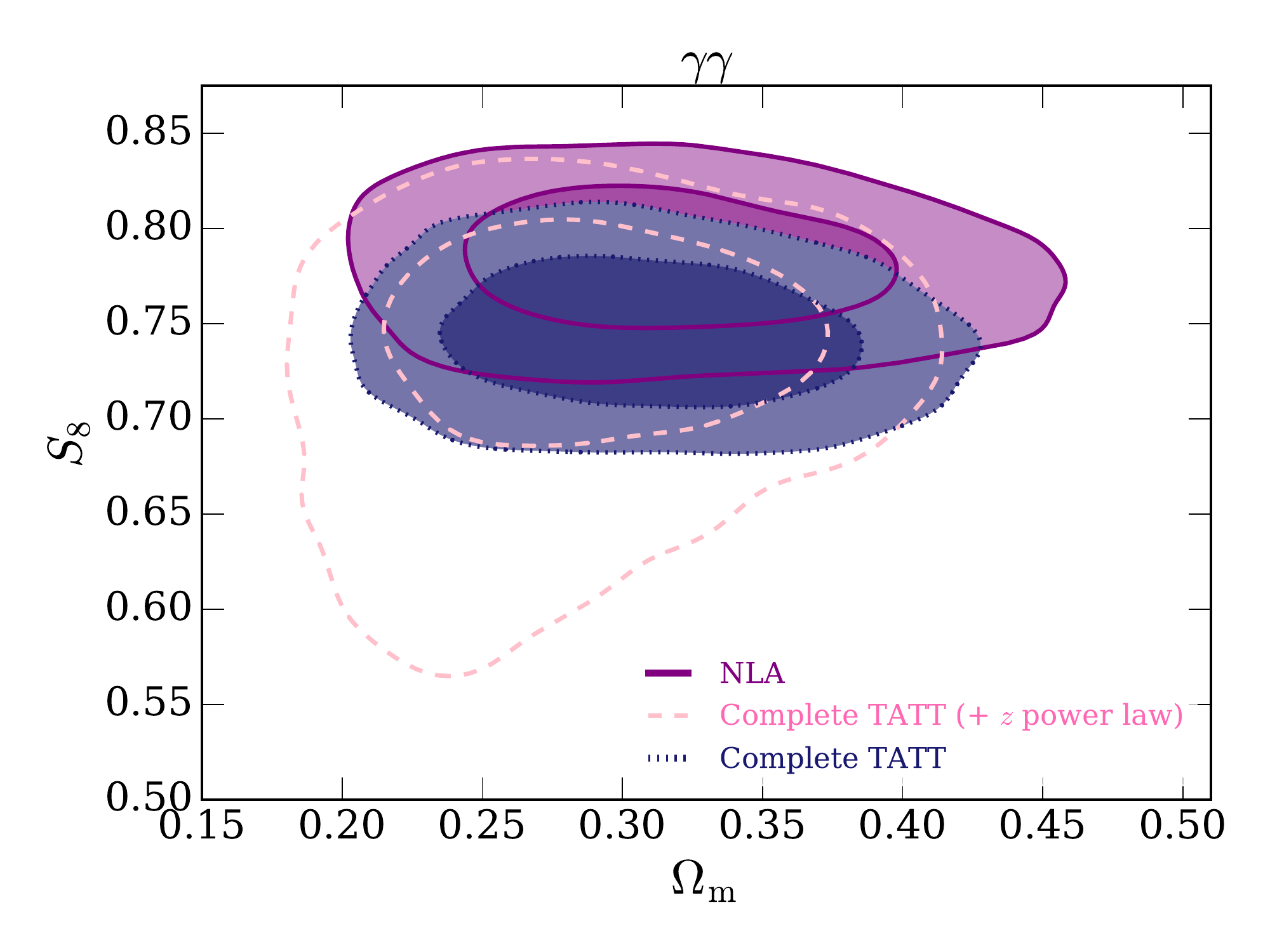}
\includegraphics[width=0.99\columnwidth]{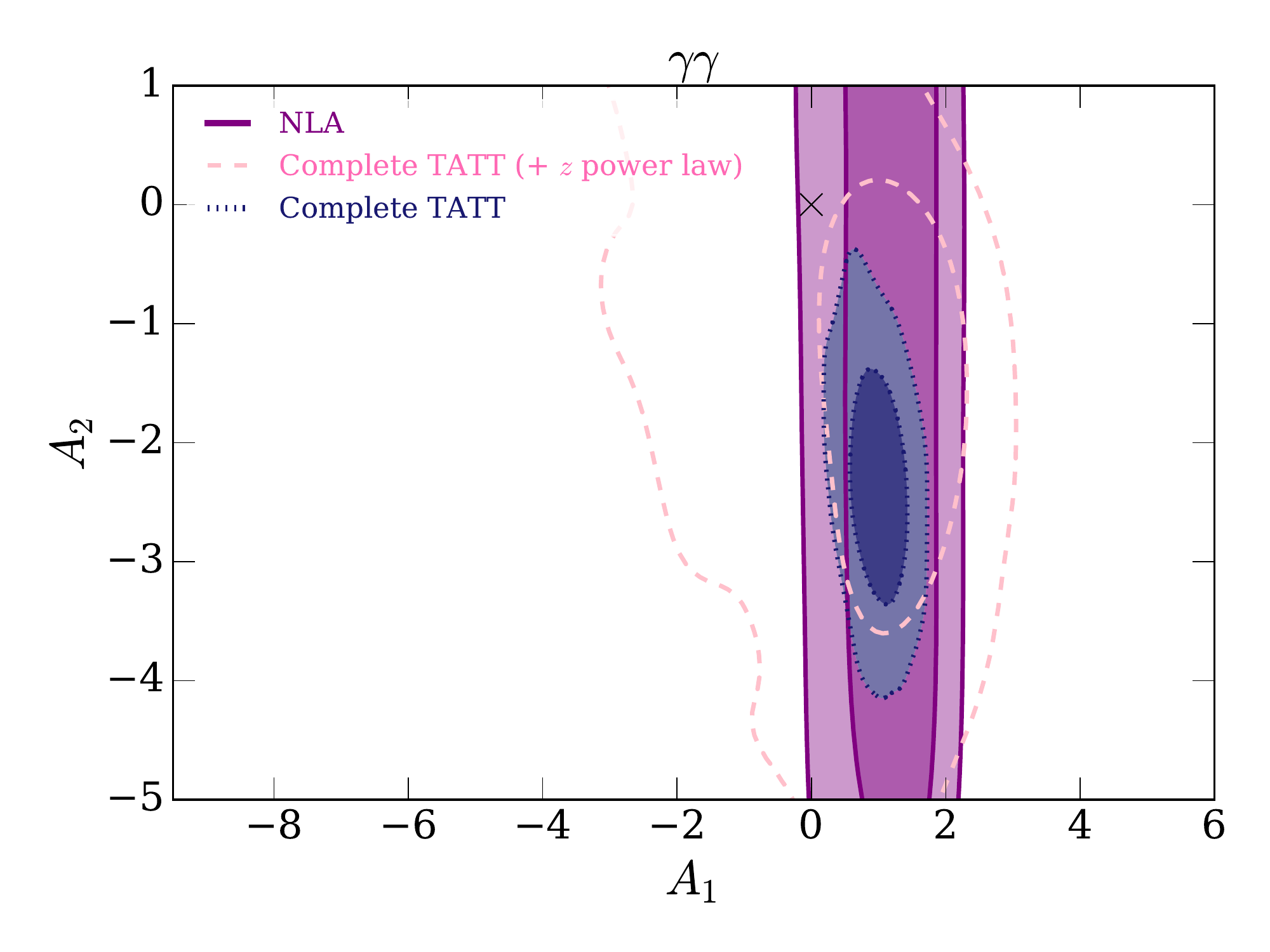}

\includegraphics[width=0.99\columnwidth]{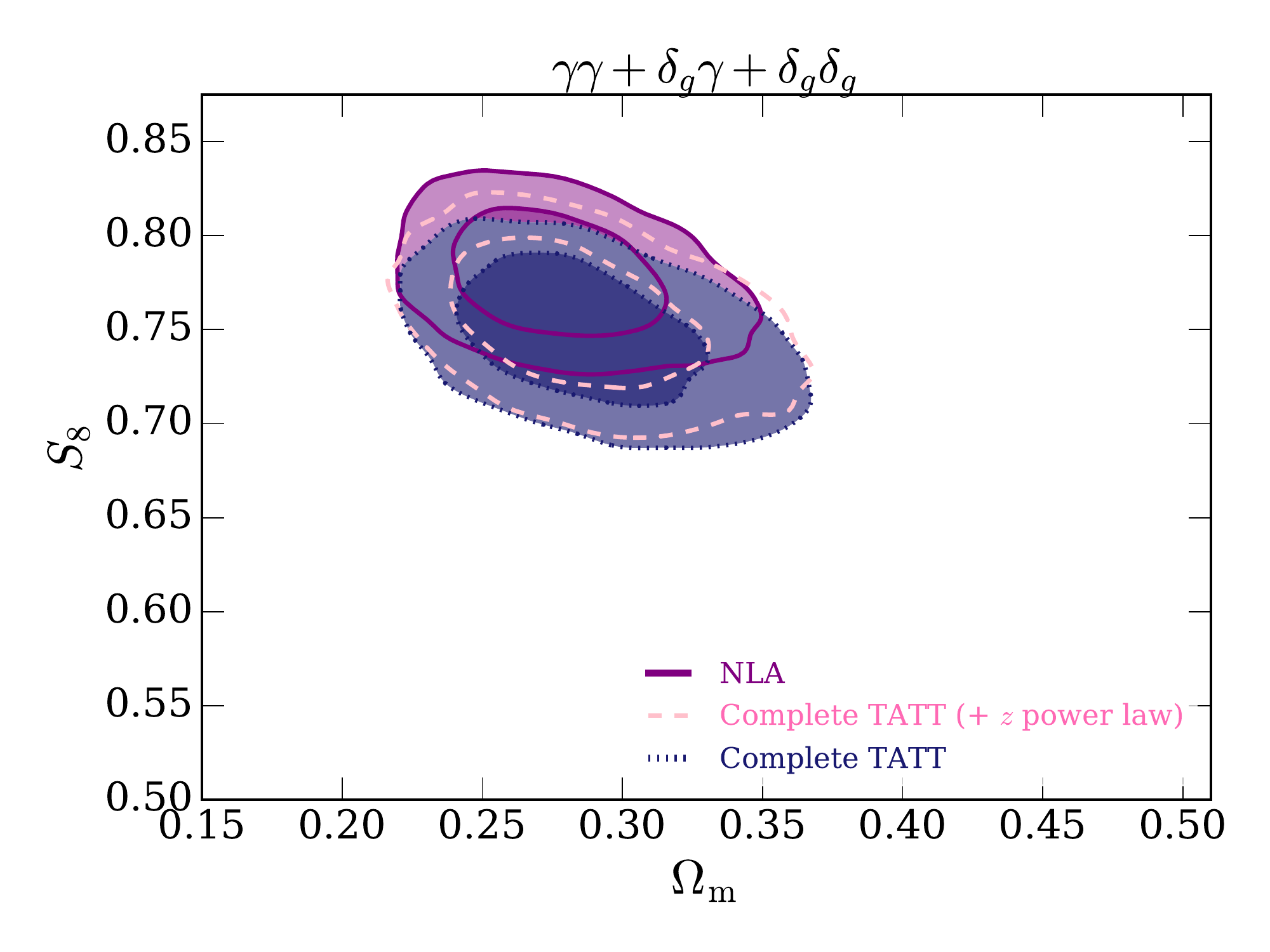}
\includegraphics[width=0.99\columnwidth]{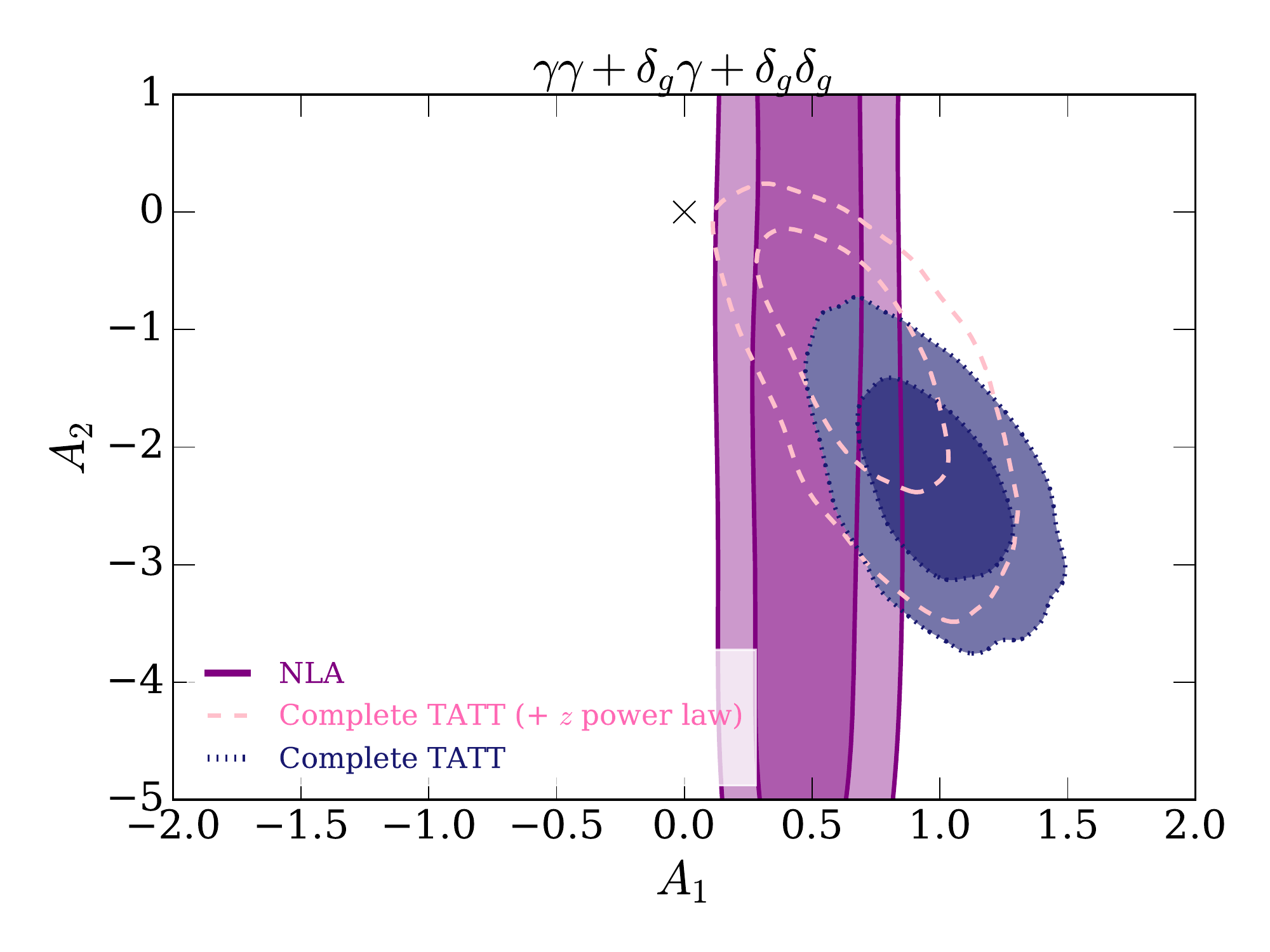}
\caption{Constraints on cosmological and TATT model parameters in a \lcdm~cosmology using the full Y1 sample.
The upper row shows constraints from cosmic shear alone, 
and the lower shows the joint constraint using the full $3\times2$pt datavector.
As labelled, we show the baseline NLA constraint (purple filled),
the TATT model (blue dotted),
and the result using the TATT power spectra,
but also marginalising over redshift dependent scaling
parameters $\eta_1$ and $\eta_2$ (pink dashed).}\label{fig:cosmology:tatt}
\end{figure*}

\begin{figure*}
\centering
\includegraphics[width=1.7\columnwidth]{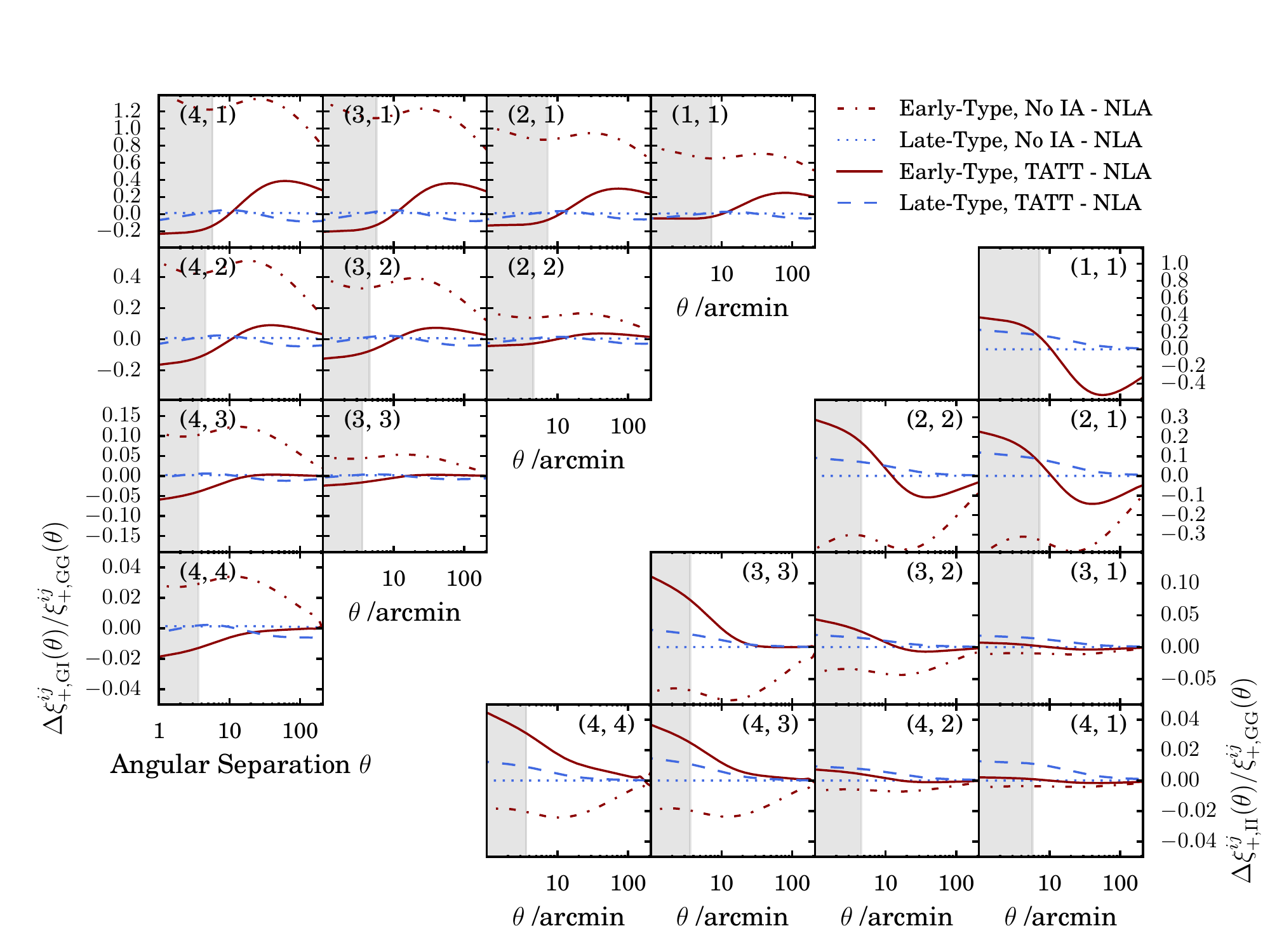}
\includegraphics[width=1.7\columnwidth]{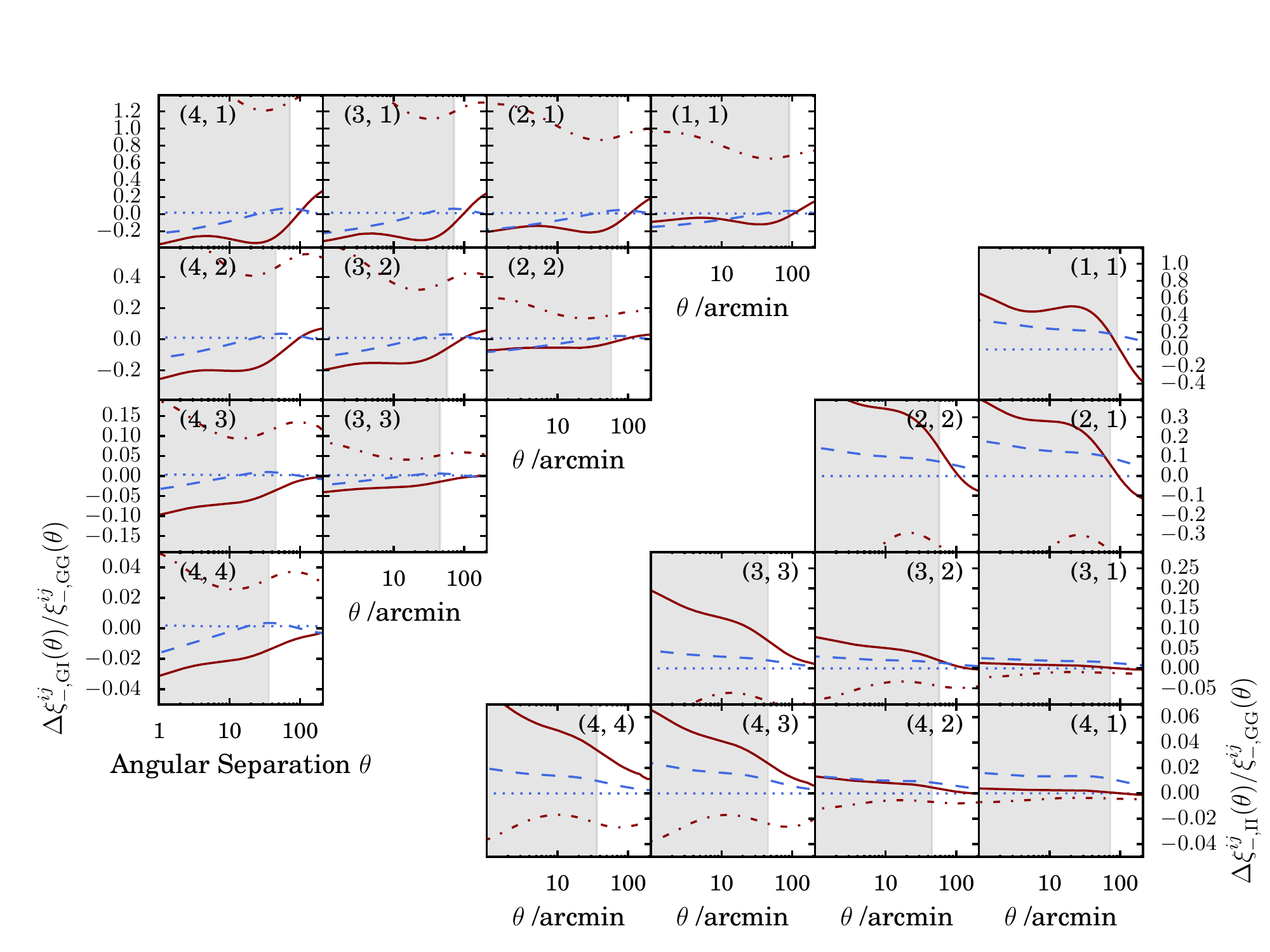}
\caption{\emph{Top}: The change in the estimated IA contributions to $\xi_+(\theta)$
when switching from the baseline NLA model to different IA scenarios.
Red and blue lines show early-type and late-type samples respectively.
We generate GI and II power spectra for each model by fixing the relevant 
parameters to the mean \emph{a posteriori} values obtained 
from the 3x2pt analysis using the same model.
The shifts between IA contributions $\Delta \xi$ under different models
are shown as a fraction of the pure cosmological signal.
Dotted and dot-dashed lines show the (No alignments - NLA) shift, while solid and dashed show the
(TATT - NLA) difference.
The upper and lower triangles show the GI and II contributions respectively.
\emph{Bottom}: The same, but for $\xi_-(\theta)$.
Each panel shows a particular pair of redshift bins (shown in parentheses). The
grey shaded regions indicate angular scales excluded 
in the likelihood calculations.
}\label{fig:cosmology:tatt_xi_residuals}
\end{figure*}

The next analysis permutation is to fit for both IA
amplitudes, $A_1$ and $A_2$ simultaneously.
Referred to as the TATT model 
(again, see Table \ref{tab:models} for reference),
this model allows for no explicit redshift evolution,
with both the indices $\eta_1$ and $\eta_2$ in 
equations \ref{eq:tatt_c1} and \ref{eq:tatt_c2}
fixed to zero.
We show the resulting split-sample IA constraints in the upper
panel of Figure \ref{fig:cosmology:tatt_sample_comparison} 
(filled red/blue contours).
The equivalent parameter fits using the unsplit Y1 shape catalogue 
are shown in Figure \ref{fig:cosmology:tatt}
(filled dark blue).

There are a number of points worth remarking on here.
First, the best fitting $A_1$ values are consistent with
those from the NLA fits previously,
with $A_1\sim2.5$ for early-types and $A_1\sim0$ for late-types.
In the split colour samples we report $A_2$ is consistent with
zero to within $1\sigma$.
The mixed Y1 sample, by contrast, favours a negative $A_2$ amplitude at
the level of a few $\sigma$.
Interestingly,
the comparison in Figure \ref{fig:cosmology:tatt} also 
suggests that the downwards shift in $S_8$
seen when switching to the TATT model 
is driven by the cosmic shear data
(compare the dark blue contours in the upper and lower right-hand
panels).

The standard physical interpretation of non-zero $A_2$
is as an IA contribution due to tidal torquing.
Under the sign convention in equation \ref{eq:tatt_c2},
$A_2 < 0$ implies intrinsic shapes of
galaxies are oriented tangentially relative to matter overdensities.
This picture is consistent with recent results 
from hydrodynamic simulations
\citep{chisari15}, 
although it is worth bearing in mind 
that there is still disagreement between
simulations with differing methods (e.g. 
the three dimensional shape-position correlation 
of disc galaxies in the Illustris and MassiveBlack-II
measured by
\citealt{hilbert17} and \citealt{tenneti16}
differ in sign from the equivalent measurement
presented in \citealt{chisari15}).
There are a number of other facts to note here, however.
As ever, mapping IA parameter constraints onto physical processes
is non-trivial, as they can very easily absorb features in the data due to
residual systematics.
We also re-iterate that, even in the absence of systematics,
possible non-zero values of both 
$A_1$ and $A_2$ in the late-type and mixed samples are not straightforward
to interpret.
As mentioned above, even in a pure TT scenario, the presence of
$A_2\neq0$ can generate an effective non-zero $A_1$ amplitude.

We also note that, as in \citet{y1cosmicshear}, the best fitting $S_8$ 
using the TATT model is shifted
down slightly relative to the NLA fits;
this shift is seen to persist in the full $3\times2$pt combination.
We echo \citet{y1cosmicshear}, however, in warning that this is not necessarily
a sign of bias in the NLA results, but could also be a result of an overly flexible
model for the constraining power of the data.
It is possible to test this idea using simulated data, and
to this end we generate a synthetic cosmic shear data vector with zero IAs.
We analyse the mock data with the maximally flexible version of TATT 
(two amplitudes and two power laws).
We confirm that, with the Y1 covariance matrix, we do indeed see a 
downward shift in $S_8$ relative to the input.
That is, switching to the more complicated model is seen to
degrade constraints in the $S_8 \omegam$ plane preferentially
towards low $S_8$, which results in a $\sim 0.5 \sigma$
downward shift in the mean $S_8$. 
This effect is seen to shrink considerably if one assumes a DES 
Y3 like covariance matrix
(both in absolute terms, and in $\sigma$).

Finally in this section, we fit a more flexible version of the
TATT model, with a parametrised redshift dependence
governed by the additional free parameters $\eta_1$ and $\eta_2$.
As above, we fit each of the early-type, late-type and mixed
samples separately.
The results can be seen in Figure \ref{fig:cosmology:tatt_sample_comparison}
(black unfilled)
and Figure \ref{fig:simultaneous_constraints:models} (dashed pink).
Note that the cosmic shear TATT + $z$ power law 
analysis of the mixed sample is 
almost\footnote{For consistency with the other analyses in this paper, 
we use photo-$z$ priors derived
from a resampled COSMOS sample, whereas \citet{y1cosmicshear} use
a combination of COSMOS and clustering cross correlations.
The difference, however, is small and will not change the conclusions presented here.}
identical to the ``Mixed Model" constraints presented by \citet{y1cosmicshear}.
In the mixed galaxy sample we find 

\begin{equation}
A^\mathrm{mixed}_1 = 0.70 ^{+0.21}_{-0.19}, \quad \quad A^\mathrm{mixed}_2 = -1.36 ^{+0.54}_{-0.70}.
\end{equation}

\noindent
For early-types we obtain the marginalised mean alignment amplitudes

\begin{equation}\label{eq:constraints_early}
A^\mathrm{early}_1 = 2.17 ^{+0.40}_{-0.38}, \quad \quad A^\mathrm{early}_2 = -0.57 ^{+2.58}_{-2.60},
\end{equation}

\noindent
and for late-type galaxies

\begin{equation}
A^\mathrm{late}_1 = 0.14 ^{+0.25}_{-0.27}, \quad \quad A^\mathrm{late}_2 = -0.66 ^{+1.88}_{-1.86}.
\end{equation}

\noindent
Our results are, again, consistent with the tidal alignment only paradigm for early-type galaxies, and the best-fitting value of the $A_1$ amplitude 
is consistent with the alignment amplitude obtained using the NLA model.

A number of notable differences become apparent when the IA signal is allowed
to vary with redshift.
First, with $\eta_2$ free, 
the favoured $A_2$ in all samples are shifted upwards
to slightly less negative values.
This is seen most strikingly in the mixed sample
(compare the purple and black solid isopleths in 
Figure \ref{fig:cosmology:tatt_sample_comparison}).
The shift results from the crescent-shaped degeneracy
seen in the middle panel in Figure \ref{fig:cosmology:tatt_sample_comparison};
fixing $\eta_2=0$ forces $A_2$ downwards to compensate,
but it appears that $A_2$ is not sufficiently degenerate with
$S_8$ for this to translate into a shift in cosmology.
Notably there is no region of this parameter space in which either
$A_2>0$ or $\eta_2>0$ is favoured.
Under the sign convention used here, 
$\eta_2<0$ implies an IA contribution that declines at high
redshift. 

In the mixed sample cosmological parameter space 
(Figure \ref{fig:simultaneous_constraints:models}, left)
the addition of the redshift scaling parameters significantly
degrades the quality of the shear-only constraint.
In the $3\times2$pt case (lower left) the extended tail is seen to
contract, 
but notably the posterior peak is not shifted back upwards
towards the NLA constraint.
It is also worth remarking that the downwards shift when switching to 
the TATT model, is driven entirely by the shear-shear data.
The cosmology constraints from the $2\times2$pt combination 
($\galaxyshear+\galaxygalaxy$; not shown in Figure \ref{fig:cosmology:tatt})
are close to identical under the NLA, TATT and TATT $+ z$ power law models.

Of the TATT model variants discussed above, a small handful 
provide clearly favourable Bayes factors relative to the NLA analysis.
The simplest one-parameter prescriptions yield $B=15.58$ in favour of
the TT model in late-types and $B=1.36$ for TA in the early-type sample
(interpreted on the Jeffreys' Scale 
as ``moderate" and ``weak" evidence respectively).
Under the more complicated models, only the early-type and unsplit
TATT fits provide $B>1$.
We interpret the low evidence ratios as an indication that the
data, including the unsplit Y1 sample, are insufficient to
support definitive statements about the relative goodness of
fit using the IA models in question. 

Finally, to understand at a more basic level how the TATT model enters the two-point observables,
we compare simulated datavectors evaluated at the means of the multivariate posterior distributions
from our multiprobe TATT and NLA analyses. 
The solid and dashed lines in Figure \ref{fig:cosmology:tatt_xi_residuals} show
the difference in the IA contribution to the 
$\shearshear$ data, as predicted by the TATT and NLA models.
The upper and lower panels show the change in $\xi_+$ and $\xi_-$ respectively.
The upper and lower triangles within each panel show the GI and II contributions. 
Since the aim here is to assess the importance of IA modelling uncertainties for cosmology, 
we show the difference as a fractional shift relative to the cosmological GG signal.
For reference we also show (dotted and dot-dashed lines) the difference between the 
baseline NLA model and a no-alignments scenario.
On the scales used in our analysis, the difference tends to be negative, particularly in
early-type galaxies. This should be interpreted as saying that the TATT model
predicts a smaller IA contribution than the NLA model. In late-type galaxies the 
reverse is true, suggesting NLA under-predicts the contamination due to IAs.
In the correlations involving the upper redshift bins
the fractional difference between the models is small 
($<10\%$ on all scales),
thanks largely to the stronger cosmological signal.
In the lowest bins we find the difference can be in excess of $50\%$ of the
GG contribution. 

\subsubsection{\wcdm~Cosmology}\label{sec:beyond_lcdm}

\begin{figure}
\centering
\includegraphics[width=1.1\columnwidth]{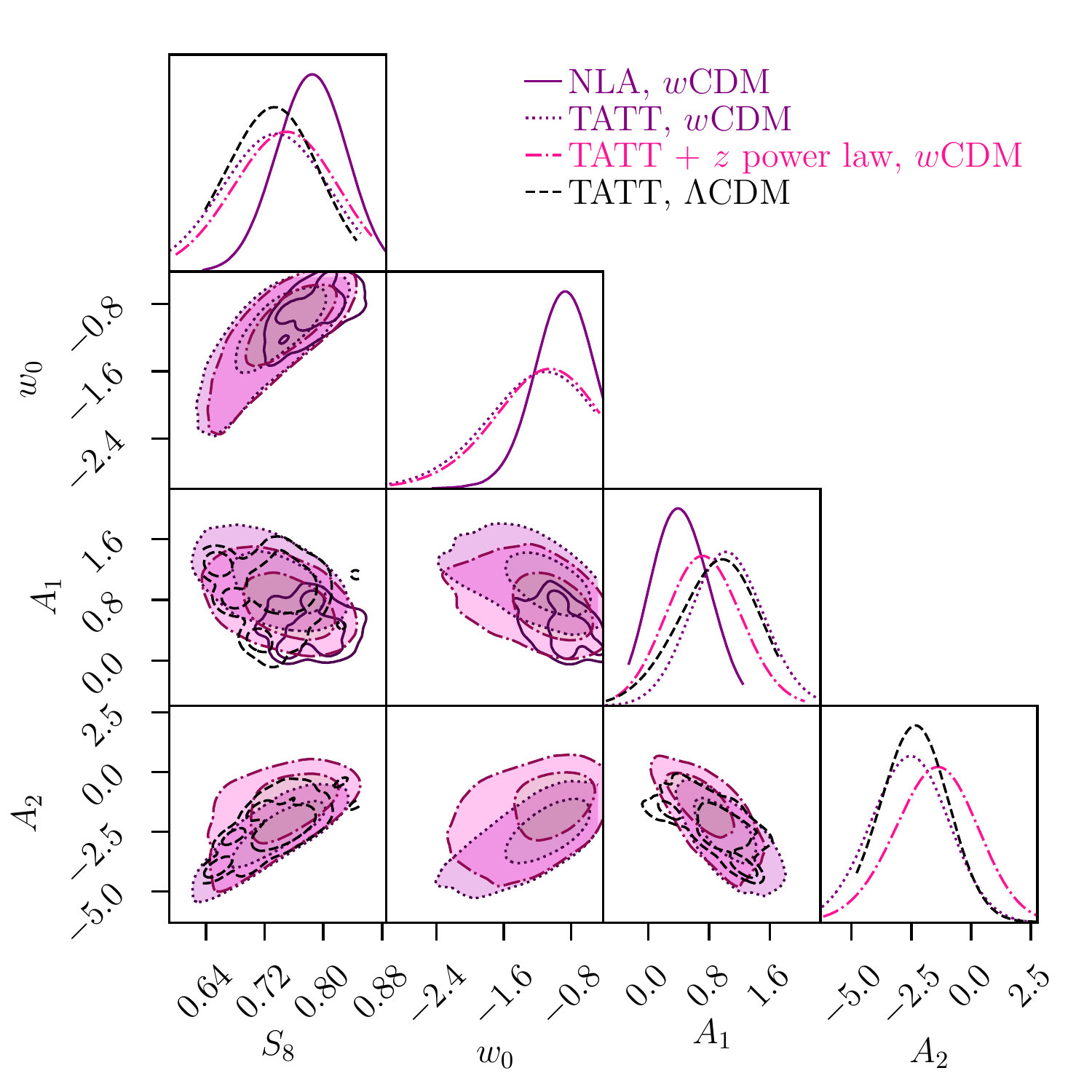}
\caption{Constraints on IA parameters in the mixed DES Y1 sample.
Dark purple (solid) contours show the fiducial NLA model, 
light purple (dotted) show the TATT model of \citet{blazek17}
and pink (dash-dotted) show the TATT model with an additional
power law in redshift, all assuming a \wcdm~cosmology. 
The TATT model constraints under \lcdm~are shown for reference in black (dashed).}\label{fig:cosmology:wcdm_tatt}
\end{figure}

The main analyses presented in this work assume a flat \lcdm~universe.
Though there is, to date, no unambiguous observational evidence for deviations from this
standard description
(see \citealt{y1extensions}),
it is still useful to test how sensitively our 
conclusions about intrinsic alignments depend upon the cosmological model. 
A simple, relatively common cosmological extension, 
is to allow the dark energy equation of state to vary with time; 
where earlier we enforced
$w_0=-1$, we now allow it to vary in the range $w_0 = [-3,-0.333]$.
In total we recompute three chains,
all using the unsplit Y1 catalogue: 
one assuming the baseline two-parameter NLA model,
and one assuming the TATT IA model with and without
redshift power laws.
The results are shown in 
Figure \ref{fig:cosmology:wcdm_tatt}.
Considering first the IA model constraints, 
shifting from \lcdm~to \wcdm~parameter space is seen to induce a
slight degradation, but no significant shift in the marginalised distributions
(compare the black dashed and purple dotted contours).
The mixed data still favour a small positive $A_1\sim1$
and a moderately negative tidal torque amplitude $A_2\sim-2.4$.
The latter is non-zero at the level of $\sim 2 \sigma$.
Conversely, if we consider how shifting to a new IA model 
affects cosmological constraints under \wcdm,
we find the following;
as evident from the two left-most columns of Figure \ref{fig:cosmology:wcdm_tatt},
switching from NLA to the more sophisticated IA model produces a 
small shift downwards in both $S_8$ and $w_0$.
It is worth bearing in mind, however, that 
the parameter constraints under the two models overlap to comfortably within $1 \sigma$.
The galaxy sample is the same in the two cases, and so both shape noise and cosmic variance
are correlated, but it is not straightforwards to assess the significance of 
such small shifts.
As noted before, it is also possible to induce such changes
by fitting a model that is too flexible for the data to properly constrain.
Computing the Bayes factor of the \wcdm~TATT analysis relative to \wcdm~NLA
we find $B=4.22$, 
which suggests the former IA model is mildly favoured by the
Y1 data under the extended cosmology.

This exercise is informative, but not exhaustive. That is, there are a 
number of other changes to the background cosmology modelling choices
(i.e. allowing for non-zero curvature, deviations from general relativity etc)
that could plausibly mimic an intrinsic alignment signal.
Exploring these degeneracies in detail is, however, an extensive
task and considered beyond the scope of the current paper. 

\subsection{Including Early $\times$ Late Cross Correlations}\label{sec:results:crosscolour}

In this section we incorporate a potential source of information
that is systematically excluded from our split-sample analyses:
cross correlations between the source galaxy samples.
The additional correlations are straightforward to measure, 
and can be incorporated into our analysis pipeline with minimal code modifications
in the same way as extra tomographic bins.
The total datavector after including all cross- and auto-correlations between colours and redshift bins
consists of 114 unique two-point functions:
\begin{itemize}
\item {five $w(\theta)$ auto correlations,}
\item {35 $\gamma_t(\theta)$ correlations $(20 \times \delta_g\gamma^\mathrm{R} + 15 \times \delta_g\gamma^\mathrm{B}$, 
excluding galaxy pairs with $\bar{z}_l>\bar{z}_s)$}
\item {74 $\xi_{+/-}(\theta)$ correlations $(2\times [ 10 \times \gamma^\mathrm{B}\gamma^\mathrm{B} + 10 \times \gamma^\mathrm{R}\gamma^\mathrm{R} + 16 \times \gamma^\mathrm{R}\gamma^\mathrm{B}])$}
\end{itemize}

\noindent
(1171 points with early $\times$ late cross correlations and 811 without,
after scale cuts).
The superscripts R and B here refer to correlations involving the
early- and late-type samples respectively.
The additional $\gamma^\mathrm{R}\gamma^\mathrm{B}$ correlations, along with the extended covariance matrix, which includes all correlations between the two samples,
are shown in Appendix \ref{app:cross_corrs}.
Modelling the intrinsic alignment contribution to the cross terms is straightforward in both the NLA and TATT models, since the IA in each sample is related to the underlying tidal field. Correlations are then simply given by linear combinations of the tidal field correlations which appear in II contributions to these models, scaled by the relevant $A_i$ pre-factors (see Sec.~IIIC of \citealt{blazek17}). 
Under this treatment, the cross terms $(\xi^\mathrm{RB}_\pm)$
are sensitive to both the early- and late-type GI power spectra,
and also to a multiplicative combination of the early- and late-type
IA amplitudes via the II contribution.

\begin{figure}
\centering
\includegraphics[width=\columnwidth]{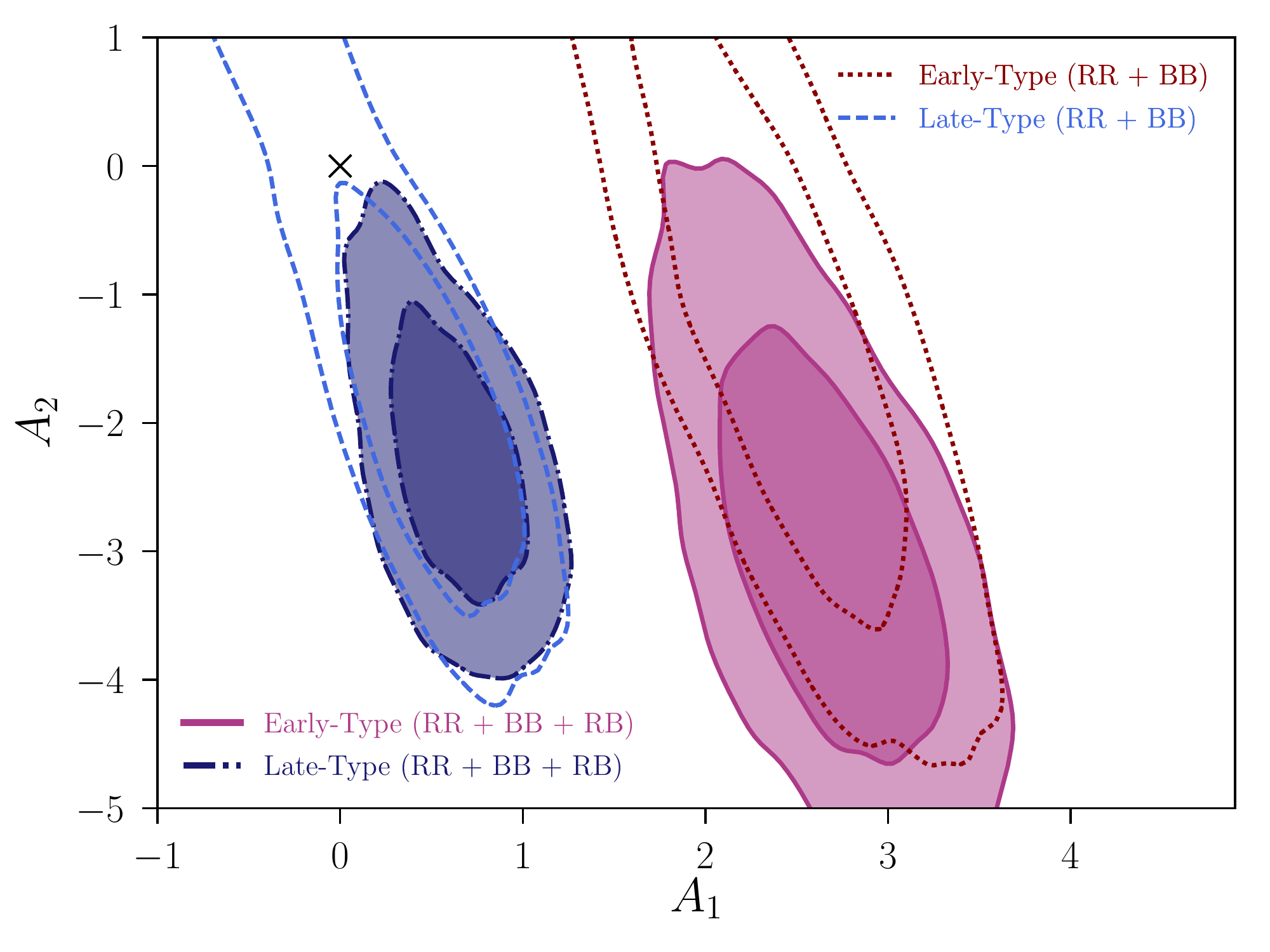}
\caption{Joint constraints on tidal alignment and tidal torque IA parameters
from a simultaneous multicolour $3\times2$pt analysis of DES Y1. As described in Section \ref{sec:results:crosscolour}, 
early- and late-type samples are analysed together using a four-parameter alignment model,
with two free amplitudes for each population. 
The unfilled contours show the constraints from such an analysis using only auto-colour
correlations, while the filled contours show the impact of also including early-late 
cross correlations $\xi^\mathrm{RB}_\pm$.
}\label{fig:cosmology:multicolour_tatt}\label{fig:results:multicolour}
\end{figure}

\begin{figure*}
\centering
\includegraphics[width=\columnwidth]{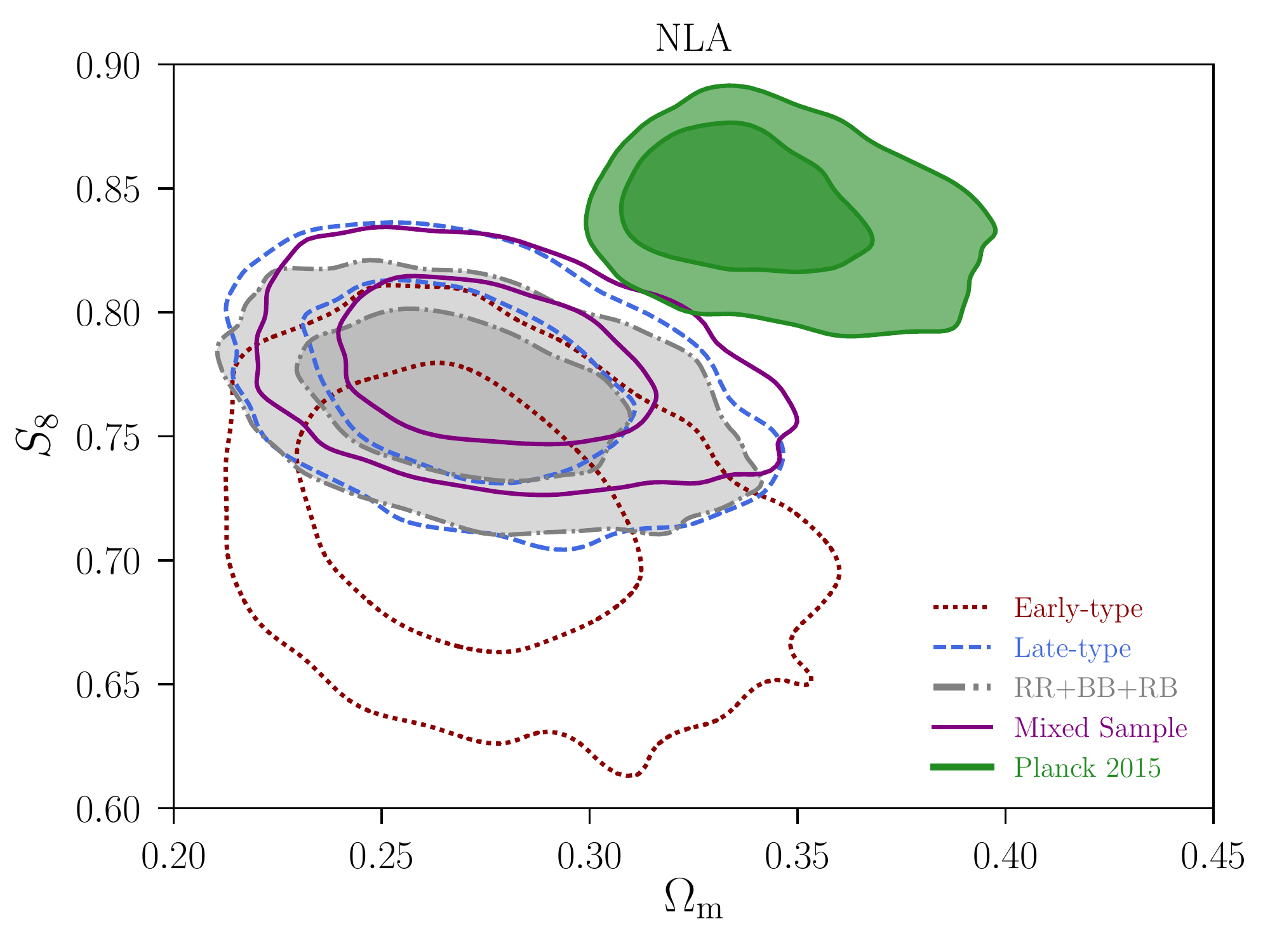}
\includegraphics[width=\columnwidth]{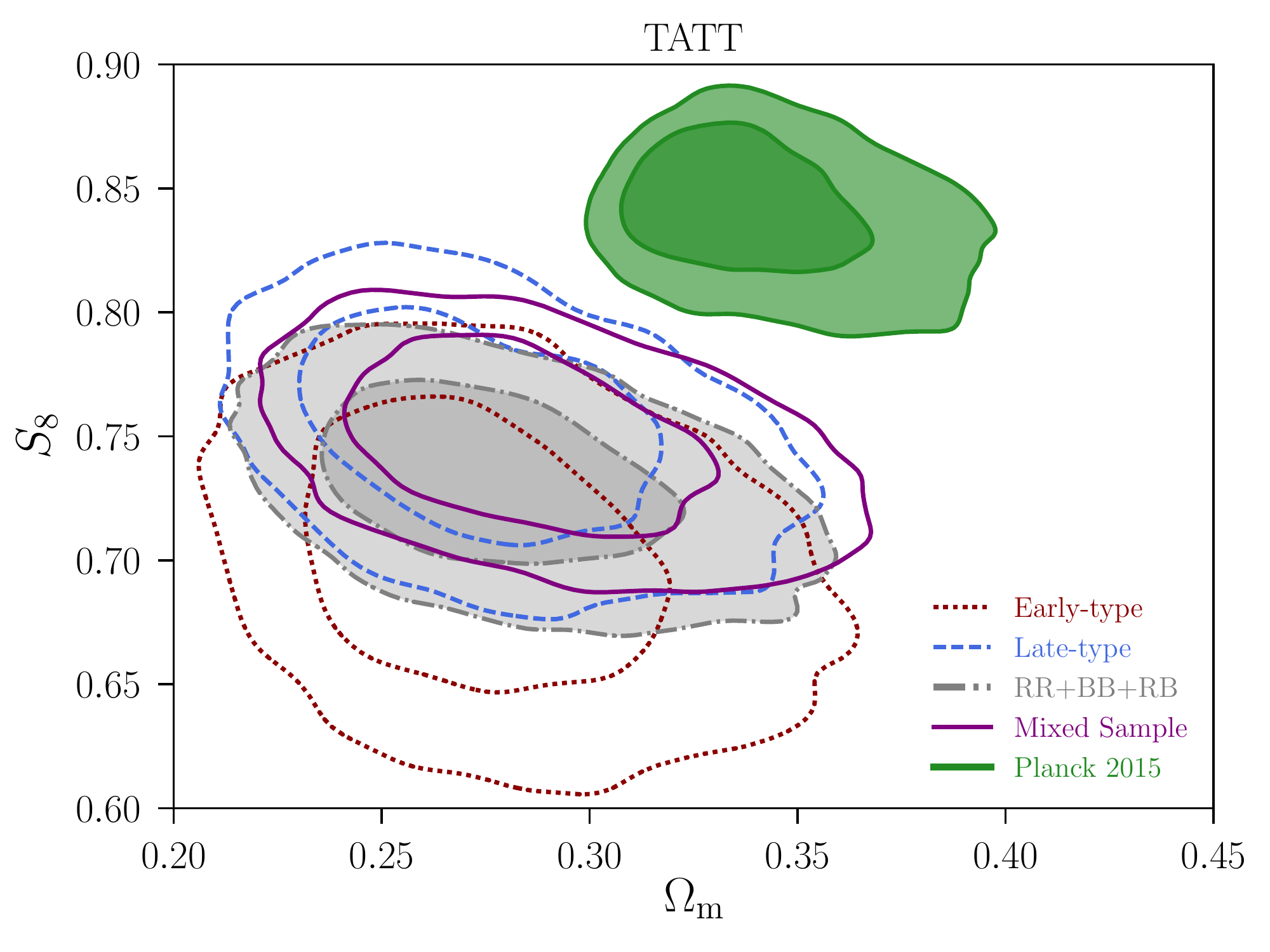}
\caption{Constraints on cosmological parameters with various intrinsic alignment 
modelling choices. \emph{Left}: Using the NLA model with two free parameters $(\aia,\eta_\mathrm{IA})$. 
The open contours show split-sample constraints using early- and late-type source galaxies,
while the purple solid show those using the unsplit Y1 source catalogue. 
Overlain in grey is the posterior distribution obtained from a joint analysis
of early-type and late-type galaxies, along with their cross-correlations. This
simultaneous fit allows the IA parameters for the 
early/late samples to vary independently, but does assume the same analytic form
for both.
\emph{Right}: As left, but using the Complete TATT model for
intrinsic alignments.
The filled green contours are identical in the two cases and show the posterior obtained 
from Planck 2015 temperature and polarization data (TT+EE+TE).
Note that, for comparability with our main results, we marginalise over the sum of the
neutrino masses here.
}\label{fig:cosmology:multicolour_cosmology}
\end{figure*}

In addition to the extra pieces to simulate,
it is also necessary to model the covariance of the extended data.
As before, we employ \cosmolike~to generate a non Gaussian covariance
matrix, the details of which can be found in Appendix A of \citet{krause16}.
Note that the non Gaussian contributions are sourced from the single-colour 
covariance matrices rather than recomputed,
an approximation that is tested and seen to have no significant impact on our results 
in Appendix \ref{app:cross_corrs}.

We repeat our earlier analysis with these multicoloured data,
fitting both early-type and late-type IAs simultaneously,
along with independent systematics parameters in the two galaxy samples.
In the TATT case, this leaves us with four IA parameters 
$\mathbf{p}_\mathrm{IA} = 
(C_1^\mathrm{R}, C_1^\mathrm{B}, C_2^\mathrm{R}, C_2^\mathrm{B})$
and a significantly expanded model space 
(36 free parameters in total, including cosmology and nuisance parameters).
We also consider a simultaneous NLA case with parameters
$\mathbf{p}_\mathrm{IA} =
(A_\mathrm{IA}^\mathrm{R}, A_\mathrm{IA}^\mathrm{B}, \eta_\mathrm{IA}^\mathrm{R}, \eta_\mathrm{IA}^\mathrm{B})$.

The combination of extra data points and greater number of free parameters
was found to increase the time required by 
\blockfont{Multinest} to converge significantly.
We instead perform this analysis using the Metropolis-Hastings
Markov Chain Monte Carlo algorithm of \blockfont{emcee}\footnote{dfm.io/emcee},
which uses the affine-invariant sampler of \citet{goodman10}.
Before accepting a chain as converged we require
the following to be true, after burn in:
(a) the samples in each parameter cannot be visibly distinguished 
from random noise of constant
mean and variance when plotted in order,
(b) no visible difference can be seen between 2D parameter 
constraints obtained from the
first and second halves of the chain and 
(c) the integrated autocorrelation scale of the chain per walker 
is much smaller than the total number of samples.   

The TATT model constraints resulting from this analysis are 
shown in Figure \ref{fig:results:multicolour}.
In this figure, the unshaded contours show the posteriors from an analysis, 
which fits cosmology and early/late IAs simultaneously, 
but excludes the RB cross correlations.
The filled contours show the equivalent from the full analysis including
all correlations.
The agreement of the former with the filled contours in
Figure \ref{fig:cosmology:tatt_sample_comparison})
is, of course, expected. 
The comparison is, however, non-trivial given differences in the sampler,
covariance matrix and theory pipeline,
in addition to the fact that one is fitting all parameters simultaneously.
Although the IA models and nuisance parameters are independent between the two galaxy samples
(e.g.\ BB correlations contain no information about early-type alignments),
the simultaneous analysis gives a slightly stronger constraint on cosmology
relative to the single-colour analyses, which in turn impacts IA constraints.
This is thought to be the source of the small differences 
in the size of the unfilled contours in
Figure \ref{fig:results:multicolour} relative to those in 
Figure \ref{fig:cosmology:tatt_sample_comparison}.

The final marginalised IA parameters obtained from the chain including cross correlations are:

\begin{equation}\label{eq:multicolour_constraints_early}
A^\mathrm{early}_1 = 2.66^{+0.67}_{-0.66}, \quad \quad A^\mathrm{early}_2 = -2.94^{+1.94}_{-1.83},
\end{equation}

\noindent
and

\begin{equation}\label{eq:multicolour_constraints_late}
A^\mathrm{late}_1 = 0.62^{+0.41}_{-0.41}, \quad \quad A^\mathrm{late}_2 = -2.26^{+1.30}_{-1.16},
\end{equation}

\noindent
Although our results hint at a possible non-zero 
IA signal in a blue galaxy sample, 
we encourage caution in interpreting this finding.
Both IA amplitudes are still consistent with zero at
the level of $2 \sigma$. It is also worth bearing in 
mind that,
as remarked upon earlier, the combination of a non-zero
$A_2$ and nonlinear growth can produce an effective $A_1$,
even if the alignment mechanism is entirely
driven by tidal torquing. We note that fixing $A_2=0$, 
as in the various NLA analyses in Section \ref{sec:results:main}, 
returns $A_1$ amplitudes consistent with zero in blue galaxies 
at all redshifts. Though our results taken at face value
might suggest both tidal alignment and tidal torquing
mechanisms at work in aligning blue 
galaxies, 
it is possible that the $A_1>0$ value is
an artefact of such interplay between the parameters
in the TATT model, rather than a genuine indication of
physical linear alignment in blue galaxies.
It is also worth noting that, as ever with 
indirect intrinsic alignment measurements,
leakage of residual photo-$z$ error into
the IA parameters is possible.
Though we have attempted to be conservative in
our analysis choices, and have tested the response of 
our baseline results to changes in the details of the $n(z)$, 
we cannot absolutely rule out photo-$z$ modelling
as the source of the apparent
non-zero IA signal in late-type galaxies. 

Notably, the shear-shear cross correlations appear to contribute a 
considerable amount of information about IAs in both samples.
This is particularly true of $A_2$, with the addition of the
RB data visibly reducing the size of the shaded contours in  
Figure \ref{fig:results:multicolour}.
That the values are consistent between early- and late-type galaxies is
also interesting; 
though $A_1$ differs significantly, 
there is no clear evidence that the amplitude $A_2$
depends on galaxy type. 

Finally, we compare the cosmology constraints obtained from the simultaneous
analysis described above with our earlier results in
Figure \ref{fig:cosmology:multicolour_cosmology}.
In each panel of this figure one IA parameterisation is used for all samples
(denoted above each).
The unfilled red dotted, blue dashed and purple solid lines show the 
constraints in this parameter space using the early-type, late-type and
unsplit source samples respectively.
Overlain in grey we show the constraints from the simultaneous 
multicolour analysis discussed.
One first conclusion to draw here is that splitting the source sample
does not bring an obvious degradation in cosmological constraining power.
In the late-type only case this is simply a result of the nature of the
data excised. The early-type sample accounts for a relatively small fraction
of the catalogue (see Table \ref{tab:data:statistics}).
Additionally, it contains a greater abundance of low redshift objects,
in which the cosmological shear signal is relatively weak.
Perhaps less intuitively, in the fully simultaneous case the additional
complexity of the IA model does not appear to be sufficient to significantly
broaden the $S_8$ posterior.

The downward shift in the TATT model relative to the NLA model is, again,
seen here. This has been remarked on previously and we will not discuss it 
further here. 
Noticeably, however, we also see a slight downward shift, under both IA models,
when one switches from a mixed sample analysis with a single set of 
effective IA parameters to an explicit colour-split simultaneous one.
The magnitude of the shift is approximately the same in both scenarios, 
and it is worth keeping in mind that the results are
still consistent to the level of $\sim 0.5 \sigma$.

\subsection{Comparison with External Data}

As noted in the introduction, cosmic shear and galaxy clustering are
far from the only usefully constraining cosmological probe available to
the community.
CMB measurements offer a particularly powerful way of probing
the high-$z$ Universe. To assess the consistency of our measurements with
existing results we 
follow \citet{y1cosmicshear} and \citet{y1keypaper}
and first recompute the Planck 2015 posterior in our fiducial \lcdm~parameter space. 
We use public\footnote{Though the Planck Collaboration have released likelihoods for 
their most recent raft of results \citep{planck18}, the code used to compute them
(the Planck Likelihood Code; PLC 
\url{https://cosmologist.info/cosmomc/readme_planck.html}) has not yet been made public. Given that our fiducial
parameter space differs slightly from theirs (by the addition of neutrino mass as a free parameter),
we opt to use the slightly older results for our comparison.} 
temperature and polarization measurements (`Planck TT + lowP'; \citealt{planck15}),
including scales $\ell = [30 – 2508]$ and $\ell = [2 – 29]$
for TT and TT+TE+EE+BB data respectively.
The full cosmological parameter space for chains including CMB data
has seven free parameters: $\as, \ns, \omegam, \omegab, h, \omeganu, \tau$.
We show the results of this reanalysis in Figure \ref{fig:cosmology:multicolour_cosmology}
(green contours).

To quantify the consistency of our DES Y1 analysis with the 
external data we use the ratio:

\begin{equation}\label{eq:evidence_ratio}
R = \frac{p({\bf D}_{\rm p15},{\bf D}_{\rm DES} | {\rm IA~model})}{p({\bf D}_{\rm p15}) p({\bf D}_{\rm DES} | {\rm IA~model})},
\end{equation}

\noindent
or the ratio of the joint DES + Planck evidence to the product
of those obtained from the independent analyses
(see Section V of \citealt{y1keypaper} for a fuller explanation).
Implicitly, all of the evidence values in Equation \ref{eq:evidence_ratio}
assume the same background model for cosmology (i.e. flat \lcdm),
even if the best fit values for its parameters differ.
The evidence ratios derived from the various analysis permutations
are listed in Table \ref{tab:evidence}.

It is worth pointing out here that the mixed NLA
entry in Table \ref{tab:evidence} does not match up with
the equivalent value reported in \citet{y1keypaper}.
This is expected, given that our analysis does not incorporate the
updates to the covariance matrix discussed in \citet{troxel18}.

Here we can see that switching to the TATT model
worsens the agreement between DES Y1 and Planck 2015,
as quantified by the evidence ratio, by a factor of $\sim3$.
This is consistent with the naive interpretation of the shift 
in $S_8$ between the purple lines in the left- and right-hand
panels of Figure \ref{fig:cosmology:multicolour_cosmology}.
Switching to the full multicolour TATT analysis we see a still
larger degradation.
One should be cautious in drawing strong conclusions from 
these numbers, however. First of all, we reiterate that we have 
seen in simulated analyses that slight shifts in $S_8$, predominantly 
downwards, can be achieved by running with an IA model that is not
well constrained by the data. Such shifts may be a modelling artefact, 
and could potentially lead to a false impression of discord.
Second, we note that there is some uncertainty on the 
\blockfont{Multinest} evidence values and, potentially, 
$R$ may have some systematic bias\footnote{Although it is claimed that these biases tend to drive one
towards agreement, not the reverse.}
which are neglected in this comparison.
With the caveats given above, we note that our reanalysis tends to drive 
DES away from Planck in cosmological parameter space.
Caution should, however, be exercised in drawing conclusions regarding 
tension from this work; we look to the code fixes and greater statistical
power in Y3 to shed light on the matter.

\begin{table}
\begin{center}
\begin{tabular}{ccc}
\hline
Sample       & IA Model &  $\mathrm{ln}R$ \\
\hhline{===} 
Mixed                   & NLA      & $-3.65$   \\
Mixed                   & TATT     & $-4.74$ \\
RR+BB+RB                & TATT     & $-5864.33$ \\
\hline
\end{tabular}
\end{center}
\caption{Evidence ratios for a subset of the analyses presented in this paper.
The evidence ratio $R$ is defined in Equation \ref{eq:evidence_ratio}.
The term `mixed' here means without colour splitting, 
and RR+BB+RB refers to the full simultaneous analysis of 
our early and late-type samples and their cross correlations.
}\label{tab:evidence}
\end{table}


\section{Discussion \& Conclusions}\label{sec:conclusion}

In this paper we have presented a follow-on study to the
Dark Energy Survey Year 1 cosmology results
of \citet{y1cosmicshear} and \citet{y1keypaper}.
Using cuts on SED type and photometric colour we have defined early- and late-type
galaxy samples. There is \emph{prima facie} reason to believe that the alignment
of intrinsic galaxy shapes should arise by different mechanisms in
these galaxy populations, and impact shear two point functions in 
different ways.
From these samples we have obtained large scale cosmic shear, 
galaxy-galaxy lensing and galaxy clustering measurements,
which we have analysed using a selection of different intrinsic alignment models,
individually and simultaneously.
Our key results are summarised below.
\begin{itemize}
   \item{We have detected a significant difference in IA amplitude between early-type
   and late-type samples, assuming the NLA model.
   Early-type galaxies were found to have 
   positive $\aia\sim2$ at $\sim 6.5\sigma$.
   Fits on late-type galaxies were consistent with no intrinsic alignments.}
   \item{We have used the split-sample DES data to impose new constraints
   on IAs and cosmology under the TATT model of \citet{blazek17}. The linear coefficient $A_1$
   in early- and late-type galaxies was found to be consistent with the amplitude obtained
   from fitting the NLA model. We have reported a null measurement of the 
   quadratic term $A_2$ in the two subsamples, and a new constraint 
   $A_2 = -1.36 ^{+1.08}_{-1.41}$ in the mixed sample.
   }
   \item{We have reported fully simultaneous constraints from the joint analysis of
   early-type and late-type correlations, plus their cross-correlations. The addition
   of $\gamma^\mathrm{R}\gamma^\mathrm{B}$ correlations is seen to tighten constraints
   on the amplitude of the quadratic term in the TATT model $A_2$ particularly. 
   This represents 
   the first hints of non-zero IAs in late-type galaxies from real data, 
   though the physical interpretation is non-trivial.}
   \item{We have assessed the differences in cosmology favoured under the
   various model and data permutations discussed in this work. We have seen
   the downwards shift in $S_8$ seen by \citet{y1cosmicshear} when switching
   to the TATT model persists in the $3\times2$pt case, but is driven by the
   cosmic shear data. In both NLA and TATT analyses we have seen another 
   downward shift at the level of $\sim 0.5 \sigma$ between the unsplit
   and split simultaneous analyses.}
\end{itemize}

\noindent
Two ongoing Stage III lensing surveys have now released 
$3\times2$pt cosmology analyses, performed by independent groups
using separate analysis pipelines, and report consistent results
\citep{y1keypaper,vanuitert17,joudaki18}.
Preliminary cosmic shear results from HSC, 
which will reach deeper than either KiDS or DES,
have also very recently been added to the literature \citep{hikage18}. 
Although they represent the state of the art
of late-time observational cosmology,
the published results from each of these surveys covers less than half
of their respective final footprints.
Understanding and correctly modelling astrophysical systematics such 
as intrinsic alignments on both large and small scales will be crucial 
to the success of these cosmology projects
and their successors.
The current paper aims to contribute to this effort,
providing a detailed study of the large scale intrinsic alignment
contamination in DES Y1.
Our results come with a number of caveats; 
notably we choose not to incorporate boost factors
into the galaxy-galaxy lensing measurements
(justified by scale cuts, which ensure 
that they have an impact of $\lesssim1\%$;
see Figure 10, \citealt{y1ggl}).
Unfortunately boost factors enter our observables on small scales and where there is a 
strong overlap between the source and lens redshift distributions,
which are precisely the regimes with the most potential for testing IA models.
It is thus likely that future analyses on similar lines to this one
will need to build these corrections into their pipelines.
Our focus here is on large angular scales, which avoids the theoretical
complexity of nonlinear growth and the interplay with between IAs and baryonic effects
or higher-order galaxy bias.
The behaviour of IAs on small scales is an important topic for future investigation, 
albeit one we consider beyond the scope of this paper.

Nonetheless, our results provide a step towards a more complete
understanding of intrinsic alignments in modern lensing surveys.
We offer the strongest 
constraints to date on physically-motivated IA models in a
number of validated, realistic lensing samples. 
Finally, this work lays the ground for future analyses using
the considerably larger datasets that will shortly become available.
Work is already underway on building lensing measurements from the 
Dark Energy Survey Year 3
data. Whether or not it proves ultimately necessary to obtain
unbiased cosmology constraints, a colour split analysis of the
type outlined here will almost certainly be required to 
test the sufficiency of the IA modelling in future lensing cosmology
studies.
Our results offer reason for cautious optimism. Though clearly challenging,
we have no reason to believe the task of adequately modelling 
large scale intrinsic alignments to be beyond the theoretical
equipment already available to the lensing community.

\section{Acknowledgements}

The authors would like to thank
Fran\c{c}ois Lanusse
and Donnacha Kirk
for many useful conversations.
We are grateful to Catherine Heymans, Elisa Chisari and Rachel Mandelbaum
for their thoughts, 
and to David Bacon, Richard Battye and Sarah Bridle for extensive
discussion of an early version of this work.
We would also like to thank our anonymous referee for their comments on this work.
The numerical analyses presented in this work made use of computing resources at
the University of Manchester and Carnegie Mellon University. 
Correlation function and covariance matrix calculations were performed on the Cori cluster at
the National Energy Research and Scientific Computing Center (NERSC; a US DoE facility).

Support for DG was provided by NASA through Einstein Postdoctoral
Fellowship grant number PF5-160138 awarded by the Chandra X-ray
Center, which is operated by the Smithsonian Astrophysical Observatory
for NASA under contract NAS8-03060.
JB is supported by an SNSF Ambizione Fellowship.

Funding for the DES Projects has been provided by the U.S. Department of Energy, the U.S. National Science Foundation, the Ministry of Science and Education of Spain, 
the Science and Technology Facilities Council of the United Kingdom, the Higher Education Funding Council for England, the National Center for Supercomputing 
Applications at the University of Illinois at Urbana-Champaign, the Kavli Institute of Cosmological Physics at the University of Chicago, 
the Center for Cosmology and Astro-Particle Physics at the Ohio State University,
the Mitchell Institute for Fundamental Physics and Astronomy at Texas A\&M University, Financiadora de Estudos e Projetos, 
Funda{\c c}{\~a}o Carlos Chagas Filho de Amparo {\`a} Pesquisa do Estado do Rio de Janeiro, Conselho Nacional de Desenvolvimento Cient{\'i}fico e Tecnol{\'o}gico and 
the Minist{\'e}rio da Ci{\^e}ncia, Tecnologia e Inova{\c c}{\~a}o, the Deutsche Forschungsgemeinschaft and the Collaborating Institutions in the Dark Energy Survey. 

The Collaborating Institutions are Argonne National Laboratory, the University of California at Santa Cruz, the University of Cambridge, Centro de Investigaciones Energ{\'e}ticas, 
Medioambientales y Tecnol{\'o}gicas-Madrid, the University of Chicago, University College London, the DES-Brazil Consortium, the University of Edinburgh, 
the Eidgen{\"o}ssische Technische Hochschule (ETH) Z{\"u}rich, 
Fermi National Accelerator Laboratory, the University of Illinois at Urbana-Champaign, the Institut de Ci{\`e}ncies de l'Espai (IEEC/CSIC), 
the Institut de F{\'i}sica d'Altes Energies, Lawrence Berkeley National Laboratory, the Ludwig-Maximilians Universit{\"a}t M{\"u}nchen and the associated Excellence Cluster Universe, 
the University of Michigan, the National Optical Astronomy Observatory, the University of Nottingham, The Ohio State University, the University of Pennsylvania, the University of Portsmouth, 
SLAC National Accelerator Laboratory, Stanford University, the University of Sussex, Texas A\&M University, and the OzDES Membership Consortium.

The DES data management system is supported by the National Science Foundation under Grant Numbers AST-1138766 and AST-1536171.
The DES participants from Spanish institutions are partially supported by MINECO under grants AYA2015-71825, ESP2015-88861, FPA2015-68048, SEV-2012-0234, SEV-2016-0597, and MDM-2015-0509, 
some of which include ERDF funds from the European Union. IFAE is partially funded by the CERCA program of the Generalitat de Catalunya.
Research leading to these results has received funding from the European Research
Council under the European Union's Seventh Framework Program (FP7/2007-2013) including ERC grant agreements 240672, 291329, and 306478.
We  acknowledge support from the Australian Research Council Centre of Excellence for All-sky Astrophysics (CAASTRO), through project number CE110001020.

This manuscript has been authored by Fermi Research Alliance, LLC under Contract No. DE-AC02-07CH11359 with the U.S. Department of Energy, Office of Science, Office of High Energy Physics. The United States Government retains and the publisher, by accepting the article for publication, acknowledges that the United States Government retains a non-exclusive, paid-up, irrevocable, world-wide license to publish or reproduce the published form of this manuscript, or allow others to do so, for United States Government purposes.

Based in part on observations at Cerro Tololo Inter-American Observatory, 
National Optical Astronomy Observatory, which is operated by the Association of 
Universities for Research in Astronomy (AURA) under a cooperative agreement with the National 
Science Foundation.

\bibliographystyle{mnras}
\bibliography{samuroff,des_y1kp_short}

\appendix

\section{Degeneracies between IA Parameters and Source Galaxy Bias}\label{app:source_bias}

In this appendix we explore the level at which uncertainty in the galaxy
bias in our lensing sample affects our resuts.
Though source galaxy bias does not enter standard NLA prescription,
the Complete TATT model includes additional terms generated by the
fact that measurements occur, by definition, where there are observable galaxies,
which leads to a density weighting that is sensitive to the galaxy bias in the 
source population \cite{blazek15}.

\begin{figure*}
\includegraphics[width=1.5\columnwidth]{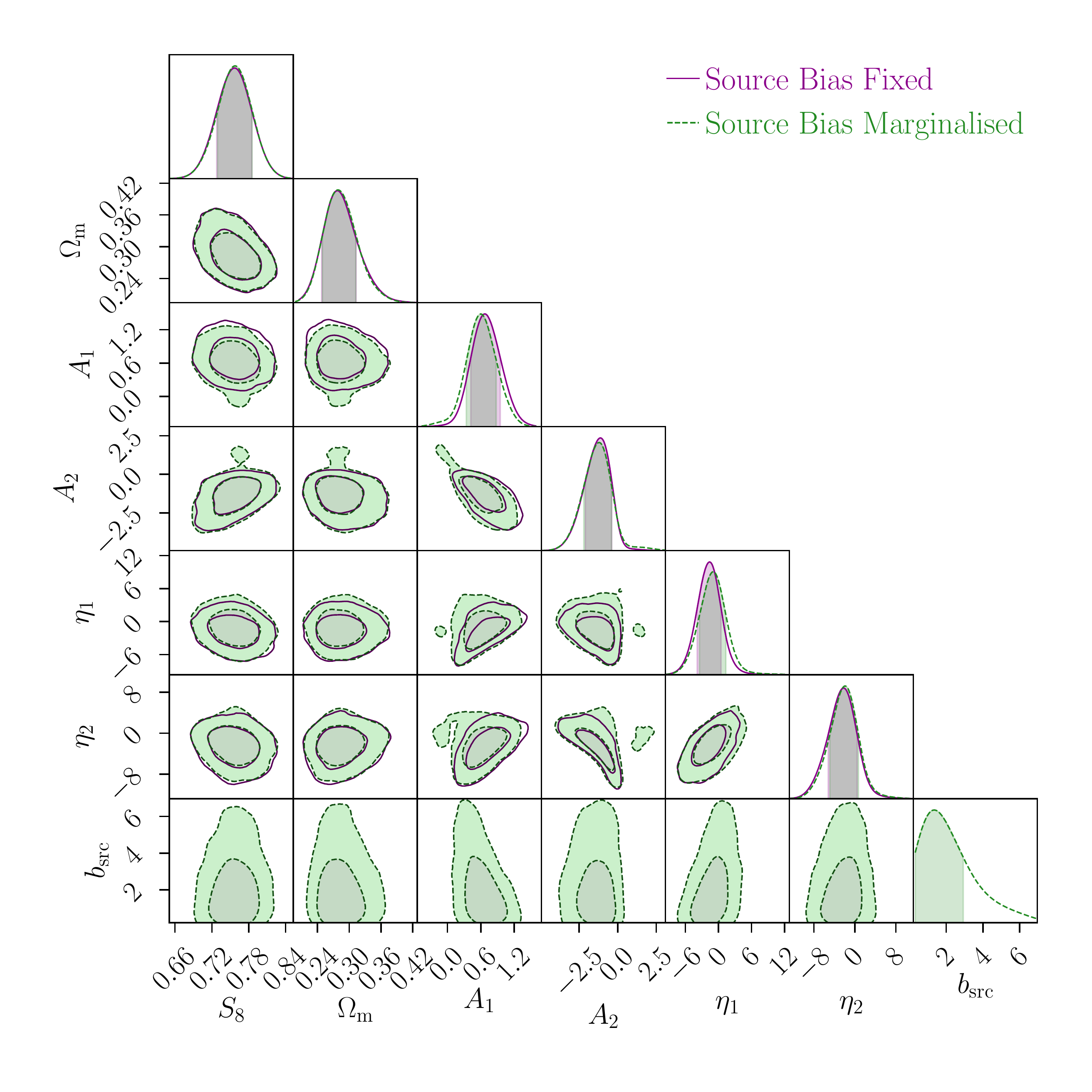}
\caption{Constraints on cosmology, intrinsic alignment parameters and 
source galaxy bias in the mixed Y1 sample, assuming a mixed tidal alignment plus tidal torque model,
and using all three correlations ($\shearshear+\galaxyshear+\galaxygalaxy$).
Solid lines show the result with source galaxy bias fixed at $b_g^\mathrm{src}=1$
and the dashed green lines show the case where $b_g^\mathrm{src}$ is allowed to vary.
}\label{fig:tatt_source_bias}
\end{figure*}

In the main body of this paper, we assume the angular scale cuts imposed
are sufficiently stringent to ensure linear galaxy bias. 
Unlike lens bias, which appears directly in our predictions of 
$\gamma_t(\theta)$ and $w(\theta)$ and is always marginalised,
our main analysis fixes the source bias 
$b_g^\mathrm{src}$ to unity, independent of scale, redshift and galaxy sample.
Although from a theoretical perspective we do not expect this to have a major impact
on our results, we test this assumption in practice here.
To this end we made a small modification to the TATT prediction code, 
which allows $b_g^\mathrm{src}$ to be varied as a single redshift independent nuisance parameter.
We then rerun the mixed TATT analysis with a flat prior on $b_g^\mathrm{src}=[0.1,8]$.
The results are shown in Figure \ref{fig:tatt_source_bias}.

One should note that we show only the TATT results here; 
mathematically $b_g^\mathrm{src}$ appears in the TA terms
and the TA/TT interaction terms (see Section III C of \citealt{blazek17}), and
does not enter the picture in a pure tidal torquing scenario.
It is apparent that the cosmological results are largely unaffected. 
We do see a slight upwards shift in the redshift index
$\eta_1$, in the mixed sample. 
This is, however, most likely an artefact of our chosen parameter volume rather 
than a sign that our main result for this sample is biased. 
Although our assumption that $b_g^\mathrm{src} > 0.1$ is 
justified by simulations and previous lensing measurements,
$b_g^\mathrm{src}$ is only weakly constrained by the $3\times 2$pt data,
and we are in effect artificially truncating the parameter space. 
Without this cut off, the joint posterior of 
$\eta_1$ and $b_g^\mathrm{src}$ would extend further into the $\eta_1<0$ regime,
which would shift the constraints slightly closer to the fiducial results in
which $b_g^\mathrm{src}$ is fixed.
We also test the sensitivity of the early-type and late-type only results 
(both TATT and individual TA and TT) to marginalising over source bias. 
We see no significant change in the findings presented in the main sections of this paper.
These analyses also provide some information (albeit weak at best) on the galaxy bias in the
two source samples. 
In early type galaxies we find
$b_g^\mathrm{src}\sim 0.1-2.5$.
The bias in late-types is very poorly constrained, providing no preferred value in
the range $b_g^\mathrm{src} = [0.1-7.0]$, thanks largely to the low amplitude
of the IA signal in these galaxies, 
and the fact that the dominant alignment, if present, is expected to come through the
tidal torquing, which is insensitive to source galaxy bias at next-to-leading order.

\section{The Impact of Neutrinos on Galaxy Bias}\label{app:bmodes}

In the main sections of this study two other notable changes were introduced
relative to \citet{y1keypaper}.
First, the impact of neutrinos on halo bias (and ultimately galaxy bias)
were omitted, whereas \citet{krause17} and \citet{y1keypaper} include an
analytic scale-dependent modification using the prescription of
\citet{loVerde14}.
Note that the first order effect of neutrino free-streaming on the shape of the 
matter power spectrum, is included in our modelling. As in the main DES Y1 analyses we
marginalise over the total neutrino mass $\Omega_\nu h^2$ using wide flat priors. 
The impact of neutrinos through a slight modification to the galaxy bias 
(referred to for the sake of brevity as ``neutrino bias") 
is small, and one would not expect their inclusion 
to have a significant bearing on results at the scale cuts and statistical 
power of DES Y1.
\citet{krause17} demonstrate this to be the case for the main Y1 results
under the NLA model.
We explicitly test their findings hold true for our split samples, under
the TATT model.  

\begin{figure}
\includegraphics[width=\columnwidth]{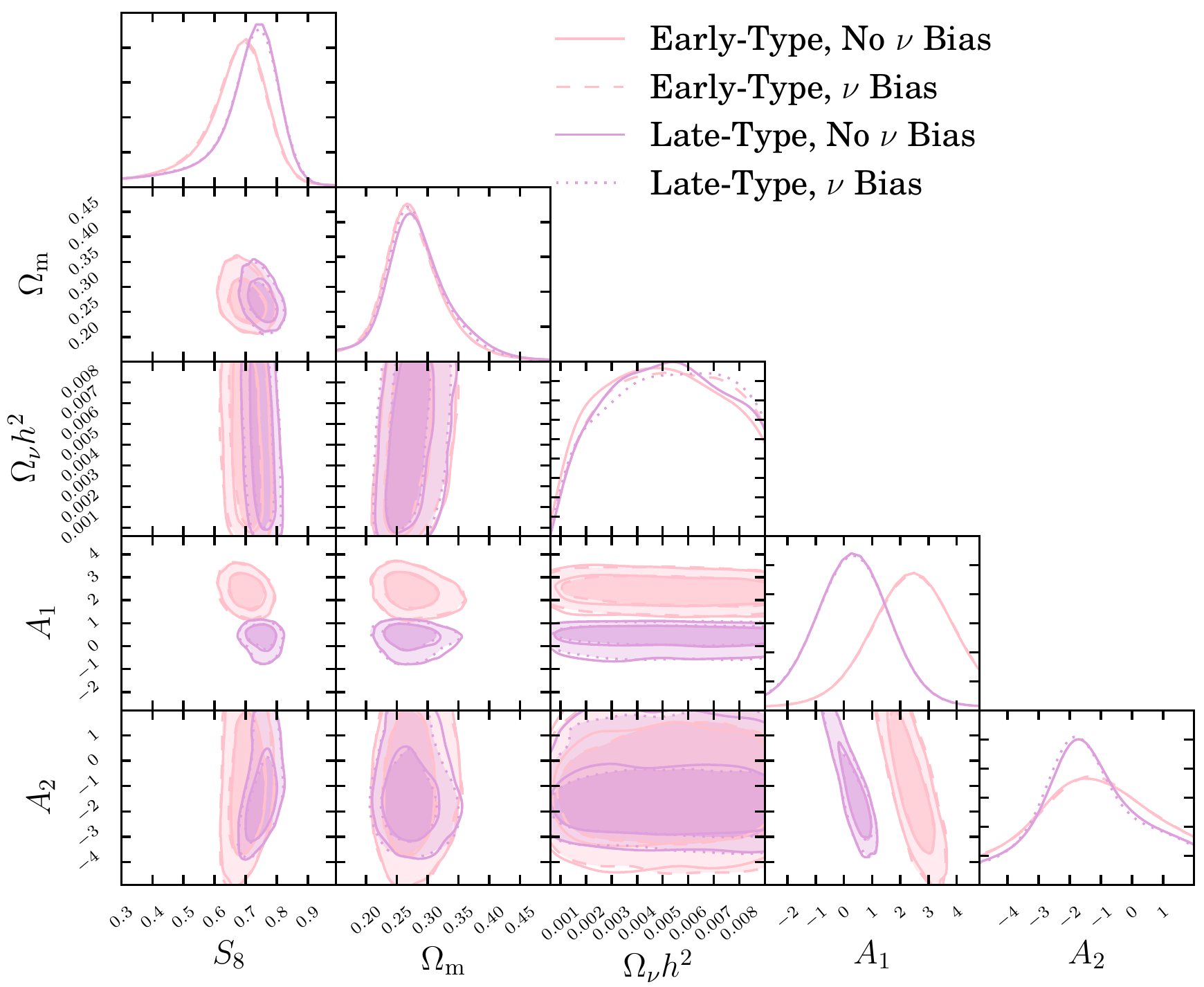}
\caption{$3\times2$pt constraints on cosmology and intrinsic alignment parameters from
DES Y1 early- and late-type samples, assuming the Complete TATT IA model 
described in Section \ref{sec:theory:ia_models}.
The unfilled solid lines show the results for the two samples presented in the main
body of this paper and the filled contours show the equivalent, but with an additional
step in the theory calculation to model the impact of massive neutrinos on galaxy bias. 
In all cases the source galaxy bias is fixed at $b_g^\mathrm{src}=1$ 
and a flat \lcdm~cosmology is assumed.
}\label{fig:tatt_nubias}
\end{figure}

We do not seek to propagate the impact of neutrinos
to the source galaxy bias,
but given the results of Appendix \ref{app:source_bias} 
this is not expected to alter our findings.
Rerunning the multiprobe TATT analysis
(with two free IA parameters $A_1$ and $A_2$ only),
we obtain the results shown in Figure \ref{fig:tatt_nubias}.
No significant change in the constraints on any particular parameter
are observed.

\section{Intrinsic Alignment Induced B-modes}\label{app:ia_bmodes} 

\begin{figure}
\includegraphics[width=\columnwidth]{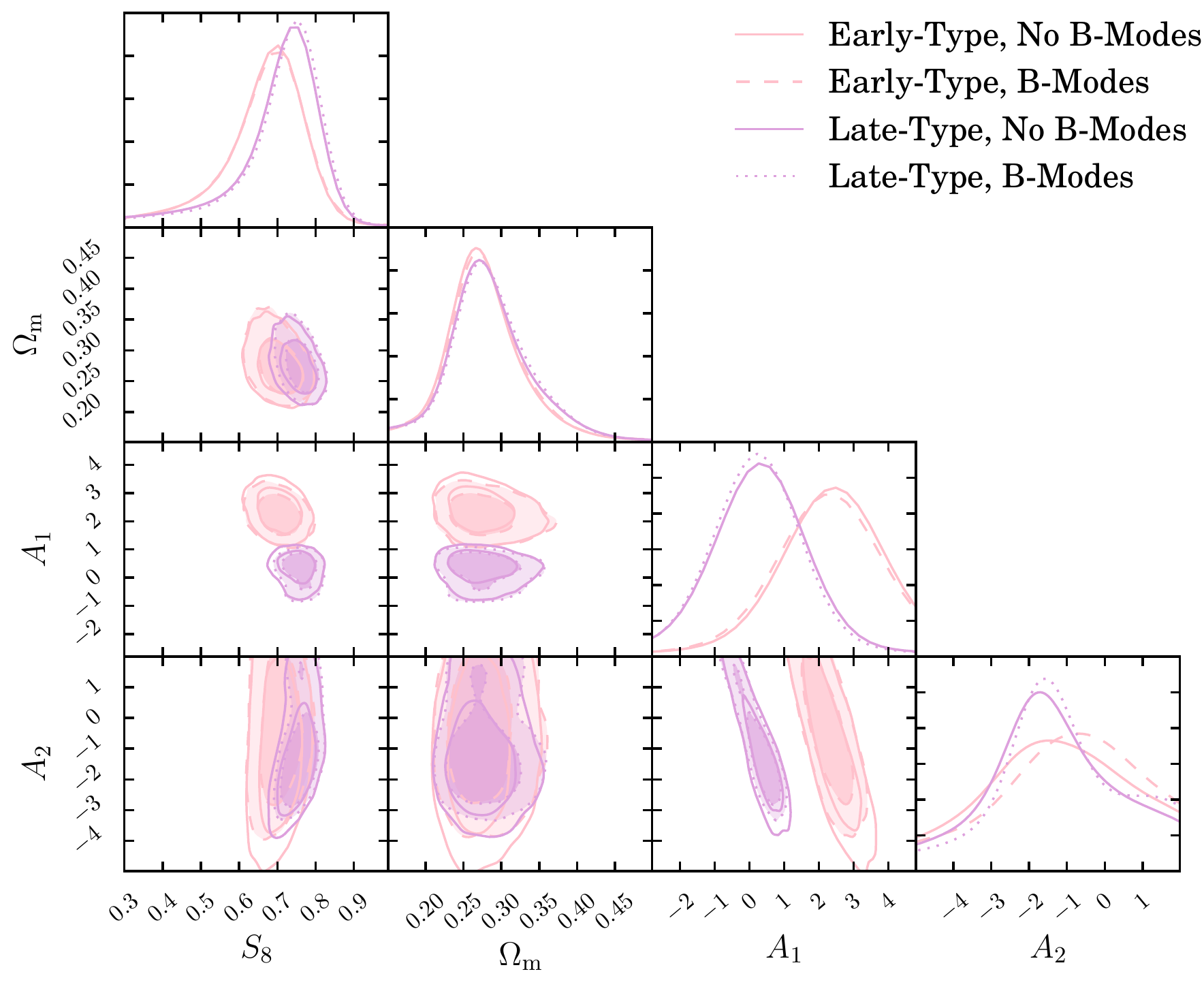}
\caption{$3\times2$pt constraints on cosmology and intrinsic alignment parameters from
DES Y1 early- and late-type samples, assuming the Complete TATT IA model 
described in Section \ref{sec:theory:ia_models}.
The unfilled solid lines show the results for the two samples presented in the main
body of this paper and the filled contours show the equivalent with additional B-modes
contributions to the IA power spectra. 
In all cases the source galaxy bias is fixed at $b_g^\mathrm{src}=1$ 
and a flat \lcdm~cosmology is assumed.
}\label{fig:tatt_bmodes}
\end{figure}

The TATT model predicts a B-mode contribution to the
II power spectrum,
which was omitted in the results shown in the main body of this work.
To test the robustness of our results to this,
we rerun the early-type and late-type TATT analyses.
The pipeline now includes an additional stage, 
in which at each step in parameter space
we compute $P_\mathrm{II}^{BB}(k)$ using 
\citet{blazek17}'s equation 39,
transform it to angular space via the Limber integral for a
particular pair of redshift bins $ij$
and add the result to $C^{ij}_\mathrm{II}(\ell)$.
The split analyses with and without B-modes are shown in Figure \ref{fig:tatt_bmodes}.
Note that the filled contours represent the two galaxy samples; 
the offset between them should not be interpreted as the bias due to the 
IA B-mode contribution.
The impact on all parameter constraints barring $A_2$ are close to
negligible. 
Though the impact on $A_2$ is non-trivial, the shift is comfortably with
the level of precision allowed by DES Y1.

\section{Multicolour Cross Correlations: Measurements \& Covariance Matrix}\label{app:cross_corrs}

\begin{figure}
\includegraphics[width=\columnwidth]{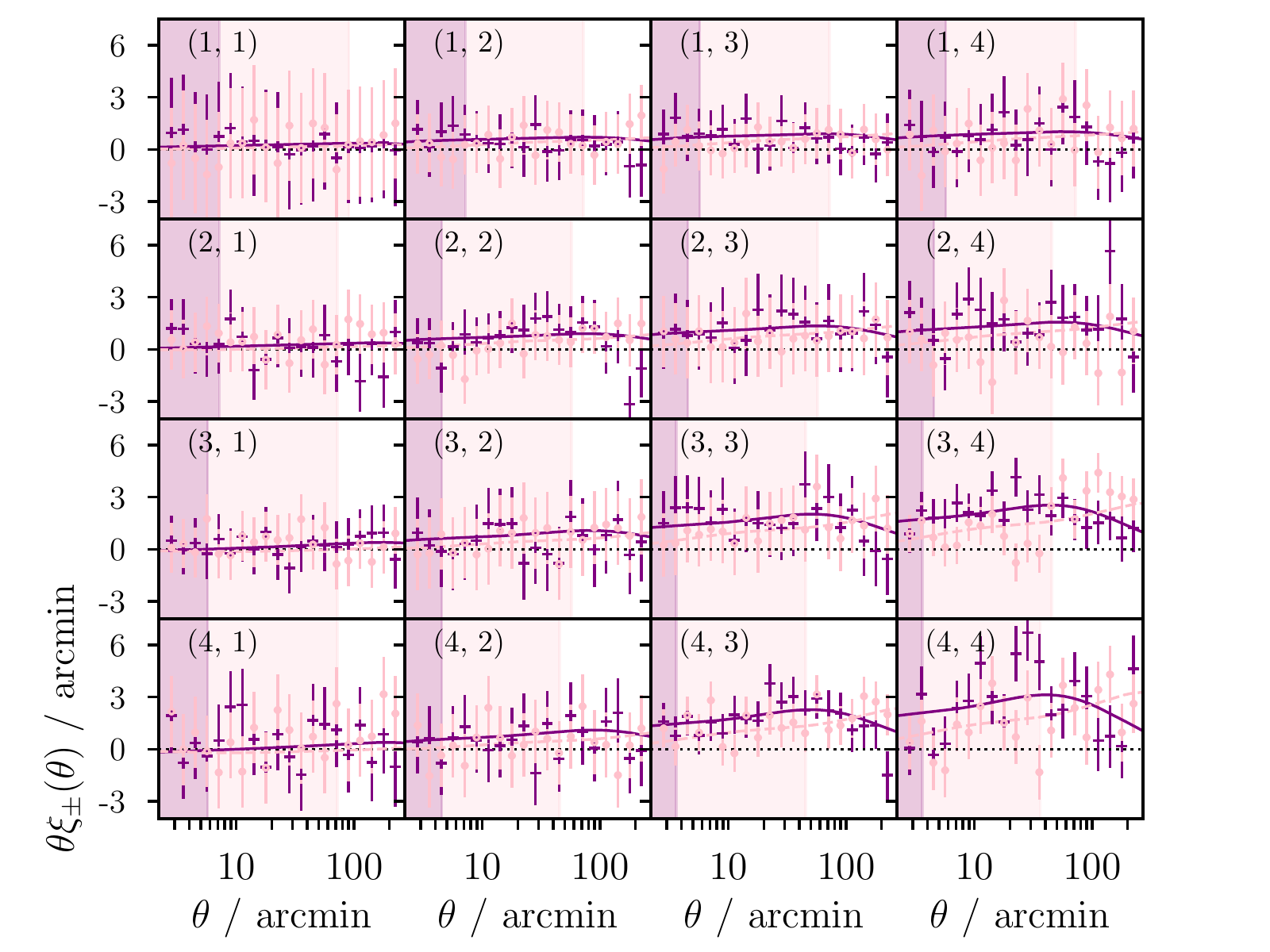}
\caption{Multicolour cross correlations, as measured from the Y1 early- and late-type samples described in this work.
Each panel shows the correlation between bin pair $(i,j)$, where $i\in(1,2,3,4)$ is the redshift bin index for the
early-type galaxies used in the correlation, and $j$ is that for late-types.
In each case purple crosses show $\xi_+$ and pink dots show $\xi_-$.
The shaded regions in the same colours indicate the scale cuts applied to each of these measurements.
}\label{fig:xipm-rbcross}
\end{figure}

The analysis of Section \ref{sec:results:crosscolour} includes a slight expansion of the total datavector.
That is, the multicolour datavector is larger than the union of the early- and late-type
datavectors due to cross-colour shear-shear correlations.
The measurements are performed using the same code pipeline with minimal modifications
to allow multiple colour bins to be handled simultaneously.
The resulting cross-correlations are shown in Figure \ref{fig:xipm-rbcross} 
($\xi_+$ in purple, $\xi_-$ in pink).

In order to obtain covariance matrices for the extended multicolour $3\times2$pt data,
we use the halo model implemented in \cosmolike (see Appendix A, \citealt{krause16}. 
In this scheme each element of the covariance matrix is evaluated as the sum of three contributions,

\begin{multline}
\mathrm{Cov}\left [ C^{ij}_{\alpha\beta}(\ell) C^{kl}_{\delta\gamma}(\ell')  \right ] =\\
\mathrm{Cov}^{\mathrm{G},ijkl}_{\alpha\beta\delta\gamma}(\ell,\ell') +
\mathrm{Cov}^{\mathrm{NG},ijkl}_{\alpha\beta\delta\gamma}(\ell,\ell') +
\mathrm{Cov}^{\mathrm{SSC},ijkl}_{\alpha\beta\delta\gamma}(\ell,\ell'),
\end{multline}

\noindent
where the upper Roman indices indicate redshift and colour bins and the lower Greek ones 
denote a particular observable probe.
The Gaussian piece is given by the standard expression

\begin{multline}
\mathrm{Cov}^{\mathrm{G},ijkl}_{\alpha\beta\delta\gamma}(\ell,\ell') = \frac{4 \pi \delta_{\ell\ell'}}{A (2\ell+1)\Delta\ell} \times \\
\left [ \tilde{C}^{ik}_{\alpha\delta}(\ell)\tilde{C}^{jl}_{\beta\gamma}(\ell) + \tilde{C}^{il}_{\alpha\gamma}(\ell)\tilde{C}^{jk}_{\beta\delta}(\ell) \right ].
\end{multline}

\noindent
This is sensitive to the total survey area $A$ and the spacing of discrete $\ell$ modes $\Delta \ell$.
The tilde indicates the sum of a cosmic variance term and a noise term, 
$\tilde{C}_{\mu\nu}^{ab} = C_{\mu\nu}^{ab} + N^{ab}_{\mu\nu}$.
The shot/shape noise is nonzero only for $\mu=\nu$ and $i=j$,
and is given by $N^{ij}_{\gamma\gamma}=\sigma_e^2/2n^{i}_g$
in the case of cosmic shear and $N^{ij}_{\delta_g\delta_g} = 1/n_g^{i}$ for galaxy clustering.
The first non Gaussian contribution is given by \citealt{krause16}'s equation A3
and is obtained by integrating the product of the trispectrum 
$T^{ijkl}_{\alpha\beta\gamma\delta}$ (sensitive to galaxy bias and cosmology)
and four lensing efficiencies (sensitive to the normalised redshift distributions)
over $\ell$ and $\chi$.
This term is, then, dependent on the shape of the reshift distributions, but not on the absolute
number densities.
Similarly, the super-sample covariance contribution is sensitive to the redshift distributions,
cosmology and the survey geometry, but is independent of the source number density.

Given the relatively high computational cost of generating the full multicolour 
non Gaussian covarance matrix we instead 
draw the non Gaussian and super sample covariance contributions
from the single colour matrices, as follows.
\cosmolike is first run to generate a base $2240\times2240$ Gaussian covariance matrix.
The non Gaussian contribution to an element $a,b$ is selected from:
\begin{itemize}
  \item{The early-type covariance matrix if $a,b$ is in blocks 
  $\mathrm{Cov}[ \xi^\mathrm{RR}_+, \xi^\mathrm{RR}_+]$,
  $\mathrm{Cov}[ \xi^\mathrm{RR}_-, \xi^\mathrm{RR}_-]$,
  $\mathrm{Cov}[ \xi^\mathrm{RR}_+, \xi^\mathrm{RR}_-]$,
  $\mathrm{Cov}[ \xi^\mathrm{RR}_+, \gamma^\mathrm{R}_t]$,
  $\mathrm{Cov}[ \xi^\mathrm{RR}_+, w]$ or
  $\mathrm{Cov}[ \gamma^\mathrm{R}_t, w]$ of the covariance matrix}
  \item{The late-type covariance matrix if $a,b$ is in any other block.}
\end{itemize}

\noindent
We emphasise here that this patching process is used only to obtain the (relatively small)
non Gaussian contribution to the multicolour covariance matrix;
the Gaussian part, which contains shot/shape noise terms is clearly sensitive to the
galaxy number densities, as so must be recomputed in full for the extended data vector. 

The validity of our approach thus relies on the qualitative similarity of the redshift distributions
of the two samples, such that the non Gaussian contributions for bin pair $ij$ are insensitive
to whether the samples being correlated are RR, BB or RB.
The resulting multiprobe matrices
include non Gaussian (up to trispectrum) contributions and supersample variance. 
We show the resulting correlation matrix in Figure \ref{fig:correlation_matrix}.
We test the impact of the approximation described above by rerunning our TATT parameter constraints
using the late-type part of the multicolour data vector only 
(i.e. applying cuts to remove all correlations involving early-type galaxies).
The results are compared with those from the single colour late-type data vector,
for which we generated the non Gaussian covariance matrix in full,
in Figure \ref{fig:cosmology_covtest}.
The purple solid and dark blue dashed contours show the constraints using the two non Gaussian
covariance matrices. For reference, the green dotted 
contour shows the impact of ignoring the non Gaussian contribution entirely.
We have chosen to show only the late-type sample here, given that its higher number density
means that it is less shape noise dominated, and so more sensitive to changes in the 
non Gaussian covariance contribution.

\begin{figure}\label{fig:correlation_matrix}
\includegraphics[width=\columnwidth]{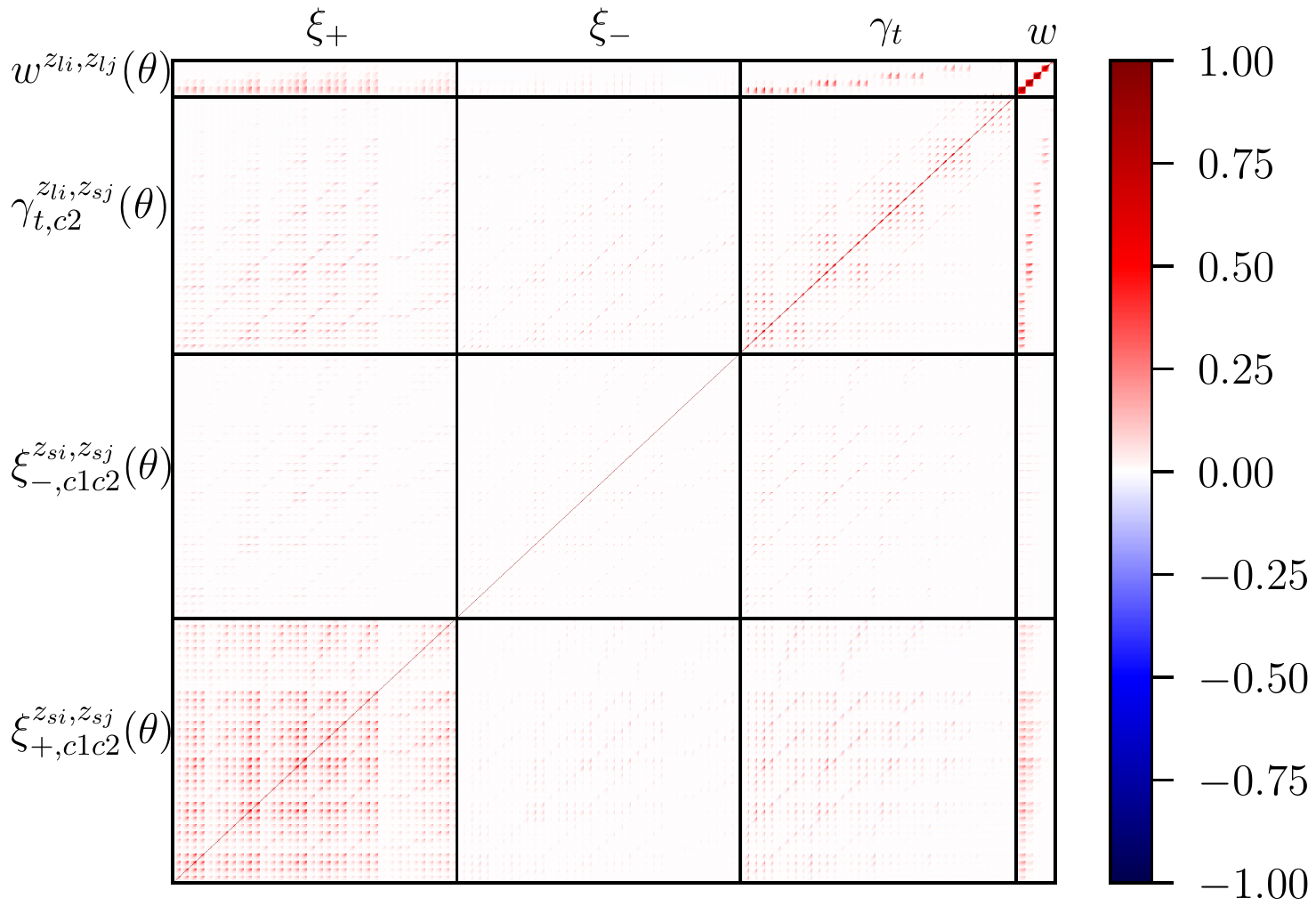}
\caption{Correlation matrix for the multiprobe multicolour datavector.
The matrix contains $2240\times2240$ elements, and includes shear $(\xi_\pm)$,
galaxy-galaxy lensing $(\gamma_t)$ and galaxy clustering $(w)$ measurements,
including all cross correlations between the various statistics.
Within each block, elements are grouped by $\theta$ bin, then by redshift bin couplet
and finally by galaxy type pairing.  
}\label{fig:correlation_matrix}
\end{figure}

\begin{figure}
\includegraphics[width=\columnwidth]{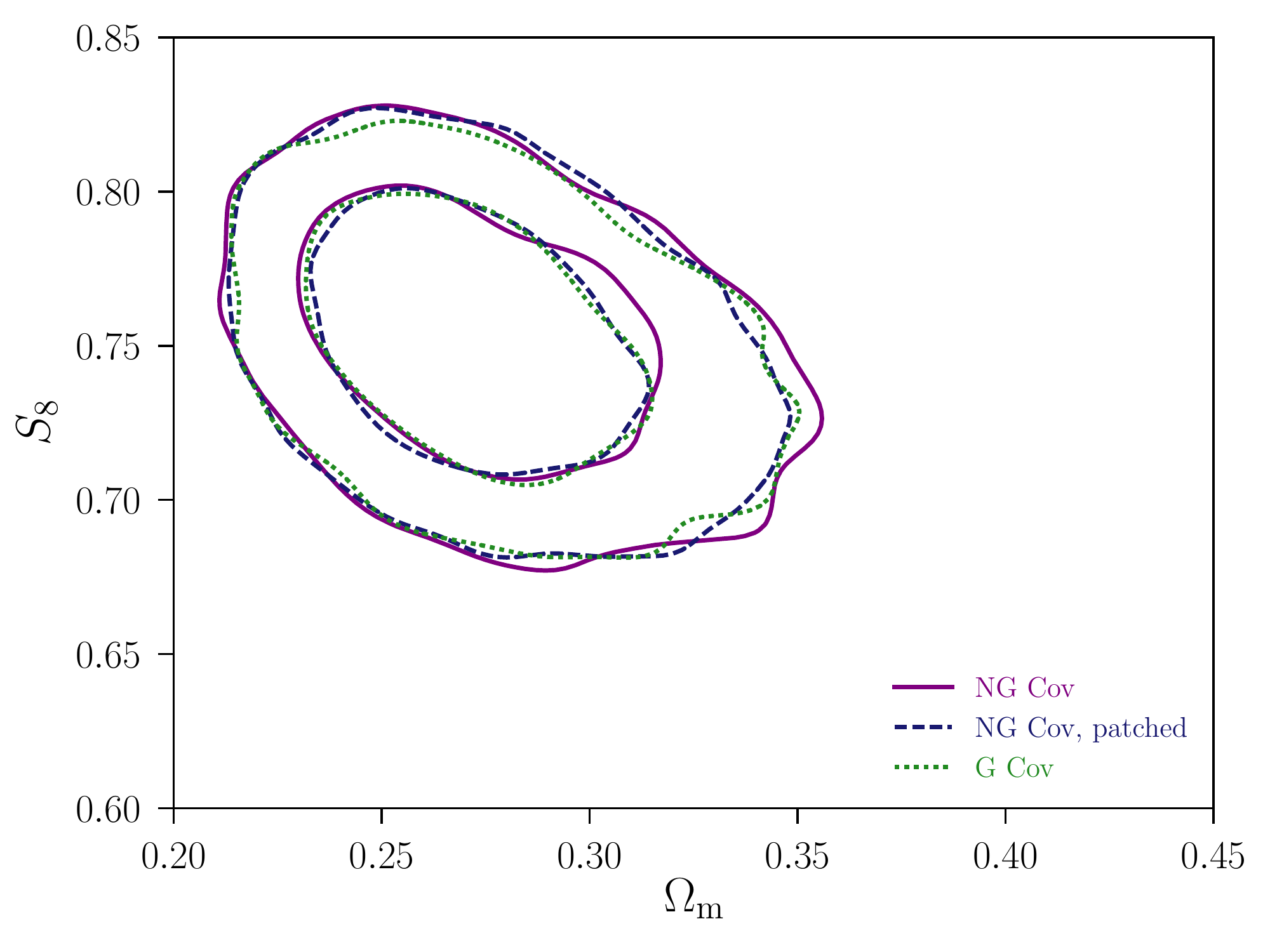}
\caption{Cosmology constraints from the late-type galaxy sample, assuming different approximations for the covariance matrix.
The solid purple line shows the $3\times2$pt constraints using a non Gaussian covariance matrix, obtained from \cosmolike~and
used in Sections \ref{sec:results:main}.
The dashed dark blue contours are obtained using the multicolour covariance matrix described in the text
and used in Section \ref{sec:results:crosscolour}
(a baseline Gaussian matrix plus non Gaussian contributions sourced from the single colour matrices).
Dotted green show the equivalent from the same data using a Gaussian covariance matrix 
(not used for any of the results presented in the main body of this paper, but included here for reference).
 }\label{fig:cosmology_covtest}
\end{figure}

Including these extra data and using the covariance matrix described above,
we obtain a reduced $\chi^2$ at the best-fitting cosmology of 1.11.
This compares with $849.1 / 774 = 1.10$ for the simultaneous early-late analysis without cross-correlations
(and with 1.36 and 1.23 respectively for the early- and late-type only TATT analyses).

The signal-to-noise of the combined $3\times2$pt multicolour data
(defined in terms of the datavector and covariance matrix
$\mathcal{S}=\left ( \sum_i \sum_j D_i C_{ij}^{-1} D_j^{T} \right ) ^{0.5}$
(see equation 15 in \citealt{chang18}) is 85.5.
Deconstructed into single two point statistic values, we obtain 30.7, 21.4 and 30.5
for $\xi_+$, $\xi_-$ and $\gamma_t$ respectively.
For early-type galaxies the corresponding values are 
15.9, 9.9 and 19.7
and for late-types we find 
20.2, 15.4 and 26.3.

\section*{Author Affiliations}
$^1$McWilliams Center for Cosmology, Department of Physics, Carnegie Mellon University, Pittsburgh, PA 15213, USA\\
$^2$Center for Cosmology and Astro-Particle Physics, The Ohio State University, Columbus, OH 43210, USA\\
$^3$Laboratory of Astrophysics, \'e Ecole Polytechnique F\'{e}d\'{e}rale de Lausanne (EPFL), Observatoire de Sauverny, 1290 Versoix, Switzerland\\
$^4$Department of Physics, The Ohio State University, Columbus, OH 43210, USA\\
$^5$Department of Astronomy/Steward Observatory, 933 North Cherry Avenue, Tucson, AZ 85721-0065, USA\\
$^6$Institut de F\'{i}sica d’Altes Energies (IFAE), The Barcelona Institute of Science and Technology, Campus UAB, 08193 Bellaterra (Barcelona) Spain\\
$^7$Kavli Institute for Particle Astrophysics \& Cosmology, P. O. Box 2450, Stanford University, Stanford, CA 94305, USA\\
$^8$SLAC National Accelerator Laboratory, Menlo Park, CA 94025, USA\\
$^{9}$Jet Propulsion Laboratory, California Institute of Technology, 4800 Oak Grove Dr., Pasadena, CA 91109, USA\\
$^{10}$ Department of Physics \& Astronomy, University College London, Gower Street, London, WC1E 6BT, UK\\
$^{11}$ Department of Physics, ETH Zurich, Wolfgang-Pauli-Strasse 16, CH-8093 Zurich, Switzerland\\
$^{12}$ Max Planck Institute for Extraterrestrial Physics, Giessenbachstrasse, 85748 Garching, Germany\\
$^{13}$ Universit\"ats-Sternwarte, Fakult\"at f\"ur Physik, Ludwig-Maximilians Universit\"at M\"unchen, Scheinerstr. 1, 81679 M\"unchen, Germany\\
$^{14}$Argonne National Laboratory, 9700 South Cass Avenue, Lemont, IL 60439, USA\\
$^{15}$Scottish Universities Physics Alliance, Institute for Astronomy, University of Edinburgh, Edinburgh EH9 3HJ, UK\\
$^{16}$ Cerro Tololo Inter-American Observatory, National Optical Astronomy Observatory, Casilla 603, La Serena, Chile\\
$^{17}$ Fermi National Accelerator Laboratory, P. O. Box 500, Batavia, IL 60510, USA\\
$^{18}$ Department of Physics and Astronomy, University of Pennsylvania, Philadelphia, PA 19104, USA\\
$^{19}$ CNRS, UMR 7095, Institut d'Astrophysique de Paris, F-75014, Paris, France\\
$^{20}$ Sorbonne Universit\'es, UPMC Univ Paris 06, UMR 7095, Institut d'Astrophysique de Paris, F-75014, Paris, France\\
$^{21}$ Jodrell Bank Centre for Astrophysics, School of Physics and Astronomy, University of Manchester, Oxford Road, Manchester, M13 9PL, UK\\
$^{22}$ Centro de Investigaciones Energ\'eticas, Medioambientales y Tecnol\'ogicas (CIEMAT), Madrid, Spain\\
$^{23}$ Laborat\'orio Interinstitucional de e-Astronomia - LIneA, Rua Gal. Jos\'e Cristino 77, Rio de Janeiro, RJ - 20921-400, Brazil\\
$^{24}$ Department of Astronomy, University of Illinois at Urbana-Champaign, 1002 W. Green Street, Urbana, IL 61801, USA\\
$^{25}$ National Center for Supercomputing Applications, 1205 West Clark St., Urbana, IL 61801, USA\\
$^{26}$ Institut d'Estudis Espacials de Catalunya (IEEC), 08034 Barcelona, Spain\\
$^{27}$ Institute of Space Sciences (ICE, CSIC),  Campus UAB, Carrer de Can Magrans, s/n,  08193 Barcelona, Spain\\
$^{28}$ Observat\'orio Nacional, Rua Gal. Jos\'e Cristino 77, Rio de Janeiro, RJ - 20921-400, Brazil\\
$^{29}$ George P. and Cynthia Woods Mitchell Institute for Fundamental Physics and Astronomy, and Department of Physics and Astronomy, Texas A\&M University, College Station, TX 77843,  USA\\
$^{30}$ Department of Physics, IIT Hyderabad, Kandi, Telangana 502285, India\\
$^{31}$ Excellence Cluster Universe, Boltzmannstr.\ 2, 85748 Garching, Germany\\
$^{32}$ Faculty of Physics, Ludwig-Maximilians-Universit\"at, Scheinerstr. 1, 81679 Munich, Germany\\
$^{33}$ Kavli Institute for Cosmological Physics, University of Chicago, Chicago, IL 60637, USA\\
$^{34}$ Instituto de Fisica Teorica UAM/CSIC, Universidad Autonoma de Madrid, 28049 Madrid, Spain\\
$^{35}$ Department of Astronomy, University of Michigan, Ann Arbor, MI 48109, USA\\
$^{36}$ Department of Physics, University of Michigan, Ann Arbor, MI 48109, USA\\
$^{37}$ Santa Cruz Institute for Particle Physics, Santa Cruz, CA 95064, USA\\
$^{38}$ Harvard-Smithsonian Center for Astrophysics, Cambridge, MA 02138, USA\\
$^{39}$ Australian Astronomical Optics, Macquarie University, North Ryde, NSW 2113, Australia\\
$^{40}$ Departamento de F\'isica Matem\'atica, Instituto de F\'isica, Universidade de S\~ao Paulo, CP 66318, S\~ao Paulo, SP, 05314-970, Brazil\\
$^{41}$ Department of Astronomy, The Ohio State University, Columbus, OH 43210, USA\\
$^{42}$ Department of Astrophysical Sciences, Princeton University, Peyton Hall, Princeton, NJ 08544, USA\\
$^{43}$ Instituci\'o Catalana de Recerca i Estudis Avan\c{c}ats, E-08010 Barcelona, Spain\\
$^{44}$ Brookhaven National Laboratory, Bldg 510, Upton, NY 11973, USA\\
$^{45}$ School of Physics and Astronomy, University of Southampton,  Southampton, SO17 1BJ, UK\\
$^{46}$ Instituto de F\'isica Gleb Wataghin, Universidade Estadual de Campinas, 13083-859, Campinas, SP, Brazil\\
$^{47}$ Computer Science and Mathematics Division, Oak Ridge National Laboratory, Oak Ridge, TN 37831\\
$^{48}$ Institute of Cosmology and Gravitation, University of Portsmouth, Portsmouth, PO1 3FX, UK\\
$^{49}$ Argonne National Laboratory, 9700 South Cass Avenue, Lemont, IL 60439, USA\\
$\dagger$ Einstein Fellow\\

\end{document}